\documentclass[a4paper,11pt]{article}
\pdfoutput=1 

\usepackage{jcappub} 

\usepackage[T1]{fontenc} 

\usepackage[latin9]{inputenc}
\usepackage{float}
\usepackage{amsmath}
\usepackage{amssymb}
\usepackage{graphicx}
\usepackage{esint}
 \usepackage{hyperref}
\usepackage{comment}
\usepackage{color}
\usepackage{microtype}
\usepackage{cleveref}
\usepackage{breakurl}
\usepackage{ulem}
\usepackage{bbm}

\makeatletter

\pdfpageheight\paperheight
\pdfpagewidth\paperwidth

\@ifundefined{textcolor}{}{%
 \definecolor{BLACK}{gray}{0}
 \definecolor{WHITE}{gray}{1}
 \definecolor{RED}{rgb}{1,0,0}
 \definecolor{GREEN}{rgb}{0,1,0}
 \definecolor{BLUE}{rgb}{0,0,1}
 \definecolor{CYAN}{cmyk}{1,0,0,0}
 \definecolor{MAGENTA}{cmyk}{0,1,0,0}
 \definecolor{YELLOW}{cmyk}{0,0,1,0}
}

\newcommand{\dd}{\mathrm{d}}
\newcommand{\mn}{{\mu\nu}}

\allowdisplaybreaks

\makeatother

\makeatletter
\DeclareRobustCommand{\rcite}[1]{%
  \rcite@aux#1,\@nil{#1}%
}
\def\rcite@aux#1,#2\@nil#3{%
  \if\relax#2\relax
    Ref.~\cite{#3}%
  \else
    Refs.~\cite{#3}%
  \fi
}
\makeatother

\hypersetup{
    colorlinks = true,
    citecolor = {red},
    linkcolor = {blue},
    urlcolor = {blue},
}

\title{On nonlocally interacting metrics, and a simple proposal for cosmic acceleration}

\author[a, b]{Valeri Vardanyan,}
\author[a]{Yashar Akrami,}
\author[c]{Luca Amendola,}
\author[a]{Alessandra Silvestri}

\affiliation[a]{Lorentz Institute for Theoretical Physics, Leiden University, PO Box 9506, Leiden 2300 RA, The Netherlands}
\affiliation[b]{Leiden Observatory, Leiden University, P.O. Box 9513, 2300 RA Leiden, The Netherlands}
\affiliation[c]{Institut f\"{u}r Theoretische Physik, Ruprecht-Karls-Universit\"{a}t Heidelberg,
Philosophenweg 16, 69120 Heidelberg, Germany}

\emailAdd{vardanyan@lorentz.leidenuniv.nl}
\emailAdd{akrami@lorentz.leidenuniv.nl}
\emailAdd{l.amendola@thphys.uni-heidelberg.de}
\emailAdd{silvestri@lorentz.leidenuniv.nl}

\abstract{
We propose a simple, nonlocal modification to general relativity (GR) on large scales, which provides a model of late-time cosmic acceleration in the absence of the cosmological constant and with the same number of free parameters as in standard cosmology. The model is motivated by adding to the gravity sector an extra spin-2 field interacting nonlocally with the physical metric coupled to matter. The form of the nonlocal interaction is inspired by the simplest form of the Deser-Woodard (DW) model, $\alpha R\frac{1}{\Box}R$, with one of the Ricci scalars being replaced by a constant $m^{2}$, and gravity is therefore modified in the infrared by adding a simple term of the form $m^2\frac{1}{\Box}R$ to the Einstein-Hilbert term. We study cosmic expansion histories, and demonstrate that the new model can provide background expansions consistent with observations if $m$ is of the order of the Hubble expansion rate today, in contrast to the simple DW model with no viable cosmology. The model is best fit by $w_0\sim-1.075$ and $w_a\sim0.045$. We also compare the cosmology of the model to that of Maggiore and Mancarella (MM), $m^2R\frac{1}{\Box^2}R$, and demonstrate that the viable cosmic histories follow the standard-model evolution more closely compared to the MM model.  We further demonstrate that the proposed model possesses the same number of physical degrees of freedom as in GR. Finally, we discuss the appearance of ghosts in the local formulation of the model, and argue that they are unphysical and harmless to the theory, keeping the physical degrees of freedom healthy.
}


\begin{document}
\maketitle

\section{Introduction}\label{sec:intro}

The question of why the late-time expansion of the Universe is accelerating is now almost twenty years old, with strong and overwhelming evidence supporting the phenomenon~\cite{Rubin:2016iqe} after its initial discovery through the observations of supernovae~\cite{Riess:1998cb,Perlmutter:1998np} (see Refs.~\cite{Caldwell:2009ix,Weinberg:2012es,Joyce:2014kja,Bull:2015stt} for recent reviews). The standard model of cosmology, $\Lambda$CDM ($\Lambda$ for the cosmological constant (CC), and CDM for cold dark matter), provides a strikingly successful description of cosmic acceleration in the arguably simplest possible way, i.e., through a single parameter, $\Lambda$. This observationally very well tested model, however, suffers from serious, theoretical issues stemming from a tremendous fine-tuning that is required for compatibility of the observed value of $\Lambda$ with the widely accepted principles of quantum field theory; see, e.g., Ref.~\cite{Martin:2012bt} for a review. Alongside the other problems with $\Lambda$CDM, different incarnations of the cosmological constant problem have been considered as strong motivations for exploring alternative cosmological models and, therefore, going beyond $\Lambda$CDM~\cite{Bull:2015stt}. One of the interesting such attempts consists of constructing alternative theories of gravity which would offer mechanisms for cosmic acceleration that are different from a simple cosmological constant in the framework of general relativity (GR); see, e.g., Refs.~\cite{2010deto.book.....A,Clifton:2011jh}. Such models must clearly be consistent with the tide of various high-quality cosmological data, at least as well as $\Lambda$CDM, be theoretically well defined and well motivated, (ideally) be simple (i.e. without introducing many free parameters), and offer predictions that are distinguishable from those of $\Lambda$CDM, making the models testable and falsifiable.

One class of interesting models of modified gravity that has attracted significant attention over the past few years is the one with the gravity sector extended by adding an extra rank-2 tensor field similar to the fundamental dynamical field of GR, the {\it metric}, describing interacting spin-2 fields. These commonly called {\it bimetric theories} have a long history, in connection to massive gravity where gravitons are assumed to possess a nonzero mass, contrary to GR which is the unique theory of massless gravitons~\cite{Gupta:1954zz,Weinberg:1965rz,Deser:1969wk,Boulware:1974sr,Feynman:1996kb}. After decades of intensive searches for theoretically consistent theories of nonlinear massive and bimetric gravity, it eventually became possible, after the discovery of {\it ghost-free} massive gravity~\cite{deRham:2010ik,deRham:2010kj,Hassan:2011vm,Hassan:2011hr,deRham:2011rn,deRham:2011qq,Hassan:2011tf,Hassan:2011ea,Hassan:2012qv,Hinterbichler:2012cn} and bigravity~\cite{Hassan:2011zd} (see Refs.~\cite{deRham:2014zqa,Hinterbichler:2011tt,Schmidt-May:2015vnx,Solomon:2015hja,Hinterbichler:2017sbd} for reviews), to explore the cosmological implications of such theories, in particular in connection to cosmic acceleration. It was quickly realized, after the successful construction of bigravity, that it admits Friedman-Lema\^{i}tre-Robertson-Walker (FLRW) cosmological solutions\footnote{See Ref.~\cite{Nersisyan:2015oha} and references therein for the cosmology of bimetric models with other choices of metrics.} which agree with observations at the background level, i.e. they successfully describe the cosmic expansion history even in the absence of an explicit cosmological constant (or vacuum energy) term~\cite{Volkov:2011an,Comelli:2011zm,vonStrauss:2011mq,Akrami:2012vf,Akrami:2013pna,Konnig:2013gxa,Enander:2014xga,Mortsell:2017fog}.\footnote{See also Ref.~\cite{Luben:2016lku} for viable background cosmologies of theories with more than two spin-2 fields.} It however turned out that the linear cosmological perturbations, investigated extensively in Refs.~\cite{Comelli:2012db,Khosravi:2012rk,Berg:2012kn,Konnig:2014dna,Solomon:2014dua,Konnig:2014xva,Lagos:2014lca,Cusin:2014psa,Yamashita:2014cra,DeFelice:2014nja,Fasiello:2013woa,Enander:2015vja,Amendola:2015tua,Johnson:2015tfa,Konnig:2015lfa,Lagos:2016gep}, suffer from either ghost or gradient instabilities.\footnote{Here, we have referred to theories where matter couples only to one of the two spin-2 fields, the {\it physical metric}, and the second metriclike field is considered only as an extra dynamical tensor field interacting with the metric. See Refs.~\cite{Hassan:2012wr,Akrami:2013ffa,Tamanini:2013xia,Akrami:2014lja,Yamashita:2014fga,deRham:2014naa,deRham:2014naa,Hassan:2014gta,Enander:2014xga,Solomon:2014iwa,Schmidt-May:2014xla,deRham:2014fha,Gumrukcuoglu:2014xba,Heisenberg:2014rka,Gumrukcuoglu:2015nua,Hinterbichler:2015yaa,Heisenberg:2015iqa,Heisenberg:2015wja,Lagos:2015sya,Melville:2015dba} for cases where matter couples to more than one metric directly, or to a composite metric.} Although a few potential ways out have been proposed (see, e.g., Refs.~\cite{Akrami:2015qga,Mortsell:2015exa}), it is still an open question whether any models of only {\it two} interacting spin-2 fields with self-accelerating solutions exist that are fully stable linearly {\it at all times}, and provide a standard {\it isotropic and homogeneous} background evolution. There have been various attempts at constructing interacting-metric theories with kinetic and/or interaction terms other than the ones in the original ghost-free nonlinear theories of massive and bimetric gravity (especially with derivative interactions)~\cite{Boulanger:2000rq,Hinterbichler:2013eza,Kimura:2013ika,deRham:2013tfa,Gao:2014jja,Noller:2014ioa,deRham:2015rxa,Li:2015izu,Li:2015iwc}, but it has proven difficult, if not impossible, to find such new terms that do not revive the so-called Boulware-Deser ghost~\cite{Boulware:1973my}. 

Another class of interesting alternative theories of gravity proposed as solutions to the cosmic acceleration problem is the one with nonlocal terms added to the Einstein-Hilbert term of GR action. No matter which definition we choose for general relativity, either the geometrical picture according to Lovelock's theorem~\cite{Lovelock1972} or the quantum-field-theoretical picture in terms of massless spin-2 fields (see Ref.~\cite{Hinterbichler:2011tt}), {\it locality} is one of the fundamental assumptions of GR. Clearly, one way of modifying GR is therefore to break this assumption. The appearance of nonlocal terms at low energies (infrared; IR) is a generic feature in effective
field theories when massless or light degrees of freedom are integrated out~\cite{Barvinsky:1987uw,Barvinsky:1990up,Barvinsky:1990uq,Barvinsky:1995jv,Barvinsky:2014lja,Codello:2015mba,Donoghue:2015nba,Maggiore:2016fbn,Maggiore:2016gpx}, and may also arise more fundamentally in Euclidean quantum
gravity \cite{Wetterich:1997bz,Barvinsky:2011rk}. Effective actions with IR nonlocal terms have also been found for theories of massive gravity~\cite{Jaccard:2013gla,Modesto:2013jea}, multimetric gravity~\cite{Cusin:2014zoa}, and post-Riemannian, affine geometry~\cite{Golovnev:2015bsa}. IR nonlocalities are usually modelled at the level of the action by adding terms that involve inverse differential operators, such as inverse d'Alembertian $\frac{1}{\Box}$ (or $\Box^{-1}$), which in Fourier space can be considered as a Feynman propagator describing the effects of the integrated-out fields. Such operators modify gravitational interactions at large temporal and spatial scales, and can therefore provide dynamical mechanisms for cosmic acceleration. Another important motivation behind these types of nonlocal modifications in the IR stems from the observation~\cite{ArkaniHamed:2002fu,Dvali:2007kt} that such operators could provide an appealing solution to the {\it old} cosmological constant problem by {\it degravitating} a large vacuum energy. Even though no consistent theory of degravitation has been found yet in this context and at the level of the action, degravitation remains an important and inspiring motivation for nonlocal modifications to gravity in the IR. And finally, the fact that many of the no-go theorems for gravity rely on the locality of the action is another motivation to relax this condition. This then opens up a large number of new possibilities for model-building. It is therefore important to try to construct simple forms of nonlocal actions and study their implications in different regimes.

The first nonlocal mechanism for cosmic acceleration was proposed by Deser and Woodard (DW)~\cite{Deser:2007jk} (see Ref.~\cite{Woodard:2014iga} for a review),\footnote{Strictly speaking, the cosmology of nonlocal models was proposed and studied first in Ref.~\cite{Wetterich:1997bz} for models with similar structures, although with no connection to cosmic acceleration, which was not yet discovered at the time.} where a simple term of the form $Rf(\frac{1}{\Box}R)$ was added to the standard Einstein-Hilbert term in GR. The function $f$ can have any arbitrary form at the phenomenological level. One very interesting feature of this model is that it does not introduce any new mass scale in the gravity sector, contrary to $\Lambda$CDM where $\Lambda$ is a dimensionful quantity with an observed value far smaller than the other scale in the theory, i.e. the Planck mass, leading to an enormous, {\it unnatural} hierarchy that requires extreme fine-tuning. The absence of such a new scale in the DW theory is therefore a highly appealing feature, if the theory would be able to explain the late-time acceleration in a way consistent with observations and without extremely fine-tuned dimensionless parameters. In addition, the model has been proven to not add extra excitations to gravity beyond those of general relativity, i.e. the number of physical degrees of freedom are the same as in GR~\cite{Deser:2013uya}. Unfortunately, though, the simplest form of the function $f$, i.e. $\alpha\frac{1}{\Box}R$, with $\alpha$ a dimensionless free parameter, does not provide viable cosmic histories, i.e. even at the cosmological background level~\cite{Koivisto:2008xfa} (see also Ref.~\cite{Koivisto:2008dh}). This particular form of $f$ is interesting not only because of its simplicity, but also because the localized formulation of the theory requires only one additional scalar field which is not a ghost; theories with other forms of the function $f$ introduce two scalar fields, one of which is a ghost. Such ghosts have however been argued to be harmless as the fields are only auxiliary and do not add to and do not affect the physical degrees of freedom (which are the same as in GR) by converting them to ghosts~\cite{Deser:2013uya}. Accepting more complicated forms of $f$, one can show that it is indeed possible to phenomenologically tune it such that any cosmic history can be reconstructed, even an exact $\Lambda$CDM background~\cite{Koivisto:2008xfa,Deffayet:2009ca}. The $\Lambda$CDM-equivalent form of the function has however turned out to be highly contrived with several free parameters that need to be fixed observationally, making the model less appealing. The model with the reconstructed $\Lambda$CDM background has been further investigated by studying linear perturbations and structure formation~\cite{Park:2012cp,Dodelson:2013sma,Park:2016jym,Nersisyan:2017mgj}, and even though it was originally claimed~\cite{Dodelson:2013sma} that the model was strongly ruled out observationally, a counter-claim has recently been made~\cite{Nersisyan:2017mgj} stating that not only is the model consistent with data, it even gives a better fit than $\Lambda$CDM. The origin of the disagreement is not yet known, but there are reasons to believe that it may be related to the framework in which the analysis has been done; the former performs all the calculations in the nonlocal formulation of the theory, while the latter studies the model in its local formulation~\cite{Jhingan:2008ym}.\footnote{In this case, the model contains two auxiliary, scalar fields, and its perturbative analysis resembles that of multi-scalar-tensor theories (see Ref.~\cite{Vardanyan:2015oha}, and references therein).} Whether or not the DW model provides an observationally viable model for cosmic evolution, the fact that it requires a highly contrived and {\it ad hoc} form of the function $f$ with several free parameters renders it both theoretically and phenomenologically difficult to accept as an interesting alternative to $\Lambda$CDM.

Independently of the DW theory, Maggiore and Mancarella (MM) proposed~\cite{Maggiore:2014sia} an alternative model of nonlocal gravity in an attempt to explain cosmic acceleration without a cosmological constant. In the MM model, a term of the form $m^2R\frac{1}{\Box^2}R$ has been added to the Einstein-Hilbert term. Such a model, contrary to the DW model, requires the introduction of a new, {\it fine-tuned}, mass scale into the gravity sector, and in addition, even the simple form of the theory with no complicated function introduced requires two scalar fields in order to localize the theory, one of which is inevitably a ghost. It is however argued, similarly to the DW case, that the presence of the ghost in the theory is not dangerous as the ghostly scalar field is only an auxiliary one with no effects on the physical degrees of freedom of the nonlocal theory, which are the same as in general relativity~\cite{Dirian:2014xoa}. This harmlessness of the ghost is guaranteed by fixing the initial conditions of the auxiliary fields such that the localized theory becomes equivalent to the original nonlocal MM theory. One can therefore work with the localized formulation as long as the initial conditions are properly chosen, and therefore the localized theory is used only as a {\it mathematical trick} for dealing with computations which otherwise prove difficult in the nonlocal formulation. Although the equivalence of the local and nonlocal formulations has been shown at the cosmological background level~\cite{Dirian:2014xoa}, it is not fully clear whether it remains so also at the level of the perturbations, as there are already reasons to suspect it to be the case by considering the similar model of DW with the seemingly discrepant results in the perturbative analyses of Refs.~\cite{Dodelson:2013sma} and~\cite{Nersisyan:2017mgj}. In addition, one needs to be cautious when trying to quantize the model in the localized form, as if the constraints on the auxiliary fields are not properly taken into account in the quantization procedure, the ghost will render the theory unviable~\cite{Dirian:2014xoa,Foffa:2013sma}. The cosmological implications of the MM model has been extensively studied and the model has proven to provide cosmic histories, as well as structure formation, consistent with observational data, although being different from that of $\Lambda$CDM~\cite{Maggiore:2014sia,Dirian:2014ara,Barreira:2014kra,Dirian:2014bma,Codello:2015pga,Dirian:2016puz,Nersisyan:2016hjh}. For that, the new mass scale of the model $m$ has to be of a similar order of magnitude as the present value of the Hubble parameter, similarly to the CC term in $\Lambda$CDM. This phenomenologically favored value has however been argued to have not emerged from perturbative quantum loop corrections due to integrating out light fields, and a more complex mechanism must be behind the generation of the nonlocalilty of the MM form~\cite{Maggiore:2016fbn}.

It should be noted that other nonlocal models with structures similar to those of the DW and MM models have also been proposed, where the nonlocal distortion term is built out of the Ricci scalar and exponential functions of $\Box^{-1}$~\cite{Conroy:2014eja}, or tensorial objects, such as the Ricci tensor $R_{\mu\nu}$ or Riemann tensor $R_{\mu\nu\alpha\beta}$, rather than the Ricci scalar $R$. Tensorial nonlocalities are theoretically very interesting, as they could, for example, help alleviate the ultraviolet divergences of GR by modifying the graviton propagator~\cite{Biswas:2011ar}, or help implement a consistent degravitation mechanism (see, e.g., Refs.~\cite{Dvali:2007kt,deRham:2007rw} for such attempts in the framework of massive gravity), which is not possible through only the introduction of scalar nonlocalities. Unfortunately, though, tensorial nonlocalities have turned out to generically contain rapidly growing modes that prevent them from providing stable background expansions~\cite{Ferreira:2013tqn,Cusin:2015rex,Tsamis:2014hra,Nersisyan:2016jta}.

Summarizing all the possibilities mentioned above, one can write down the most general nonlocal action, quadratic in the curvature invariants, as~\cite{Biswas:2011ar,Nersisyan:2016jta}
\begin{equation}
S=\frac{M_{\rm Pl}^2}{2}\int\text{d}^{4}x\sqrt{-g}\left(R+Rh_1(\triangle)R+
R^{\alpha\beta}h_2(\triangle)R_{\alpha\beta}+R^{\mu\nu\alpha\beta}h_3(\triangle)
R_{\mu\nu\alpha\beta}\right)+S_{\text{matter}}\,,\label{eq:NLAction}
\end{equation}
where $\triangle$ is some differential operator, usually assumed to be the operator $\Box$, $M_{\rm Pl}$ is the reduced Planck mass, and $S_{\text{matter}}$ is the matter action. $h_1$, $h_2$ and $h_3$ are arbitrary functions of the operator $\triangle$ involving negative powers of $\triangle$.\footnote{Note that the functions can involve positive powers, in which case the nonlocal modifications to GR are in the ultraviolet. See Refs.~\cite{Talaganis:2016ovm,Modesto:2016max} for recent progress in the construction of ghost-free, ultraviolet, nonlocal gravity.} For $\triangle=\Box$, action (\ref{eq:NLAction}) is the most general parity-invariant quadratic curvature action.\footnote{This action contains the MM model with the choice of $h_1(\Box)=m^2\frac{1}{\Box^2}$ and the simplest DW model with $h_1(\Box)=\alpha\frac{1}{\Box}$. Although the DW models with sophisticated forms of the function $f(\frac{R}{\Box})$ are not included in action (\ref{eq:NLAction}), we do not consider them here, as we choose to keep the structure of the theory as simple as possible.} As discussed above, unfortunately a large (and the most interesting) part of the action has been proven to be problematic, for one reason or another. Although specific forms of the terms are phenomenologically viable and theoretically consistent, such as the MM model of $m^2R\Box^{-2}R$, these are only in the scalar sector, and for only restricted forms of the free function $h_1$. It is therefore natural, and important, to ask whether there are ways to modify the action and expand its viability to more (and specifically tensorial) models, while keeping the structure of the action and all its interesting features intact.
\vspace{0 mm}\\
\indent In this paper, we propose one way to do this, in the framework of bimetric theories of gravity, where the above action is modified by only assuming that one of the two curvature quantities in each pair of $R$, $R_{\mu\nu}$, and $R_{\mu\nu\alpha\beta}$ in the nonlocal terms corresponds to an extra spin-2 field $f_{\mu\nu}$. This is in a sense a minimal modification, as the structure of the action is kept almost the same as the original one, and only the field content is changed through a minimal, single, additional tensor field that accounts, in a unified way, for the scalar and tensorial structures required for the curvature quantities in all the terms. Such models are natural to construct in the framework of interacting spin-2 theories, and they, therefore, reside at the interface of bimetric and nonlocal theories. From the point of view of bimetric theories, given the stringent constraints on the form of possible, consistent, {\it local} interactions between two metrics (or, more correctly, a metric and an extra spin-2 field), as we discussed above, here it is natural to ask whether new consistent interactions are possible if we relax the locality condition in constructing theories of gravity. There are various no-go theorems about interacting spin-2 fields, massive or massless, but in all of those theorems locality has been assumed, one way or another. There have already been studies in the literature, see e.g. Refs.~\cite{Taronna:2011kt,Porrati:2012rd}, where there are hints that such no-go theorems could be evaded by including nonlocal effects. The aim of the present paper is however not to study such no-go theorems in detail and to investigate whether (and how) they can be evaded in nonlocal models. We only take a phenomenological approach here and simply try to study the cosmological implications of a simple, phenomenological model of nonlocally interacting spin-2 fields. We hope though that our work will trigger more theoretical work in the future on such possibilities, where the no-go theorems and the mathematical consistency will be studied rigorously for such theories. Although perhaps the most interesting class of nonlocal models constructed this way are the ones with tensorial terms in the action, as we explained above, we start our investigation of this new direction by only considering the interaction terms that contain only the Ricci scalars, i.e. the terms corresponding to the function $h_1$ in the action (\ref{eq:NLAction}), and devote the present paper to only this sector of the full action. We see this as only a first and the simplest step towards the construction of theories of nonlocally interacting metrics with potentially very interesting implications, and leave a detailed investigation of the tensorial terms for future work.
\vspace{0 mm}\\
\indent We start the paper with investigating a simple model in which two metrics interact nonlocally through their Ricci scalars only, while each has its own Einstein-Hilbert kinetic term. The structure of the nonlocal interaction is inspired by arguably one of the simplest possible nonlocal terms that one could construct out of the curvature $R$ and the operator $\Box$, i.e., the simplest version of the DW model, $\alpha R\frac{1}{\Box}R$, with all its interesting properties mentioned above. We derive all the field equations for the model, and study the implications of the Bianchi constraints and the conservation of the matter stress-energy tensor. We show that the consistency of the field equations and Bianchi conditions places strong constraints on the properties of the extra spin-2 field in the simplest model considered in this paper, suggesting, strongly, that tensorial interactions must also be added if the theory is to be considered as a bimetric setup with the tensorial properties of an extra spin-2 field involved. That being said, the two-metric theory that we start with leads us to a new, {\it single parameter}, {\it single-metric}, nonlocal modification to general relativity as another alternative to $\Lambda$CDM, which provides a simple mechanism for cosmic acceleration. The model adds a term of the form $m^2\frac{1}{\Box}R$ to GR, and although it is inspired by our first attempt at constructing a nonlocal model of interacting metrics, we can consider it as a purely phenomenological model that could originate from other theoretical frameworks. We show that, contrary to the DW model with also the simple, nonlocal operator $\frac{1}{\Box}$ in its structure, ours provides viable cosmological expansion histories in its simple form. One interesting feature of the model is that, similarly to the DW and MM cases, the nonlocal interaction in the model does not add any new physical degrees of freedom to the noninteracting theory. In addition, even though our model now needs two auxiliary fields to be localized (as opposed to the DW $\alpha R\frac{1}{\Box}R$ model which needs one) and one of the fields is a ghost, they do not affect the physical degrees of freedom by converting them to ghosts. In this respect, our model behaves similarly to the MM nonlocal model of $m^2R\frac{1}{\Box^2}R$. The model, although resembling the DW model $\alpha R\frac{1}{\Box}R$ with the simplest form of the function $f$, where the quantity $\alpha R$ is replaced by a constant $m^2$, has a very different phenomenology, consistent with the observed cosmic evolution while avoiding issues such as sudden future singularities present in the DW models~\cite{Koivisto:2008xfa}. In this respect, the model is more appealing than the DW models where the $f$ function is constructed in a contrived way, with several free parameters, to describe the cosmic evolution; our model introduces only one free parameter just as in $\Lambda$CDM. In comparison to the MM model, on the other hand, it is arguably a simpler model, as it includes the operator $\frac{1}{\Box}$, rather than $\frac{1}{\Box^2}$. Even though the localized formulation of the model requires two auxiliary fields, just as in the MM model, we show that it provides a different cosmic evolution, still different from that of $\Lambda$CDM. It is easier to see the connection of our model to both DW and MM models if we write the nonlocal term in each case in terms of the quantity $X\equiv\frac{1}{\Box}R$. In that case, the ``viable" DW model is of the form $Rf(X)$ with $f(X)=a_1[\tanh(a_2(X+a_5)+a_3(X+a_5)^2+a_4(X+a_5)^3)-1]$, where $a_1,...,a_5$ are free parameters to be set by observations~\cite{Koivisto:2008xfa,Deffayet:2009ca}, and the MM model is of the form $m^2X^2$. The model proposed in this paper is then of the simpler form $m^2X$. We show that this simple model is able to provide a cosmic history that is consistent with the observed one while being different from both MM and $\Lambda$CDM models. In the present paper, we study only the background evolution of the Universe, and leave the investigation of perturbations for future work.
\vspace{0 mm}\\
\indent This paper is organized as follows. In Sec.~\ref{sec:model}, we start our investigation of a nonlocal model with two metrics interacting through their Ricci scalars by first presenting the action. We derive the gravitational field equations in Sec.~\ref{sec:NLeom} in the original, nonlocal formulation. We then localize the theory in Sec.~\ref{sec:localization} by introducing two auxiliary fields, and write down the action, as well as the field equations, in their local forms. Bianchi constraints for both metrics (without assuming any specific forms) are presented in Sec.~\ref{sec:Bianchi}, which will be shown to place a strong constraint on the structure of the model, implying that the reference metric (or the second spin-2 field of the model) should have a constant scalar curvature, both spatially and temporally. This leads us to the simple and phenomenologically interesting, single metric, $m^2\frac{1}{\Box}R$ model, which we introduce in Sec.~\ref{sec:m2model}. After presenting all equations of motion, we begin our investigation of the cosmological implications of the model in Sec.~\ref{sec:cosmology}, by first deriving in Sec.~\ref{sec:background} the general evolution equations for background dynamics of the Universe, i.e. the equations equivalent to Friedmann equations in $\Lambda$CDM. We then discuss in Sec.~\ref{sec:acceleration} the ability of the model in providing cosmic histories and expansion evolutions consistent with observations, in particular how cosmic acceleration can be obtained with no need for a cosmological constant term. We also compare the implications of the model to those of $\Lambda$CDM and Maggiore-Mancarella models, and in Sec.~\ref{sec:DW}, we further compare its cosmology to that of the Deser-Woodard $\alpha R\frac{1}{\Box}R$ model. In Sec.~\ref{sec:solutions4f}, we discuss the cosmological solutions for the original, two-metric model, focusing on the existence of background solutions for the reference metric, taking into account the implications of the Bianchi constraint. After presenting the cosmology of our model, we discuss in Sec.~\ref{sec:ghostsdof} the appearance of ghosts in the local formulation of the model, and argue that, like in other nonlocal models, such ghosts are of no harm to the single-metric, nonlocal, $m^2\frac{1}{\Box}R$ model. In particular, we prove explicitly in Sec.~\ref{sec:ghosts} that the local formulation contains a ghost, but we show in Sec.~\ref{sec:dof}, by analyzing the model in the nonlocal formulation, that the number of physical degrees of freedom is the same as in GR, and their healthiness is not affected by the presence of the nonlocal terms. We also discuss in Sec.~\ref{sec:ghostsdof} the number of degrees of freedom and the issue of ghosts in the nonlocaly-interacting-metric model, and argue that these are more subtle in that case compared to the single-metric, $m^2\frac{1}{\Box}R$ model. We argue that this two-metric model possesses 2+2 degrees of freedom linearly and around the cosmological solutions studied in the present work, and it may be the case that the full, nonlinear model contains more degrees of freedom, implying the existence of ghosts in the theory or that it is infinitely strongly coupled around cosmological backgrounds, which may be considered as a major issue for the model. Finally, our conclusions and some discussions are presented in Sec.~\ref{sec:conclusions} with suggestions for future work. Appendix~\ref{sec:appendix} briefly discusses some generalizations of the model, and presents the ghost-free condition in such models.

\section{Nonlocally interacting spin-2 fields and $\alpha R_{f}\frac{1}{\Box}R$ term}
\label{sec:model}

As described above, our goal is to construct a model of gravity where two metrics (or spin-2 fields) interact nonlocally. This inevitably means that the main ingredients of our model should be two metrics that we call $g_{\mu\nu}$ and $f_{\mu\nu}$. The {\it action} of the model should then generically include three main pieces, as in other gravity models: {\it kinetic} terms for the metrics, {\it interaction} terms between the two, and their {\it couplings} to matter. Adhering to the {\it standard} recipe for constructing gravity theories, and in regard to our discussion in Sec.~\ref{sec:intro} on only considering scalar interactions in this paper, we start building our model using three (simplest possible) types of ingredients, i.e. the Ricci scalars $R_{g}$ and $R_{f}$ for $g_{\mu\nu}$ and $f_{\mu\nu}$, respectively, the volume elements $\dd^4x\sqrt{-g}$ and $\dd^4x\sqrt{-f}$ with $g$ and $f$ being determinants of the two metrics, and the nonlocal operators $\Box_{g}^{-1}$ and $\Box_{f}^{-1}$, where $\Box_{g}$ and $\Box_{f}$ are the d'Alembertian operators corresponding to $g_{\mu\nu}$ and $f_{\mu\nu}$, respectively. With these elements, we suggest an action of the form
\begin{equation}\label{action_non_local}
S=\frac{M_{g}^{2}}{2}\int \dd^{4}x\sqrt{-g}R_{g}+\frac{M_{f}^{2}}{2}\int \dd^{4}x\sqrt{-f}R_{f}-I_{g,f}+S_{\text{matter}}[g,\Psi],
\end{equation}
where the first two terms are the Einstein-Hilbert terms, in charge of giving dynamics to the metrics, and $I_{g,f}$ is the interaction term. Before we introduce the form of $I_{g,f}$, let us discuss our choices for the $g_{\mu\nu}$ and $f_{\mu\nu}$ kinetic terms and the matter action. First of all, following the common recipe in constructing bimetric theories, we assume that $g_{\mu\nu}$ is the {\it physical metric}, which couples to matter and is used for measuring distances and time intervals, and that the {\it reference metric} $f_{\mu\nu}$ is only an extra spin-2 field which interacts directly only with $g_{\mu\nu}$ and not with matter. That is why the matter action involves only $g_{\mu\nu}$ and matter fields (collectively denoted by $\Psi$). In addition, we have assumed the standard Einstein-Hilbert form for the kinetic terms of $g_{\mu\nu}$ and $f_{\mu\nu}$ with the volume elements $\dd^4x\sqrt{-g}$ and $\dd^4x\sqrt{-f}$, respectively. There are various reasons for considering such terms, which follow the arguments in the literature for the consistency and healthiness of kinetic terms~\cite{Boulanger:2000rq,Hassan:2011zd,deRham:2013tfa,Noller:2014ioa,deRham:2015rxa}. $M_g$ and $M_f$ are the two (reduced) Planck masses corresponding to $g_{\mu\nu}$ and $f_{\mu\nu}$, respectively, and since we have coupled only $g_{\mu\nu}$ to matter, we assume that $M_g$ is equivalent to the standard Planck mass in GR, i.e. $M_{\text{Pl}}$.\footnote{We show later in Sec.~\ref{sec:localization} that $M_{f}$ is redundant and is not a free parameter of the model.}

Let us now turn to the interaction term $I_{g,f}$. Following our guiding principle of building the simplest possible nonlocal terms for the interaction of the two metrics using the ingredients mentioned above, we construct the $I_{g,f}$ term out of structures of the form $R_{f,g}\frac{1}{\Box_{f,g}}R_{f,g}$. These terms resemble the simplest form of the DW nonlocal theory, and are the simplest, scalar, nonlocal interactions one could construct along the lines of the structure of the action (\ref{eq:NLAction}). We emphasize here again that although the interaction terms involving tensorial quantities $R_{\mu\nu}$ and $R_{\mu\nu\alpha\beta}$ are more interesting and more natural to add for the effects of the reference metric as a tensor field to be fully present, we restrict our investigation in the present paper to only the scalar sector, and leave the analysis of the tensorial nonlocalities for future work. That being said, even the set of simple operators chosen here gives us several possibilities, depending on which combinations of $R_{f,g}$ and $\Box_{f,g}$ we choose. Clearly, one possibility is to include all the terms at this phenomenological level, however since we are interested in the simplest possible model with the least number of free parameters, here we pick only a specific subset of the operators based on some phenomenological reasons.

As discussed above, nonlocal infrared modifications are believed to generically emerge from integrating out light degrees of freedom. For example, if one considers a scalar field with a canonical kinetic term and a nonminimal coupling to gravity, nonlocal structures of the DW and MM forms appear in the effective theory after integrating out the scalar field (see, e.g., Ref.~\cite{Maggiore:2016gpx}). In order to generate nonlocal terms involving the Ricci scalars of both $g_{\mu\nu}$ and $f_{\mu\nu}$, we can assume that a light scalar field $\varphi$ couples nonminimally to both metrics, giving rise to terms of the above structure when the scalar field is integrated out. In addition, since $g_{\mu\nu}$ is the spin-2 field coupled to matter, it seems more natural to assume that the kinetic term of the scalar field is defined through $g_{\mu\nu}$, i.e. is of the form $g^{\mu\nu}\nabla^{(g)}_{\mu}\varphi\nabla^{(g)}_{\nu}\varphi$. Integrating this field out then leads to the appearance of the inverse of the $\Box_g$ operator, rather than $\Box_f$. Independently of this particular way of generating nonlocalities, i.e. through integrating out light fields, this choice can also be motivated purely phenomenologically by noticing again that $g_{\mu\nu}$ is the physical metric, suggesting the appearance of differential operators in the action in terms of $g_{\mu\nu}$ rather than $f_{\mu\nu}$.\footnote{One may argue that since $g_{\mu\nu}$ is the ``physical
metric,'' it is more natural to trace $(R_{f})_{\mu\nu}$ with $g^{\mu\nu}$ instead of $f^{\mu\nu}$, and therefore construct the action with terms like $(R_{f})_{\mu\nu}g^{\mu\nu}$ instead of $R_{f}$. The argument is based on the fact that the Ricci tensor is the most basic geometrical object as it contains only the connection and not the metric directly. This is certainly a possibility, and perhaps even more natural, but here we have chosen to work with $R$ and $R_{f}$ in the present work purely in order to keep the structure of the gravity sector of the theory as symmetric as possible in terms of the two metrics $g_{\mu\nu}$ and $f_{\mu\nu}$; this clearly need not be the case, and we leave the investigation of such possibilities for future work.' Similarly, we have assumed that both connections of the metrics are torsion-free, which again is only a simplicity assumption. One should however keep in mind that since here there is a curvature interaction between the two tensors, this is no longer necessarily the natural choice.} This then eliminates half of the possibilities for the nonlocal interaction terms. Keeping only the terms that involve both $R_{g}$ and $R_{f}$ (for explicit interactions between the two metrics), we are left with the two terms $R_{g}\frac{1}{\Box_{g}}R_{f}$ and $R_{f}\frac{1}{\Box_{g}}R_{g}$. These terms are equivalent through ``integration by parts"\footnote{Note that the variations made at the level of the action treat all Green's functions as some formal Green's functions, without yet specifying whether they are retarded or advanced. This choice comes either in {\it in-in} computations, or, by choosing the retarded boundary conditions at the end of the calculations (i.e. at the level of the equations of motion)~\cite{Woodard:2014iga}. Then, as explained in detail in e.g. Ref.~\cite{Mitsou:2015yfa}, all $\Box^{-1}$ occurrences inside a nonlocal action should be treated formal, i.e. undetermined linear inverses of $\Box$. When the equations of motion are computed, all the $\Box^{-1}$ should be turned into retarded ones by hand. One implication of treating all the $\Box^{-1}$ as equivalent during the variation is that one can effectively {\it integrate} $\Box^{-1}$ {\it by parts}, as follows:
\begin{align}
\int \dd^D x \phi(x) \Box^{-1} \psi(x) & \equiv \int \dd^D x \dd^D y \phi(x) G(x,y) \psi(y) \nonumber \\
& = \int \dd^D x \dd^D y \psi(y) G^T(y,x) \phi(x) \nonumber  \\
& = \int \dd^D y \psi(y) (\Box^{-1})^T \phi(y) \nonumber  \\
& \equiv \int \dd^D y \psi(y) \Box^{-1} \phi(y) , 
\end{align}
since the transposed $(\Box^{-1})^T$ is also a right-inverse $\Box (\Box^{-1})^T = {\rm id}$~\cite{Mitsou:2015yfa}. Here $D$ is the number of dimensions, $\phi$ and $\psi$ are arbitrary fields, and $G(x,y)$ is the Green's function appearing in the equations of motion.} if the boundary terms are assumed to vanish,\footnote{Looking at the integration procedure given in the previous footnote, we notice that going from the left-hand side of the first line to the right-hand side we have assumed that the homogeneous solution is vanishing, which corresponds to the minimal choice of the boundary conditions as we will discuss later. If we did not discard this homogeneous solution, we would have the homogeneous solution added to the entire $y$-integral, still within the $x$-integral.} and we can therefore construct our action with only one of them and without loss of generality.\footnote{Note, however, that this choice introduces a subtlety in the model when dealing with nonzero initial conditions. Unless explicitly stated, we assume vanishing boundary conditions everywhere in this paper, which is consistent with our choice of initial conditions in the cosmological studies of the single-metric model that we introduce later, as well as with the common choices for the DW and MM models. We comment on nonzero initial conditions later.} However, in order to keep the structure of the model symmetric, which significantly simplifies the calculations, we include both terms in our action, which now becomes
\begin{equation}\label{eq:action}
S=\frac{M_{\text{Pl}}^{2}}{2}\int \dd^{4}x\sqrt{-g}R+\frac{M_{f}^{2}}{2}\int \dd^{4}x\sqrt{-f}R_{f}-\frac{M_{\text{Pl}}^{2}}{2}\int \dd^{4}x\sqrt{-g}\alpha (R_{f}\frac{1}{\Box}R+R\frac{1}{\Box}R_{f})+S_{\text{matter}}[g,\Psi],
\end{equation}
where $\alpha$ is a free, dimensionless parameter, to be constrained observationally. In addition, we have omitted the index $g$ in the operator $\Box_{g}$, as well as in the Ricci scalar $R_{g}$, in order to keep the notation simple. From now on, and throughout the paper, all differential operators and curvature quantities with no metric indices are defined with respect to the physical metric $g_{\mu\nu}$; we use a label $f$ when an operator or a quantity is defined with respect to $f_{\mu\nu}$.

As discussed above, in this paper we study only the minimal action (\ref{eq:action}) for nonlocally interacting metrics, and leave the investigation of the more complete set of possible terms, scalar and tensorial, as well as a detailed and rigorous construction of consistent and theoretically sound models for future work. We believe that such properly constructed models should be different and more sophisticated than the simple and phenomenologically constructed model (\ref{eq:action}) studied here.

\subsection{Nonlocal equations of motion}\label{sec:NLeom}

Given our model (\ref{eq:action}), the first step is to derive the modified Einstein field equations, i.e. the equations of motion corresponding to $g_{\mu\nu}$ and $f_{\mu\nu}$, by varying the nonlocal action with respect to the two metrics. We obtain
\begin{align}
G_{\mu\nu}+\Delta G_{\mu\nu} & =\frac{1}{M_{\text{Pl}}^{2}}T_{\mu\nu},\label{eq:einstein1}\\
G_{\mu\nu}^{f}+\Delta G_{\mu\nu}^{f} & =0,\label{eq:einstein2}
\end{align}
where $G_{\mu\nu}$ and $G_{\mu\nu}^{f}$ are Einstein tensors corresponding to the physical and reference metrics $g_{\mu\nu}$ and $f_{\mu\nu}$, respectively. $\Delta G_{\mu\nu}$ and $\Delta G_{\mu\nu}^{f}$ are nonlocal distortion terms, i.e. nonlocal corrections to the Einstein tensors, for both metrics, with the forms\footnote{Here $\nabla$ is a covariant derivative, and $(\mu\nu)$ denotes symmetrization over the indices.}
\begin{align}
\Delta G_{\mu\nu}=&-2\alpha[(\frac{1}{\Box}R_{f})G_{\mu\nu}+g_{\mu\nu}R_{f}(1-\frac{1}{2\Box}R)-\nabla_{\mu}\nabla_{\nu}(\frac{1}{\Box}R_{f})-\frac{1}{2}g_{\mu\nu}\nabla^{\rho}(\frac{1}{\Box}R)\nabla_{\rho}(\frac{1}{\Box}R_{f})\nonumber\\
&+\nabla_{(\mu}(\frac{1}{\Box}R_{f})\nabla_{\nu)}(\frac{1}{\Box}R)],\label{eq:DeltaGgNL}\\
\Delta G_{\mu\nu}^{f}=&-2\alpha\frac{M_{\text{Pl}}^{2}}{M_{f}^{2}}[\sqrt{f^{-1}g}(\frac{1}{\Box}R)R_{\mu\nu}^{f}+f_{\mu\nu}\Box_{f}(\sqrt{f^{-1}g}\frac{1}{\Box}R)-\nabla_{\mu}^{f}\nabla_{\nu}^{f}(\sqrt{f^{-1}g}\frac{1}{\Box}R)],\label{eq:DeltaGfNL}
\end{align}
and $T_{\mu\nu}$ is the stress-energy tensor for matter computed in the usual way through the variation of the matter action with respect to $g_{\mu\nu}$. Note that the $f$-metric equations of motion are expectedly not sourced by matter, as the reference metric $f_{\mu\nu}$ does not couple to matter directly.

\subsection{Localization}\label{sec:localization}

As mentioned in Sec.~\ref{sec:intro}, a powerful technique for dealing with nonlocal equations is to rewrite them in a localized form, by introducing some auxiliary fields. While this provides the possibility of solving and interpreting the equations using regular local methods, one should be cautious that the local versions of the theory are equivalent to the original nonlocal theory {\it only if} some conditions are applied to the fields in such a way that the physical degrees of freedom of the theory are kept intact. Otherwise, the ``artificial" local fields can behave as ``regular" fields which may affect the implications of the theory, both classically and quantum-mechanically, especially since in most cases some of the extra fields (or their combinations) are of ghost behavior. We discuss this issue in Sec.~\ref{sec:ghosts}, and here only introduce the localized formulation of our two-metric model.

In order to do that, let us introduce the two scalar fields $U$ and $V$,
\begin{align}
U & \equiv\frac{1}{\Box}R,\label{eq:Udef}\\
V & \equiv\frac{1}{\Box}R_{f}.\label{eq:Vdef}
\end{align}
The action (\ref{eq:action}) can then be written in the local form
\begin{align}\label{eq:localaction}
S=&\frac{M_{\text{Pl}}^{2}}{2}\int \dd^{4}x\sqrt{-g}R+\frac{M_{f}^{2}}{2}\int \dd^{4}x\sqrt{-f}R_{f}-\frac{M_{\text{Pl}}^{2}}{2}\int\dd^4x\sqrt{-g}\alpha(R_{f}U+RV)+\nonumber\\
&+\int\dd^4x\sqrt{-g}\lambda_{1}(R-\Box U)+\int\dd^4x\sqrt{-g}\lambda_{2}(R_{f}-\Box V)+S_{\text{matter}}[g,\Psi],
\end{align}
where we have added the two terms $\lambda_{1}(R-\Box U)$ and $\lambda_{2}(R_{f}-\Box V)$ in order to impose the two conditions (\ref{eq:Udef}) and (\ref{eq:Vdef}), ensuring that the local and nonlocal actions describe the same equations of motion; $\lambda_{1}$ and $\lambda_{2}$ are the corresponding Lagrange multipliers.

First of all, the variation of action (\ref{eq:localaction}) with respect to the Lagrange multipliers $\lambda_{1}$ and $\lambda_{2}$ expectedly gives Eqs. (\ref{eq:Udef}) and (\ref{eq:Vdef}). Let us now vary the action with respect to the fields $U$ and $V$. These give, respectively,
\begin{align}
\lambda_{1} & =-\frac{M_{\text{Pl}}^{2}}{2}\alpha V,\label{eq:lambda1}\\
\lambda_{2} & =-\frac{M_{\text{Pl}}^{2}}{2}\alpha U,\label{eq:lambda2}
\end{align}
which fix the two Lagrange multipliers $\lambda_{1}$ and $\lambda_{2}$ in terms of the fields $U$ and $V$. Plugging Eqs. (\ref{eq:lambda1}) and (\ref{eq:lambda2}) back into the action yields
\begin{align}\label{eq:localaction2}
S=&\frac{M_{\text{Pl}}^{2}}{2}\int \dd^{4}x\sqrt{-g}R+\frac{M_{f}^{2}}{2}\int \dd^{4}x\sqrt{-f}R_{f}-\frac{M_{\text{Pl}}^{2}}{2}\int\dd^4x\sqrt{-g}2\alpha(R_{f}U+RV)+\nonumber\\
&+\frac{M_{\text{Pl}}^{2}}{2}\int\dd^4x\sqrt{-g}2\alpha V\Box U+S_{\text{matter}}[g,\Psi].
\end{align}

Before deriving the field equations in the local formulation by varying the localized action with respect to $g_{\mu\nu}$ and $f_{\mu\nu}$, we note that the rescaling
\begin{align}
\alpha & \rightarrow (\frac{M_{f}}{M_{\text{Pl}}})^{-2}\alpha,\\
f_{\mu\nu} & \rightarrow (\frac{M_{f}}{M_{\text{Pl}}})^{-2}f_{\mu\nu}\Rightarrow\sqrt{-f}\rightarrow (\frac{M_{f}}{M_{\text{Pl}}})^{-4}\sqrt{-f},\;R_{f}\rightarrow (\frac{M_{f}}{M_{\text{Pl}}})^{2}R_{f},\\
V & \rightarrow (\frac{M_{f}}{M_{\text{Pl}}})^{2}V,
\end{align}
leaves the action, and therefore the equations of motion, invariant. This means that the quantity $M_{\star}\equiv \frac{M_{f}}{M_{\text{Pl}}}$ is redundant and is not a free parameter. We therefore use this freedom to set $M_{\star}=1$ without loss of generality.

The variation of the action (\ref{eq:localaction2}) with respect to $g_{\mu\nu}$ and $f_{\mu\nu}$ then leads to
\begin{align}
\Delta G_{\mu\nu} &=-2\alpha [VG_{\mu\nu}+g_{\mu\nu}R_{f}(1-\frac{1}{2}U)-\nabla_{\mu}\nabla_{\nu}V-\frac{1}{2} g_{\mu\nu}\nabla^{\rho}V\nabla_{\rho}U+\nabla_{(\mu}V\nabla_{\nu)}U],\label{eq:DeltaGglocal}\\
\Delta G_{\mu\nu}^{f} &=-2\alpha[\sqrt{f^{-1}g}UR_{\mu\nu}^{f}+ f_{\mu\nu}\Box_{f}(\sqrt{f^{-1}g}U)-\nabla_{\mu}^{f}\nabla_{\nu}^{f}(\sqrt{f^{-1}g}U)],\label{eq:DeltaGflocal}
\end{align}
for $\Delta G_{\mu\nu}$ and $\Delta G_{\mu\nu}^{f}$, the nonlocal corrections to Einstein tensors for the two metrics, introduced in Eqs. (\ref{eq:einstein1}) and (\ref{eq:einstein2}). Eqs. (\ref{eq:DeltaGglocal}) and (\ref{eq:DeltaGflocal}) are identical to Eqs. (\ref{eq:DeltaGgNL}) and (\ref{eq:DeltaGfNL}) when we localize the latter directly at the level of the equations of motion, as expected. This yields another confirmation of the equivalence of our nonlocal and local formulations of the model, as far as the field equations for $g_{\mu\nu}$ and $f_{\mu\nu}$ are concerned (note that here $M_{\star}=1$).

\subsection{Bianchi constraints}\label{sec:Bianchi}

In addition to the equations of motion, i.e. the modified Einstein field equations for the metrics, we need to know which extra constraints are imposed on the fields when the Bianchi identities are used for $G_{\mu\nu}$ and $G_{\mu\nu}^{f}$, as well as the conservation of the matter energy-momentum tensor $T_{\mu\nu}$. Imposing $\nabla^{\mu}G_{\mu\nu}=\nabla^{\mu}T_{\mu\nu}=0$ for the $g$-metric field equations (\ref{eq:einstein1}), and using the expression (\ref{eq:DeltaGgNL}) for $\Delta G_{\mu\nu}$, we obtain\footnote{Note that here we derive the Bianchi constraints in the nonlocal formulation of the model. Imposing the Bianchi identities and the conservation of $T_{\mu\nu}$ on the equations of motion in the local formulation results in exactly the same Bianchi constraints. The procedure is identical, and we do not repeat it here.}
\begin{align}
&G_{\mu\nu}\nabla^{\mu}(\frac{1}{\Box}R_f)+[1-\frac{1}{2}(\frac{1}{\Box}R)]\nabla_{\nu}R_{f}-[R_{\rho\nu}\nabla^{\rho}(\frac{1}{\Box}R_f)+\nabla_{\nu}R_{f}]+\frac{1}{2}R\nabla_{\nu}(\frac{1}{\Box}R_f)\nonumber\\
&=-\frac{1}{2}(\frac{1}{\Box}R)\nabla_{\nu}R_{f}=0.\label{Bianchig}
\end{align}
Assuming $\frac{1}{\Box}R\ne0$, which we need for the nonlocal modification of gravity in our model, this implies
\begin{equation}
\nabla_{\nu}R_{f}=0.\label{eq:Bianchiconst_g}
\end{equation}
By performing similar calculations for the $f$-metric field equations (\ref{eq:einstein2}), imposing $\nabla_f^{\mu}G_{\mu\nu}^f=0$ as well as using the expression (\ref{eq:DeltaGfNL}) for $\Delta G_{\mu\nu}^{f}$, we obtain the constraint
\begin{align}
&\nabla_f^{\mu}(\sqrt{f^{-1}g}(\frac{1}{\Box}R)R_{\mu\nu}^{f})+\nabla_{\nu}^{f}\Box_{f}(\sqrt{f^{-1}g}(\frac{1}{\Box}R)) -[R_{\rho\nu}^{f}\nabla^{\rho}(\sqrt{f^{-1}g}(\frac{1}{\Box}R))+\nabla_{\nu}\Box_{f}\sqrt{f^{-1}g}(\frac{1}{\Box}R)]\nonumber \\
 & =\nabla_f^{\mu}(\sqrt{f^{-1}g}(\frac{1}{\Box}R)R_{\mu\nu}^{f})-R_{\rho\nu}^{f}\nabla^{\rho}(\sqrt{f^{-1}g}(\frac{1}{\Box}R))=\sqrt{f^{-1}g}(\frac{1}{\Box}R)\nabla_f^{\mu}R_{\mu\nu}^{f}=0.\label{Bianchif}
\end{align}
Note that here a subscript or superscript $f$ indicates that the corresponding quantity or operator is defined in terms of $f_{\mu\nu}$. Now, requiring the prefactor $\sqrt{f^{-1}g}(\frac{1}{\Box}R)$ in Eq. (\ref{Bianchif}) to be nonvanishing (otherwise it would yield trivial and uninteresting results) implies $\nabla_f^{\mu}R_{\mu\nu}^{f}=0$. On the other hand, we have
\begin{equation}
\nabla_f^{\mu}G_{\mu\nu}^{f}=0\Rightarrow\nabla_f^{\mu}R_{\mu\nu}^{f}-\frac{1}{2}\nabla_{\nu}^{f}R_{f}=0.
\end{equation}
Combining the two conditions, we obtain
\begin{equation}
\nabla_{\nu}^{f}R_{f}=0.\label{eq:Bianchiconst_f}
\end{equation}
Since $R_{f}$ is a scalar quantity, the covariant derivatives $\nabla_ {\mu}$ and $\nabla_ {\mu}^{f}$ are independent of the metrics, and both conditions (\ref{eq:Bianchiconst_g}) and (\ref{eq:Bianchiconst_f}) imply the {\it Bianchi constraint}
\begin{equation}\label{Bianchiconstraint}
\partial_{\mu} R_{f}=0.
\end{equation} 

This is a surprising result, as the constraint is very strong; let us understand its implications. The constraint (\ref{Bianchiconstraint}) tells us that the Ricci scalar of the reference metric must be temporally and spatially constant. This means that the form of the reference metric $f_{\mu\nu}$ is highly constrained, as it has to be a metric with specific dynamics, for example, a Minkowski metric for $R_{f}=0$, or a de Sitter metric for a constant and nonzero $R_{f}$. This would in principle be of no problem if the full set of field equations (\ref{eq:einstein1}) and (\ref{eq:einstein2}) could always be satisfied consistently, for all the interesting choices of the physical metric $g_{\mu\nu}$. The problem however is that the Bianchi constraint (\ref{Bianchiconstraint}) is a condition independent of the physical system under consideration, and independent of the chosen $g_{\mu\nu}$. This then means that not only should the modified Einstein equations (\ref{eq:einstein1}) for $g_{\mu\nu}$ be solved, but also the extra set of Einstein equations (\ref{eq:einstein2}) for $f_{\mu\nu}$ should additionally be satisfied for the chosen $g_{\mu\nu}$ and with an $f_{\mu\nu}$ that has already been fixed to a metric with a constant curvature. This may overconstrain the system, which means that it may not be possible to always solve the set of equations. An explicit example is cosmology. It may be possible to find a solution for $f_{\mu\nu}$ with a constant curvature for the background evolution, i.e. with $g_{\mu\nu}$ having an FLRW form, but the Bianchi constraint (\ref{Bianchiconstraint}) may not allow us to perturb such a metric when we perturb $g_{\mu\nu}$, as a perturbed $f_{\mu\nu}$ may not allow a constant curvature. In this paper we will show that a consistent solution for $f_{\mu\nu}$ exists for background cosmology, and we leave the nontrivial question of the existence of perturbative solutions for future work. Although we do not know the answer at this stage, it is possible that the $f$-metric equations cannot be satisfied in all interesting cases. This would then mean that Eqs. (\ref{eq:einstein2}) should be discarded, which in turn would mean that we would not be allowed to vary the action (\ref{eq:action}) with respect to the reference metric $f_{\mu\nu}$. In that case, the kinetic term for $f_{\mu\nu}$ would have no effects on any physical quantities, and can also be dropped. The situation then would be very similar to the ghost-free theory of massive gravity, in which the reference metric is not dynamical, as opposed to bimetric theories with both metrics dynamical, and is fixed for the theory independently of the form of the physical metric $g_{\mu\nu}$. There is however a very important difference between our case and massive gravity, and that is the fact that the reference metric here affects the physical sector {\it only} through its Ricci scalar, or in other words, the curvature of $f_{\mu\nu}$ enters the $g_{\mu\nu}$ equations {\it only} through the Ricci scalar. In massive gravity, the reference metric, being a rank-2 field, is required in order to give mass to graviton, while in our nonlocal model all of its tensorial properties are lost. Clearly, the situation could change if the structure of the model were extended to include other possible scalar terms. The same could be the case if tensorial nonlocalities were (also) considered, for which either the Bianchi constraint would not be as strong as in our simple model, or the theory would become similar to massive gravity with a fixed reference metric affecting the gravity sector as a full, tensor field, and not only through a scalar contribution. Another possibility in that case would be to add the local, ghost-free interactions of massive or bimetric gravity to our simple nonlocal model, which may also violate the strong Bianchi constraint (\ref{Bianchiconstraint}). These are all interesting and exciting possibilities, but are beyond the scope of the present paper, and we leave their investigation for future work. For now, we take a closer look at our model (\ref{eq:action}) with only the simple scalar interaction terms, and see whether the structure of the model when the Ricci scalar is assumed to be fixed to a constant would suggest an interesting model. This is the subject of the next section.

\section{The $m^{2}\frac{1}{\Box}R$ model}\label{sec:m2model}

Let us look again at action (\ref{eq:action}) for our model of nonlocally interacting metrics, this time taking into account the condition required for the consistency of the solutions, i.e. the Bianchi constraint (\ref{Bianchiconstraint}), which, as discussed in the previous section, forces the Ricci scalar of the reference metric, $R_{f}$, to be a constant. Calling the constant combination $2\alpha R_{f}$ simply $m^2$,\footnote{Recall that $R_{f}$ has dimension $[M^2]$, and $\alpha$ is dimensionless.} we can now impose the condition (\ref{Bianchiconstraint}) at the level of the action. As argued above, as long as we are interested only in the dynamics of the physical metric $g_{\mu\nu}$, we can assume that $f_{\mu\nu}$ is a {\it fixed} metric, and therefore, no longer vary the action with respect to $f_{\mu\nu}$.\footnote{The reference metric $f_{\mu\nu}$ is, for example, of a de Sitter form, which is determined purely through the constant curvature $R_{f}$, completely independently of $g_{\mu\nu}$ and matter.} We can thus fully ignore the $f$-metric kinetic (Einstein-Hilbert) term in the action. The action then reads
\begin{equation}
S=\frac{M_{\text{Pl}}^{2}}{2}\int \dd^{4}x\sqrt{-g}[R+\frac{1}{2}(m^{2}\frac{1}{\Box}R+R\frac{1}{\Box}m^{2})]+S_{\text{matter}}[g,\Psi],\label{eq:m2model}
\end{equation}
which is a {\it single-metric} model with some nonlocal distortion terms added to GR. Before we discuss the implications of this observation, let us quickly obtain the field equations for action (\ref{eq:m2model}) and find the relations between the quantities in this model and the ones in our original two-metric model. It is important to note, as we discussed in Sec.~\ref{sec:model}, that the two terms $m^{2}\frac{1}{\Box}R$ and $R\frac{1}{\Box}m^{2}$ are identical through ``integration by parts" if the boundary terms vanish. In that case, the action simplifies to
\begin{equation}
S=\frac{M_{\text{Pl}}^{2}}{2}\int \dd^{4}x\sqrt{-g}(R+m^{2}\frac{1}{\Box}R)+S_{\text{matter}}[g,\Psi].\label{eq:m2model2}
\end{equation}
Looking at the two actions (\ref{eq:m2model}) and (\ref{eq:m2model2}), we notice one more subtlety, and that is for the zero value of the parameter $m$. In this case, clearly the action (\ref{eq:m2model2}) reduces to the standard action of GR, as expected, while the second nonlocal term in the action (\ref{eq:m2model}) can remain nonvanishing. The reason is that an equation like $\Box X=0$ can have nonvanishing solutions for $X$, and therefore, the quantity $\frac{1}{\Box}m^{2}$ is not identically zero for $m=0$. However, this difference between the two actions also stems from the subtleties in choosing the initial and boundary conditions for quantities like $\frac{1}{\Box}m^{2}$ in nonlocal models. If the quantity is set to zero initially and at the boundary, it remains vanishing everywhere and at all times, and the two forms of the action become identical. In this paper, we only consider vanishing initial conditions, and it therefore does not matter for our considerations which form of the action to choose. We therefore use the two forms (\ref{eq:m2model}) and (\ref{eq:m2model2}) interchangeably. We comment on this again in the next section when we discuss vanishing and nonvanishing values of $m$ in our study of the background cosmology for the model.

Let us now derive the equations of motion for the gravity sector of the model. The procedure is similar to the one for the original two-metric model, and, working in the localized formulation of the model, we first introduce the auxiliary field
\begin{equation}
U\equiv\frac{1}{\Box}R.\label{eq:Phidef}
\end{equation}
Plugging this into the action, and adding a Lagrange multiplier $\lambda$ in order to impose (\ref{eq:Phidef}), we obtain the localized action
\begin{equation}
S=\frac{M_{\text{Pl}}^{2}}{2}\int \dd^{4}x\sqrt{-g}(R+m^{2}U)+\int \dd^{4}x\sqrt{-g}\lambda(R-\Box U)+S_{\text{matter}}[g,\Psi].\label{eq:m2modellocalaction}
\end{equation}
By varying this local action with respect to $g_{\mu\nu}$, we obtain the modified Einstein equations
\begin{equation}\label{eq:m2modelEoM}
(\frac{M_{\text{Pl}}^{2}}{2}+\lambda)G_{\mu\nu}+\frac{M_{\text{Pl}}^{2}}{2}m^{2}g_{\mu\nu}(1-\frac{U}{2})-\nabla_{\mu}\nabla_{\nu}\lambda-\frac{1}{2}g_{\mu\nu}\nabla^{\rho}\lambda\nabla_{\rho}U+\nabla_{(\mu}\lambda\nabla_{\nu)}U=\frac{1}{2}T_{\mu\nu},
\end{equation}
which are identical to Eqs. (\ref{eq:einstein1}) in combination with Eqs. (\ref{eq:DeltaGglocal}), after performing the transformations
\begin{align}
\lambda & \rightarrow-M_{\text{Pl}}^{2}\alpha V,\nonumber\\
m^{2} & \rightarrow-2\alpha R_{f}.\nonumber
\end{align}
The modified Einstein equations (\ref{eq:m2modelEoM}) now take the form
\begin{equation}\label{eq:m2modelEoM2}
(1-2\alpha V)G_{\mu\nu}+m^{2}(1-\frac{U}{2})g_{\mu\nu}+2\alpha\nabla_{\mu}\nabla_{\nu}V+\alpha\nabla^{\rho}V\nabla_{\rho}Ug_{\mu\nu}-2\alpha\nabla_{(\mu}V\nabla_{\nu)}U=\frac{1}{M_{\text{Pl}}^{2}}T_{\mu\nu},
\end{equation}
which can be solved together with the extra equations
\begin{align}
\Box U &=R,\label{eq:Um2model}\\
\Box V &=-\frac{1}{2\alpha}m^{2},\label{eq:lambdam2model}
\end{align}
which are obtained through varying the action (\ref{eq:m2modellocalaction}) with respect to $\lambda$ and $U$, respectively.

We can see explicitly from the structure of Eq. (\ref{eq:lambdam2model}) for $V$ (or, equivalently, for the Lagrange multiplier $\lambda$) that this model, although involving only one operator $\Box$ in the nonlocal term, still needs two auxiliary fields for localization, in contrast to the DW model studied in Sec.~\ref{sec:DW}. The reason is that Eq. (\ref{eq:lambdam2model}) does not determine $V$ in terms of the other fields that already exist in the model, or the local operators. It involves the nonlocal operator $\Box^{-1}$ acting on the parameter $m^2$, and therefore, after plugging $V$ (or $\lambda$) back into the action (\ref{eq:m2modellocalaction}), the nonlocalilty remains. We therefore need an extra auxiliary field to take care of this. The fact that one needs two auxiliary fields for localization, even though the nonlocal operator is of the form $\Box^{-1}$, is interesting also in comparison to the MM model with $\Box^{-2}$. This can be understood by noticing the asymmetric structure of the term $m^2\frac{1}{\Box}R$, and the fact that $\Box^{-1}$ acts on both sides when the action is varied.

In addition, we should note that although we obtained the model (\ref{eq:m2model}) through our original two-metric model, the connection to a bimetric setup is now lost, and from a purely phenomenological point of view the model can simply be taken as a simple nonlocal modification of gravity not necessarily related to a model of interacting metrics, with the nonlocality generated by a completely different mechanism. In this respect, for phenomenologists who are not necessarily interested in the fundamental theory behind the model, the action (\ref{eq:m2model2}) can be considered as an standalone, consistent model of modified gravity, and the starting point for any phenomenological studies. The fact that it provides a viable background cosmology, contrary to the similar DW model $\alpha R\frac{1}{\Box}R$ (as we show in the next sections), while being simpler in structure than the MM model $m^2R\frac{1}{\Box^{2}}R$, makes the model appealing. It is however important to note that the $m^2\frac{1}{\Box}R$ model introduces a new mass scale, similarly to the MM and differently from the DW models. As discussed in Sec.~\ref{sec:intro}, $m^2\frac{1}{\Box}R$ is nothing but a model with the structure $m^2X$, where $X\equiv\frac{1}{\Box}R$, in comparison to the MM model with the structure $m^2X^2$. It is quite interesting that such a simple model provides a well-behaved cosmology, at least at the background level, while introducing only one free parameter, just as in $\Lambda$CDM and the MM model. In the next section, we perform a detailed study of the background cosmology of the model, and leave the exploration of possible theoretical origins of the model, not necessarily based on a theory of interacting metrics, for future work.

\section{Cosmology and expansion histories}
\label{sec:cosmology}

Now that we have the field equations (\ref{eq:m2modelEoM2}) for the metric $g_{\mu\nu}$, as well as the the constraints relating the auxiliary fields $U$ and $V$ to the metric and the parameter $m$, i.e. Eqs. (\ref{eq:Um2model}) and (\ref{eq:lambdam2model}), we can start the investigation of background cosmology in the $m^{2}\frac{1}{\Box}R$ model.

Let us follow the standard recipe for modelling the background dynamics of the Universe, and assume that the Universe is described by an FLRW metric. Specializing to a spatially flat universe and working in cosmic time $t$, we have
\begin{align}
g_\mn \dd x^\mu \dd x^\nu &= -\dd t^2 + a^2(t)\delta_{ij}\dd x^i \dd x^j. \label{eq:FLRWg}
\end{align}
Here, $a(t)$ is the scale factor, being a function of time only. We furthermore take a perfect-fluid form for the matter source, and therefore assume $T^{\mu}_{~\nu}=\mathrm{diag}(-\rho,p,p,p)$, with $\rho$ and $p$ being the matter+radiation energy density and pressure, respectively.

\subsection{Background equations}\label{sec:background}

Plugging the FLRW expressions for the metric (\ref{eq:FLRWg}) into the $(0,0)$ components of the field equations (\ref{eq:m2modelEoM2}), we obtain the (modified) Friedmann equation
\begin{align}
(1-2\alpha V)H^{2}-\frac{1}{3}m^{2}(1-\frac{1}{2}U)+\frac{2}{3}\alpha\ddot{V}-\frac{1}{3}\alpha\dot{V}\dot{U} & =\frac{1}{3M_{\text{Pl}}^{2}}\rho,\label{eq:friedmanndS1}
\end{align}
where $H\equiv\frac{\dot{a}}{a}$ is the Hubble rate, and an overdot denotes a derivative with respect to cosmic time. In order to solve this equation, we also need the equations for the auxiliary fields $U$ and $V$, i.e. Eqs. (\ref{eq:Um2model}) and (\ref{eq:lambdam2model}), which now take the forms
\begin{align}
\ddot{U}+3H\dot{U} & =-R,\label{eq:U_eom_dS}\\
\ddot{V}+3H\dot{V} & =\frac{m^{2}}{2\alpha}.\label{eq:V_eom_dS}
\end{align}
As is common in similar analyses of modified gravity models, let us introduce a derivative with respect to the number of $e$-folds $N\equiv \text{ln}a$, which we denote by
a prime. The cosmic-time derivatives $\dot{U}$, $\ddot{U}$, $\dot{V}$, and $\ddot{V}$ can now be written in terms of the derivatives with respect to $N$,
\begin{align}
\dot{U} & =Hu^{\prime},\\
\ddot{U} & =H^{2}u^{\prime\prime}+H^{2}\xi u^{\prime},\\
\dot{V} & =\frac{H_{0}}{\alpha h}(v^{\prime}-2\xi v),\\
\ddot{V} & =\frac{H_{0}^{2}}{\alpha}(v^{\prime\prime}-3\xi v^{\prime}+2(\xi^{2}-\xi^{\prime})v),
\end{align}
where we have introduced $v\equiv \alpha h^{2}V$, $u\equiv U$, and $\xi \equiv h^{\prime}/h$, with $h\equiv H/H_{0}$ and $H_0$ being the present value of the Hubble rate. Eqs. (\ref{eq:U_eom_dS}) and (\ref{eq:V_eom_dS}) now read
\begin{align}
u^{\prime\prime}+(\xi+3)u^{\prime} + 6(\xi+2) &=0,\label{eq:U_eom_dS_N}\\
v^{\prime\prime}-3(\xi-1)v^{\prime}+2(\xi^{2}-3\xi-\xi^{\prime})v & =\frac{m^{2}}{2H_{0}^{2}},\label{eq:V_eom_dS_N}
\end{align}
where we have additionally used $R=6(\xi+2)H^2$. The Friedmann equation (\ref{eq:friedmanndS1}) in terms of these new variables takes the form
\begin{align}
h^{2} & =\Omega_{\text{M}}^{0}e^{-3N}+\Omega_{\text{R}}^{0}e^{-4N}+2v-\frac{1}{6}\frac{m^{2}}{H_{0}^{2}}u-(2\xi v-v^{\prime})(2+\frac{1}{3}u^{\prime}),\label{eq:Friedmann_dS}
\end{align}
where the Universe is assumed to be filled with matter and radiation, with the present density parameters $\Omega_{\text{M}}^{0}$ and $\Omega_{\text{R}}^{0}$, respectively. From this expression we can then read off the effective dark-energylike contribution in our model, with the density parameter
\begin{equation}
\Omega_{\text{NL}}\equiv\frac{\rho_{\text{NL}}}{\rho_{\text{tot}}}=h^{-2}(2v-\frac{1}{6}\frac{m^{2}}{H_{0}^{2}}u-(2\xi v-v^{\prime})(2+\frac{1}{3}u^{\prime})),\label{eq:rho_DE}
\end{equation}
where $\rho_{\text{NL}}$ is the nonlocal contribution to the total energy density of the Universe $\rho_{\text{tot}}$ at any given time. The evolution of the density parameters $\Omega_{\text{M}}$, $\Omega_{\text{R}}$, and $\Omega_{\text{NL}}=1-\Omega_{\text{M}}-\Omega_{\text{R}}$ can be obtained by solving the continuity equations for matter, radiation, and the nonlocal contribution,
\begin{equation}
\dot{\rho}_{\text{M,R,NL}}+3H\rho_{\text{M,R,NL}}(1+w_{\text{M,R,NL}})=0,
\end{equation}
where $w_{\text{M}}\equiv p_{\text{M}}/\rho_{\text{M}}=0$, $w_{\text{R}}\equiv p_{\text{R}}/\rho_{\text{R}}=1/3$, and $w_{\text{NL}}\equiv p_{\text{NL}}/\rho_{\text{NL}}$ are the equations of state for matter, radiation and the dark-energylike nonlocal contribution, respectively. The density parameters $\Omega_{\text{M}}$, $\Omega_{\text{R}}$, and $\Omega_{\text{NL}}$ then evolve, in terms of $N$, through the equation
\begin{equation}\label{eq:Omegaevol}
\Omega^\prime_{\text{M,R,NL}}+\Omega_{\text{M,R,NL}}(3+2\xi+3w_{\text{M,R,NL}})=0.
\end{equation}
Note that $\xi=-1.5$ and $\xi=-2$ for a universe filled with only matter or radiation, respectively, and we therefore recover the evolution equations for matter- and radiation-domination epochs from the general equation (\ref{eq:Omegaevol}). The nonlocality equation of state $w_{\text{NL}}$ as a function of $N$ can be obtained using Eq. (\ref{eq:rho_DE}) in combination with
\begin{equation}\label{eq:contNL}
\rho^\prime_{\text{NL}}+3\rho_{\text{NL}}(1+w_{\text{NL}})=0.
\end{equation}   
Finally, another important quantity for the background study of any cosmological model is the effective equation of state~\cite{2010deto.book.....A}
\begin{equation}\label{eq:weff_def}
w_{\text{eff}}=-1-\frac{2}{3}\frac{h^{\prime}}{h}=-1-\frac{2}{3}\xi,
\end{equation}
which parametrizes the evolution of the total energy density, and is the key quantity in comparing the model's predicted background dynamics to observations.

Before we study the implications of these equations for the real Universe, and see whether the model is able to describe the cosmic evolution correctly, let us briefly discuss the curious case of $m=0$. Setting $m$ to zero in the Friedmann equation (\ref{eq:friedmanndS1}), as well as in Eqs. (\ref{eq:U_eom_dS}) and (\ref{eq:V_eom_dS}) for $U$ and $V$, we obtain
\begin{align}
(1-2\alpha V)H^{2}+\frac{2}{3}\alpha\ddot{V}-\frac{1}{3}\alpha\dot{V}\dot{U} & =\frac{1}{3M_{\text{Pl}}^{2}}\rho,\label{eq:Friedmann_Mink}\\
\ddot{U}+3H\dot{U} & =-R,\label{eq:U_eom_Mink}\\
\ddot{V}+3H\dot{V} & =0.\label{eq:V_eom_Mink}
\end{align}
These equations may seem surprising, as $m$ being zero should mean that we are back to standard cosmology, with the standard Friedmann equations, whereas here we seem to have several extra terms remaining in Eq. (\ref{eq:Friedmann_Mink}). As discussed in Sec.~\ref{sec:m2model}, the reason is that we have derived the Friedmann equation using the symmetrized form of the action, i.e. the expression (\ref{eq:m2model}), in which the second nonlocal term is not necessarily vanishing for a vanishing $m$. However, as we argued before, such a term does vanish if the initial and boundary conditions are set to zero, as in that case the quantity $\frac{1}{\Box}m^2$ is always zero. This can be seen here explicitly, as Eq. (\ref{eq:V_eom_Mink}) forces $V$ (which is proportional to $\frac{1}{\Box}m^{2}$) to always be zero, if it is initially set to zero. This then in turn forces all the nonstandard terms in Eq. (\ref{eq:Friedmann_Mink}) to also vanish, and we recover standard cosmology with no cosmological constant term, and therefore with no viable solutions. We can of course use nonzero initial conditions for the auxiliary fields $U$ and $V$, and check whether this $m=0$ case could yield any viable cosmologies, along the lines of the analysis performed in Ref.~\cite{Nersisyan:2016hjh}. The numerical investigation of Eq. (\ref{eq:Friedmann_Mink}), together with Eqs. (\ref{eq:U_eom_Mink}) and (\ref{eq:V_eom_Mink}), shows however that the set of equations does not provide an evolution for the Universe consistent with observations. This can be seen qualitatively by looking into the structure of the equations. Note that in Eq. ($\ref{eq:Friedmann_Mink}$) all the nonstandard terms are proportional to $V$, $\dot{V}$, or $\ddot{V}$. On the other hand, Eq. ($\ref{eq:V_eom_Mink}$) for $V$ does not have any source, and therefore, whatever initial values we start with for $V$ and its derivatives, the Hubble friction term will reduce $V$ with time, forcing it to vanish eventually. This means that the nonlocal terms in Friedmann equation (\ref{eq:Friedmann_Mink}) will vanish at late times and, therefore, there will be no dark-energylike behavior in the asymptotic future. One could however think that by appropriately setting the initial values for the auxiliary fields $U$ and $V$, and their derivatives, the nonlocal terms would remain nonvanishing for a sufficiently long period of time over the history of the Universe, with viable behavior. This may seem to work especially because there is a compensating term, $\frac{1}{3}\alpha\dot{V}\dot{U}$, in Eq. ($\ref{eq:Friedmann_Mink}$), which might remain nonvanishing, as $U$ is sourced and can compensate for the decaying behavior of $V$. However, a detailed investigation of the dynamical equations shows that $V$ decays very quickly, and the nonlocal terms vanish rapidly, after a short period of time, independently of the initial conditions. We therefore conclude that the model with $m=0$ does not provide a viable cosmological solution, and from now on assume that $m$ is nonzero.

\subsection{Cosmic acceleration, and comparison with Maggiore and Mancarella's $m^2R\frac{1}{\Box^2}R$ model}\label{sec:acceleration}

Having all the equations needed to study the background solutions for our nonlocal model with a nonzero $m$, here we investigate the implications of the solutions, and explore whether the model can provide a viable cosmological evolution. In particular, we study the possibility of obtaining self-accelerating solutions in the absence of a cosmological constant term.

In order to obtain the cosmic evolution for the model, we need to know how the auxiliary fields $u$ and $v$ evolve with time. For that, we solve Eqs. ($\ref{eq:U_eom_dS_N}$)
and ($\ref{eq:V_eom_dS_N}$), for which we need to know how $\xi$ evolves with time. The evolution of $\xi$ can, on the other hand, be obtained using Eq. ($\ref{eq:Friedmann_dS}$) and its derivative (remember that $\xi=\frac{h^{\prime}}{h}$). This yields the quadratic equation
\begin{align}
&4v\left(u^{\prime}+6\right)\xi^{2}+2[-3(\Omega_{\text{M}}^{0}e^{-3N}+\Omega_{\text{R}}^{0}e^{-4N})+\frac{m^{2}}{2}u+6(4v-v^{\prime}+u^{\prime}v)-u^{\prime}v^{\prime}]\xi\nonumber\\
&-3(3\Omega_{\text{M}}^{0}e^{-3N}+4\Omega_{\text{R}}^{0}e^{-4N})-6[u^{\prime}v^{\prime}+4v^{\prime}-\frac{m^{2}}{2}] =0\label{eq:xi}
\end{align}
for $\xi$. This equation, combined with Eqs. ($\ref{eq:U_eom_dS_N}$) and ($\ref{eq:V_eom_dS_N}$), gives a closed system of differential equations for $u$, $v$, and $\xi$. 

We solve all the equations numerically with the integration initiated at $N=-15$, i.e. well inside the radiation-domination epoch in the early Universe. For the initial conditions
we simply set $u_0=u^{\prime}_0=v_0=v^{\prime}_0=0$.\footnote{Note that the indices $0$ here denote the initial values of the quantities, not their present values.} Note that, as discussed before, different initial conditions for the auxiliary fields correspond to different homogeneous solutions of the equations that relate the fields to $R$ and $m^{2}$, defining the operator $\Box^{-1}$ in the nonlocal formulation of the model. The choices we have made here are the simplest ones, commonly made for nonlocal models, and correspond to a vanishing homogeneous solution.\footnote{Let us for example assume that $\phi$ is the homogeneous solution for $U$ in the equation
\begin{equation}
U=\frac{1}{\Box}R \iff \Box U=R,
\end{equation}
meaning that $\phi$ satisfies $\Box\phi=0$. It is easy to show that a vanishing homogeneous solution $\phi=0$ corresponds to the minimal choice of the initial conditions that we made, i.e. $U_0=U^{\prime}_0=0$ in this case. Assuming that the nonlocal effects kicked in at some initial time, and that they were absent before that, terms like $\frac{1}{\Box} R$ are nonvanishing only after that initial time. This means that the homogeneous solution $\phi$ and its derivative are initially zero as they are equal to $U_0$ and $U^{\prime}_0$, respectively. Since $\Box \phi=0$ does not have a source, if $\phi$ and $\phi^{\prime}$ are initially zero, they will always remain zero.} In principle, though, one can relax the constraints on the initial conditions, as done in Ref.~\cite{Nersisyan:2016hjh}, and explore the effects of different choices on the dynamics, which corresponds to exploring various local formulations of the model. The evolution we obtain in the present paper corresponds to an attractor solution, and even though we tried a few different initial conditions by hand, we always ended up with the same solution at late times. It however remains to see whether other attractor solutions exist by properly choosing the initial conditions, as has been done in Ref.~\cite{Nersisyan:2016hjh} for the MM model. This requires a proper phase-space analysis of the system, which is beyond the scope of the present paper, and we leave it for future work.

In all the numerical calculations and results presented below, we have set $\Omega_{\text{M}}^{0}\approx0.31$ and $\Omega_{\text{R}}^{0}\approx9.2\times10^{-5}$ for the abundance of matter and radiation at present, consistent with the latest cosmological measurements~\cite{Ade:2015xua}. One should however note that these values have been obtained for the $\Lambda$CDM model, and one therefore clearly needs to perform a proper statistical analysis of our nonlocal model in order to find the best-fit values for these parameters within the framework of the model. As we will see later, the background evolution of the Universe in this model is close to the $\Lambda$CDM one, and therefore, the chosen values should be close to the values one would obtain through a detailed statistical analysis of the model in comparison to the data. We further set $m^{2}=0.232H_{0}^2$, which, as we will see, provides a well-behaved and viable cosmic history.

\begin{figure}[t]
 \includegraphics[height=5cm]{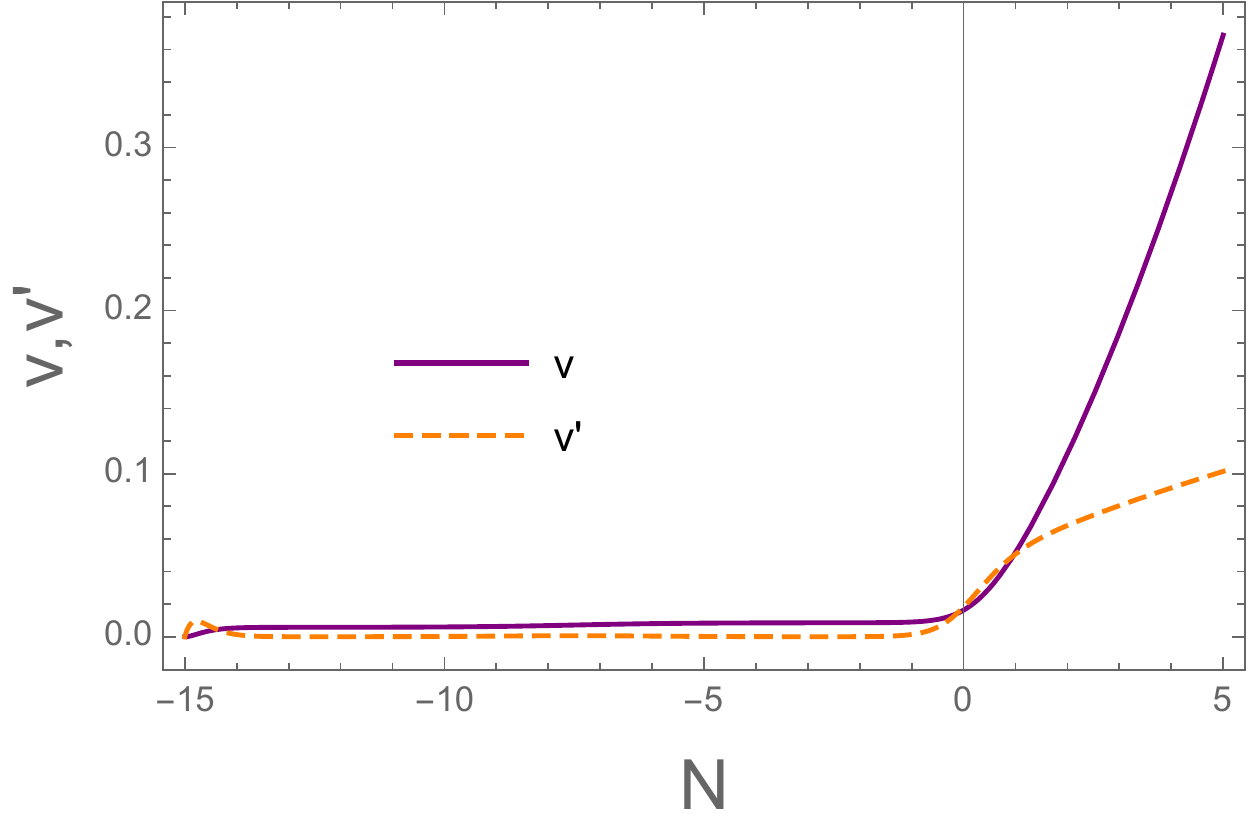}
  \includegraphics[height=5cm]{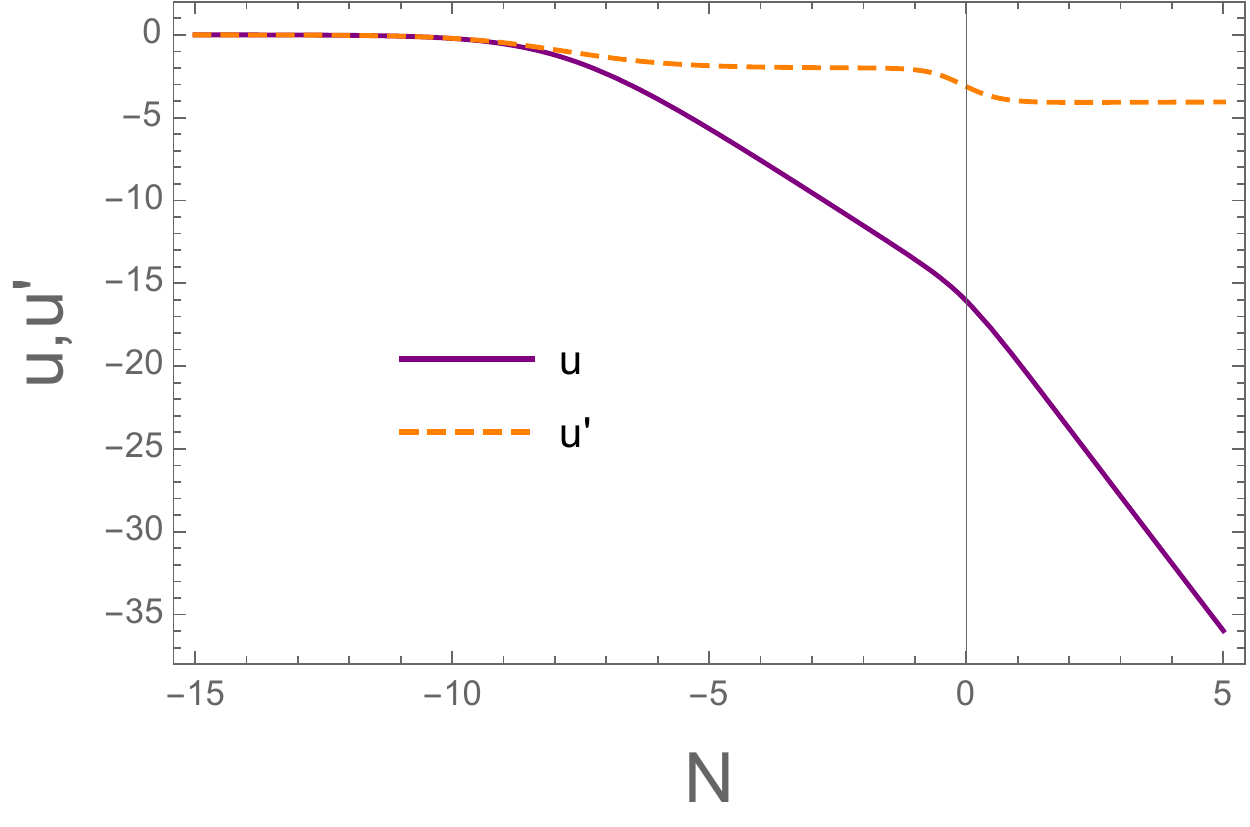}
\caption{Evolution of the auxiliary fields $v$ and $u$, and their derivatives $v^\prime$ and $u^\prime$, as functions of the number of $e$-folds $N$.}\label{fig:scalar_fields}
\end{figure}

\begin{figure}
  \includegraphics[height=5cm]{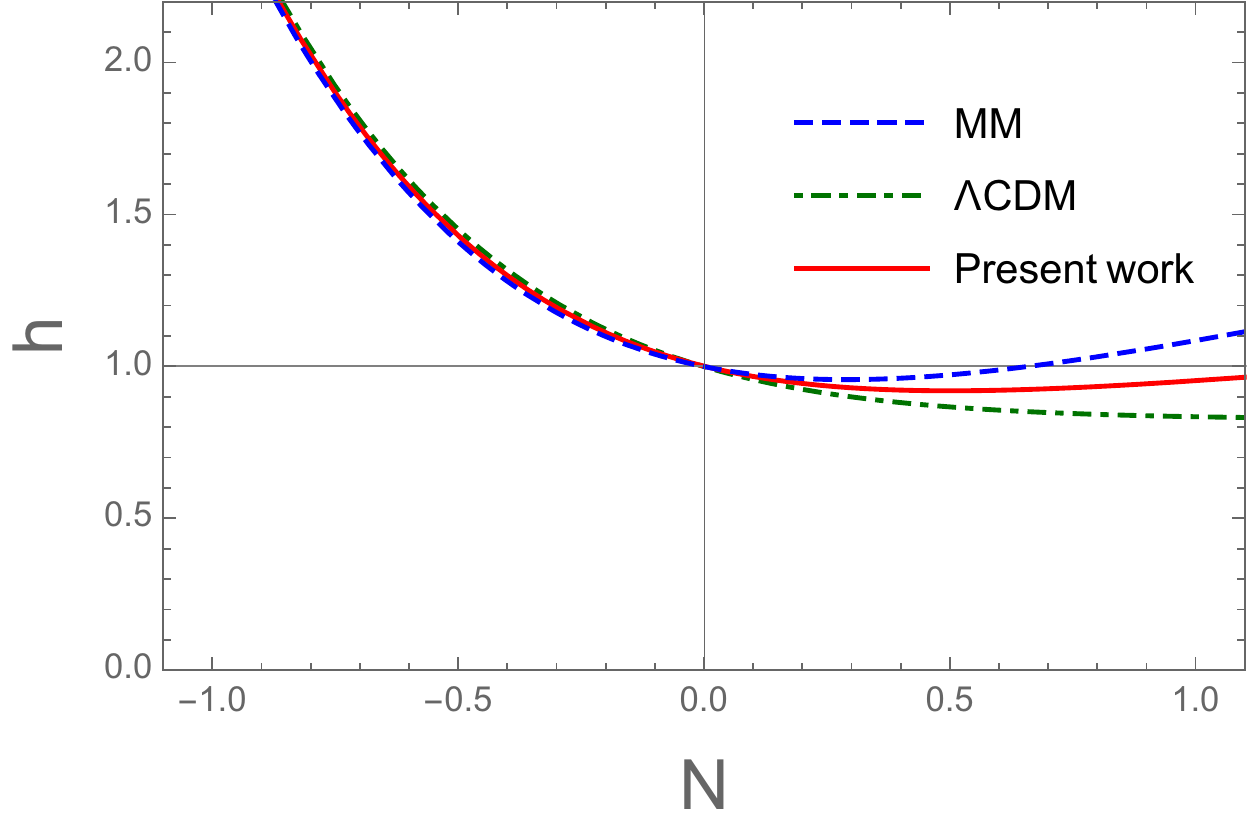}
  \includegraphics[height=5cm]{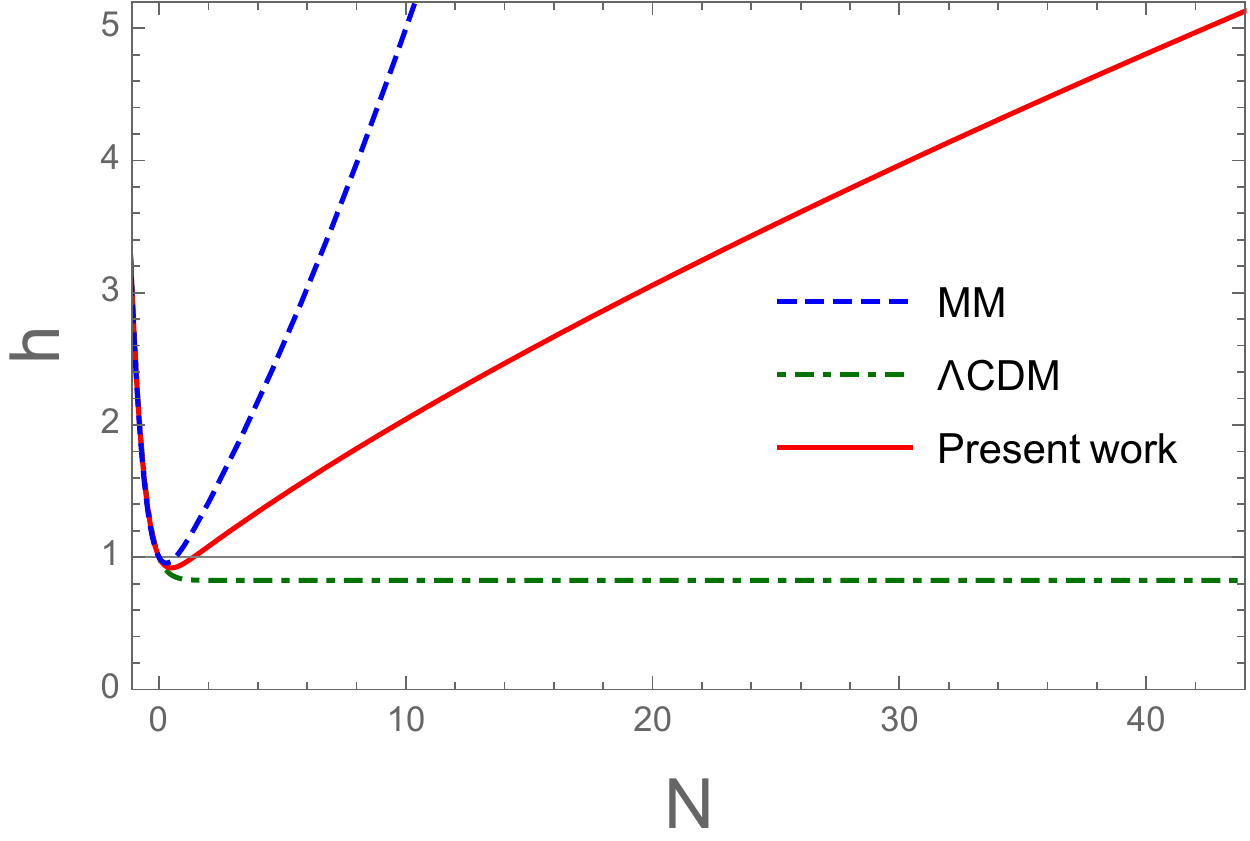}
\caption{\label{fig:h} {\it Left panel:} Evolution of the normalized Hubble rate $h\equiv H/H_{0}$ as a function of the number of $e$-folds $N$ for the present work (red, solid curve), as well as for the MM model (blue, dashed curve) and $\Lambda$CDM (green, dot-dashed
curve). {\it Right panel:} The same as in the left panel but for a longer period of time in the future.}
\end{figure}

Our numerical solutions for the auxiliary fields $u$ and $v$ in terms of the number of $e$-folds $N$ are shown in Fig.~\ref{fig:scalar_fields}. Fig.~\ref{fig:h} then shows the evolution of the normalized Hubble rate $h$ ($\equiv H/H_0$) computed through Eq. (\ref{eq:Friedmann_dS}). For comparison, we have also shown the evolution of $h$ for the standard $\Lambda$CDM and the MM nonlocal models, where we have assumed the same values for $\Omega_{\text{M}}^{0}$ and $\Omega_{\text{R}}^{0}$ as in our model. We have set $\gamma=0.00891$ for the free parameter of the MM model~\cite{Maggiore:2014sia}. The figure shows that $h$ behaves almost identically in the past ($N < 0$) for all three models, while behaving significantly differently in the (far) future. In contrast to $\Lambda$CDM, with $h$ becoming almost constant in the future, both MM and our models predict Hubble rates that increase with time, albeit much faster in the MM case. Despite this behavior of $h$ in our model and the fact that it does not become a constant in the future, implying that the Universe does not evolve into a de Sitter phase\footnote{Here, by a de Sitter phase, we mean the Universe being dominated by a nonzero, finite, cosmological constant, with $h$ becoming a finite constant. One could consider cases like ours also a de Sitter phase, with an asymptoticly infinite cosmological constant, and therefore an infinite Hubble rate.} as opposed to $\Lambda$CDM, we still obtain an accelerating solution with an effective equation of state $w_{\text{eff}}$ close to $-1$, as we will see later. The reason is the slower increase in $h^{\prime}$ compared to $h$, making $\xi$ and $w_{\text{eff}}$ approach $0$ and $-1$, respectively, in the future; see Eq. (\ref{eq:weff_def}). This is similar to what happens in the MM model.

\begin{figure}
\center
\includegraphics[height=5cm]{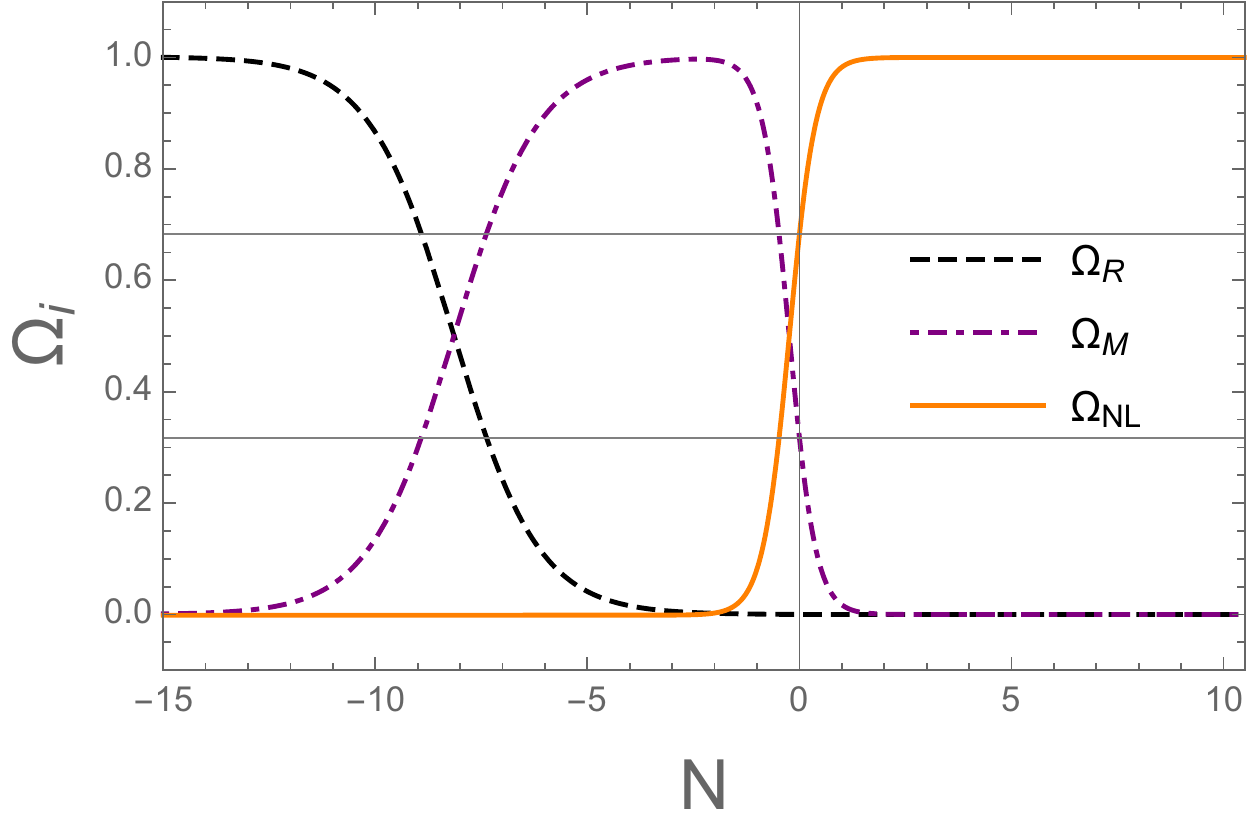}
\caption{\label{fig:Omegas} Evolution of the density parameters $\Omega_{\text{R}}$ (black, dashed curve), $\Omega_{\text{M}}$ (violet, dot-dashed curve), and $\Omega_{\text{NL}}$ (orange, solid curve) as functions of the number of $e$-folds $N$ for the present work. The lower and upper, grey, horizontal lines represent $\Omega^{0}_{\text{M}}$ and $1-\Omega^{0}_{\text{M}}-\Omega^{0}_{\text{R}}$, respectively.}
\end{figure}

Fig.~\ref{fig:Omegas} depicts how the density parameters $\Omega_{\text{M}}$, $\Omega_{\text{R}}$, and $\Omega_{\text{NL}}$ evolve in our nonlocal model. These have been calculated by solving Eq. (\ref{eq:Omegaevol}) numerically. The figure clearly shows the three epochs in the history of the Universe, i.e. the radiation- and matter-domination eras, as well as the final period in which the nonlocalities dominate. Comparing the same curves with those from $\Lambda$CDM, as well as the MM model, (not shown in the figure) shows an almost exact match between all three models.

\begin{figure}
 \includegraphics[height=5cm]{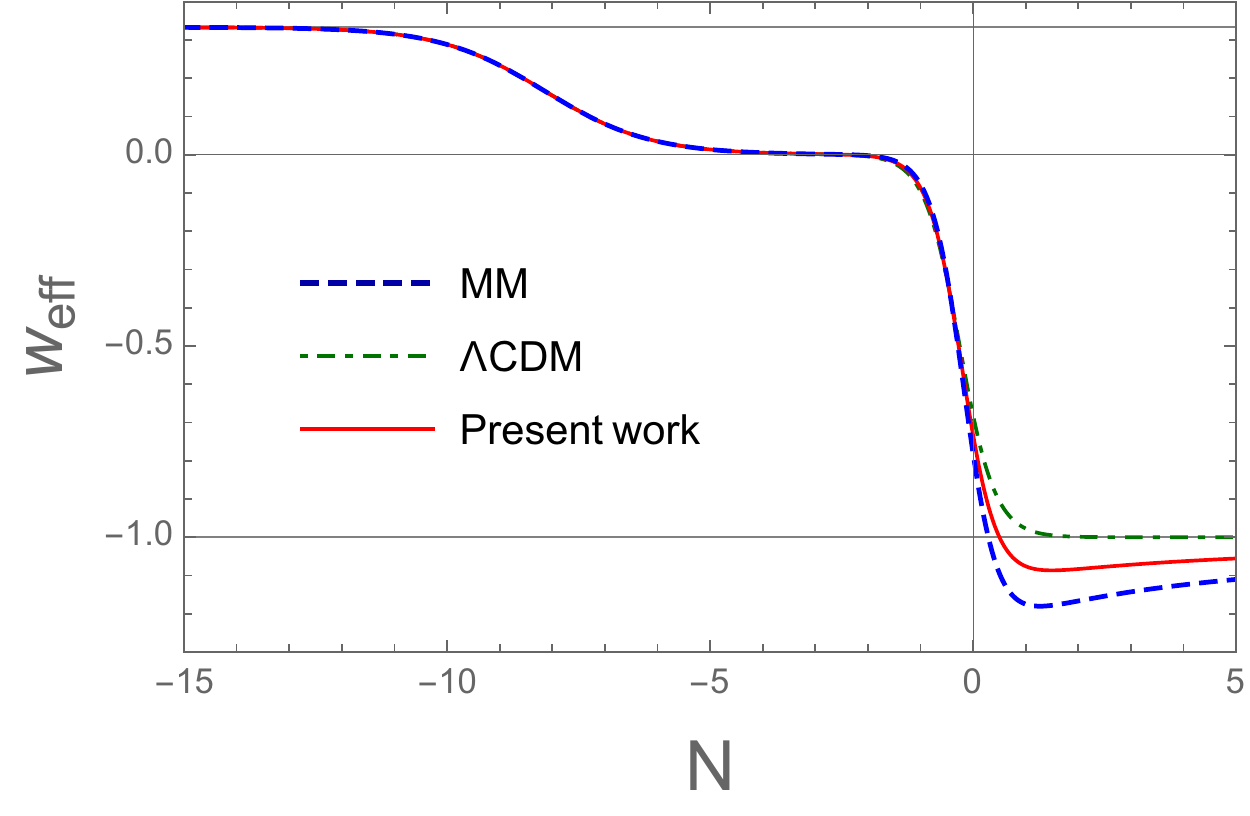}
 \includegraphics[height=5cm]{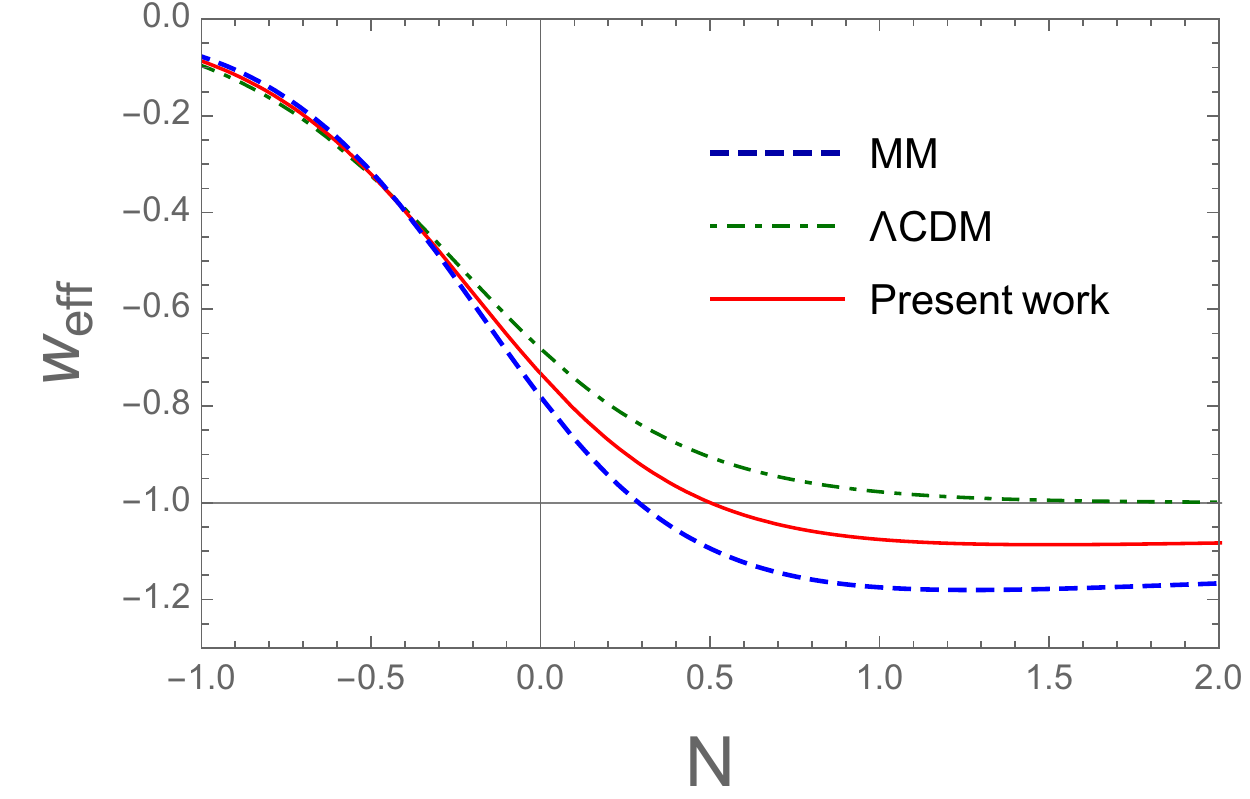}
\caption{\label{fig:w_eff} {\it Left panel:} Evolution of the effective equation of state $w_\text{eff}$ as a function of the number of $e$-folds $N$ for the present work (red, solid curve), as well as for the MM model (blue, dashed curve) and $\Lambda$CDM (green, dot-dashed curve). {\it Right panel:} Zoomed version of the same curves as in the left panel for shorter periods of time in the past and in the future.}
\end{figure}

Although our model seems to successfully describe the three phases in the expansion history as observations require in terms of the evolution of the density parameters, this is not sufficient for the viability of the model; we further need to study the properties of the energy density at each epoch in terms of the evolution of the effective equation of state $w_{\text{eff}}$, which is shown in Fig.~\ref{fig:w_eff}. This has been calculated simply through Eq. (\ref{eq:weff_def}) and the evolution of $h$ and $h^{\prime}$. The figure clearly shows that $w_{\text{eff}}$ evolves very closely to its evolution in both $\Lambda$CDM and the MM model. The three epochs of radiation, matter, and dark-energy domination can now be seen with $w_{\text{eff}}$ starting with $w=1/3$ at very early times (radiation domination), decreasing to $w=0$ (matter domination), and then becoming negative at late times (dark-energy domination). Although there are differences in $w_{\text{eff}}$ predicted by the three models at late times, the differences are more pronounced in the future. The asymptotic values of $w_{\text{eff}}$ are the same in all three models ($=-1$), but contrary to $\Lambda$CDM, where $w_{\text{eff}}$ always remains larger than $-1$, both the MM model and ours show phantom behavior in the future, with $w_{\text{eff}}$ decreasing, crossing $-1$, and then increasing again towards $-1$. It is however interesting to notice that our model predicts an effective equation of state that is closer to the $\Lambda$CDM behavior compared to the MM model. In order to understand the source of this observation in an intuitive (but handwaving) way, let us notice that the action of our model includes a nonlocal term of a lower order in $\frac{1}{\Box}R$ compared to the MM one; this makes the effects of the nonlocalities less pronounced. The nonlocal term in our model is of the form $m^{2}X$, with $X\equiv\frac{1}{\Box}R$, while the nonlocality in the MM model is of the $m^{2}X^{2}$ form. $X$ in an FLRW cosmology is a double time-integral of $R$~\cite{Maggiore:2014sia},
\begin{equation}\label{iBoxDW}
X(t)=-\int_{t_*}^{t} \dd t'\, \frac{1}{a^3(t')}
\int_{t_*}^{t'}\dd t''\, a^3(t'') R(t'')\, ,
\end{equation}
where $t_*$ is some initial time at which the nonlocal effects kick in. Since `integral' is a cumulative quantity, $X$ generically increases with time. Considering the cosmological constant in $\Lambda$CDM as a term of zeroth order in $X$, i.e. of the form $m^{2}X^0$, it can give a qualitative explanation for why the evolution curves in our model lie between the ones for $\Lambda$CDM and the MM model. One can consider both nonlocal models as cases with some time-dependent and growing cosmological constants, with the growth rate higher in the MM model compared to ours. This suggests that the higher the power in $\frac{1}{\Box}R$ in the action, the larger the deviation from $\Lambda$CDM.

\begin{figure}
 \includegraphics[height=5cm]{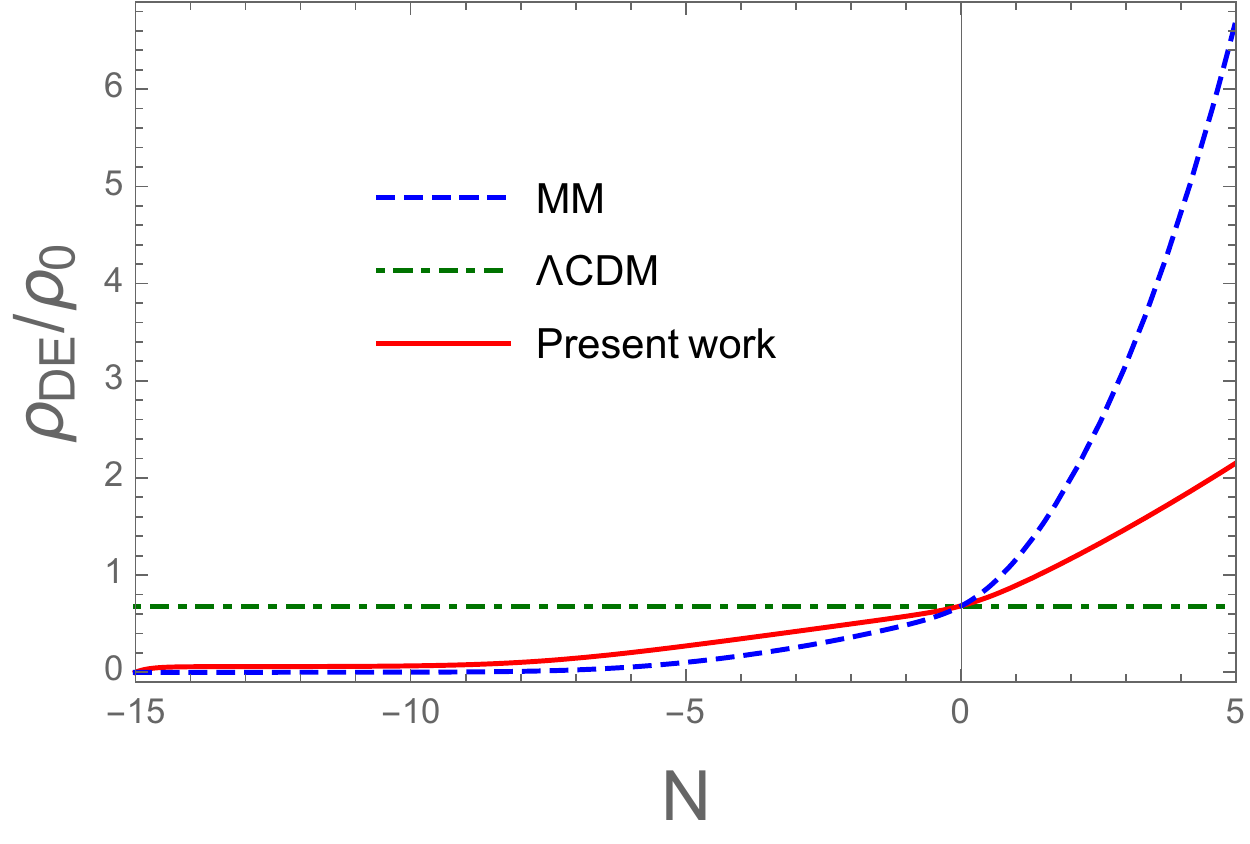}
 \includegraphics[height=5cm]{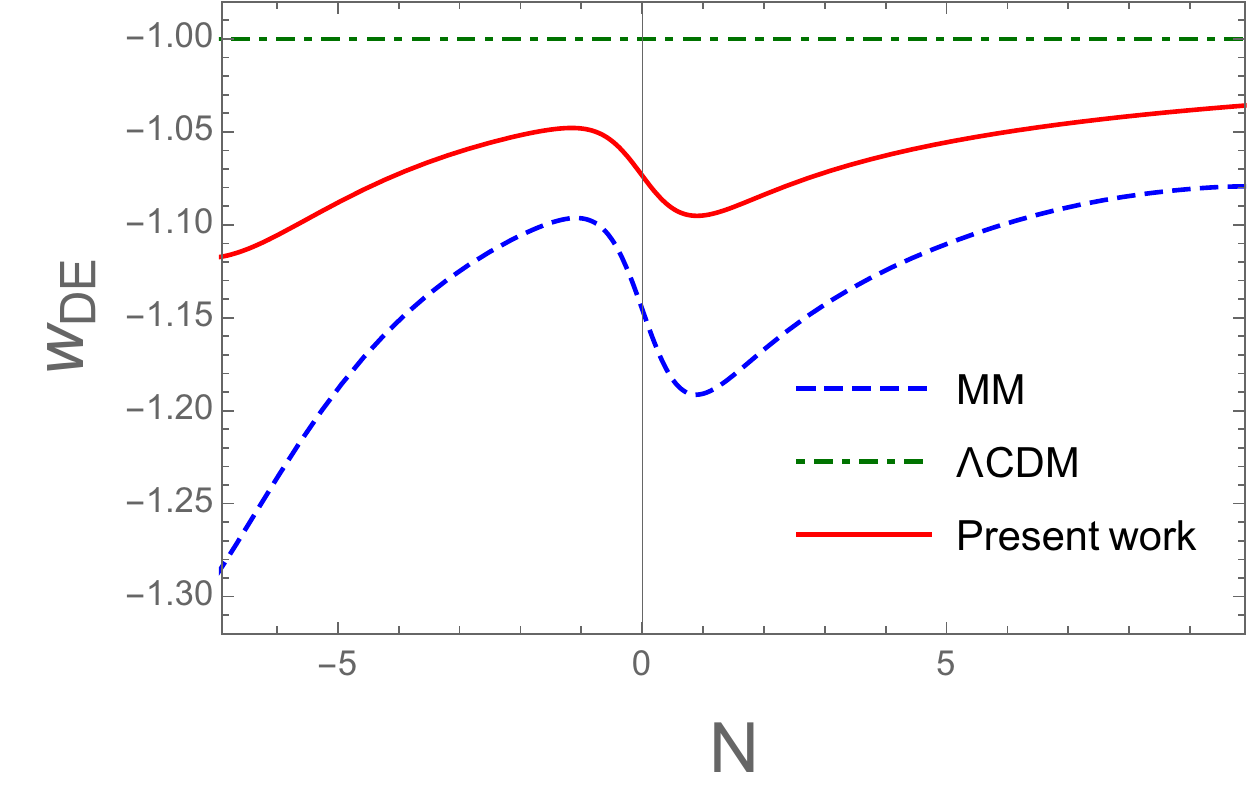}
\caption{\label{fig:wrhoDE} {\it Left panel:} Dark-energylike contribution to the energy density of the Universe $\rho_{\text{DE}}$ divided by the total energy density today $\rho_0$ as a function of the number of $e$-folds $N$ for the present work (red, solid curve), as well as for the MM model (blue, dashed curve) and $\Lambda$CDM (green, dot-dashed curve). {\it Right panel:} The same as in the left panel, but for the dark-energy equation of state $w_{\text{DE}}$.}
\end{figure}

The effective equation of state is the key quantity in constraining the model by background observations, but it is interesting to also investigate the properties of the dark-energy contribution itself through the study of the evolution of its equation of state $w_{\text{DE}}$ and energy density $\rho_{\text{DE}}$. $w_{\text{DE}}$ in our model, i.e. the same quantity as $w_{\text{NL}}$, can be obtained by solving the continuity equation (\ref{eq:contNL}) using the evolution of $\rho_{NL}$ and its time derivative. Both of these quantities are shown in Fig.~\ref{fig:wrhoDE} for our nonlocal model, as well as for the $\Lambda$CDM and MM models, as functions of the number of $e$-folds $N$. As expected, in both MM model and ours, $\rho_{\text{DE}}$ evolves from zero in the past and during radiation domination to a nonzero value today (for dark-energy domination), in contrast to $\Lambda$CDM with a constant $\rho_{\text{DE}}$. This is because the nonlocal terms behave as integration terms over time, and do not exist in the far past. Another interesting observation is the behavior of $\rho_{\text{DE}}$ in the future. While it remains constant in $\Lambda$CDM, it increases with time in our model, as it does so in the MM model. The rate of increase is however lower in the former, implying again that the future behavior of the Universe is now closer to $\Lambda$CDM. This can be seen also from the evolution of $w_{\text{DE}}$ in Fig.~\ref{fig:wrhoDE}, where the nonlocal terms in both nonlocal models have phantomlike behavior over the entire history of the Universe (i.e. with $w_{\text{DE}}< -1$). The deviation from the $\Lambda$CDM value of $w_{\Lambda}=-1$ is however smaller in our nonlocal model compared to MM. In addition, although $w_{\text{DE}}$ changes its behavior around the present time in both models, the change is more dramatic in the MM case.

Using the Chevallier-Polarski-Linder (CPL) parametrization \cite{2001IJMPD..10..213C,2003PhRvL..90i1301L},
\begin{equation}
w(z)= w_0 + w_az/(1+z),\label{eq:CPL}
\end{equation}
where $z$ is redshift, $w_{\text{DE}}$ in our model is best fit by $w_0=-1.075$ and $w_a=0.045$. This parameterization is however valid only near the present time (in the region $-1<N<0$), and therefore, Eq. (\ref{eq:CPL}) cannot be used to fit the equation of state at early times or in the future. These values can be compared with the ones for the MM model, i.e. $w_0=-1.144$ and $w_a=0.084$~\cite{Maggiore:2014sia}, which again shows that our nonlocal model gives an expansion history closer to $\Lambda$CDM compared to the MM model.

\subsection{Comparison with Deser and Woodard's $\alpha R\frac{1}{\Box}R$ model}\label{sec:DW}

We have seen so far that, similarly to the MM model, our $m^{2}\frac{1}{\Box}R$ nonlocal model provides a viable cosmic expansion history. It is interesting to now take a more detailed look, in terms of the background cosmology, into the other nonlocal model that we mentioned in Sec.~\ref{sec:intro}, i.e. the simplest version of the DW model~\cite{Deser:2007jk}, with $f(\frac{1}{\Box}R)=\frac{1}{\Box}R$, which is very similar to our model in terms of the structure of the nonlocal term. The action for this model is of the form
\begin{equation}
S_{\text{DW}}=\frac{M_{\text{Pl}}^{2}}{2}\int \dd^{4}x\sqrt{-g}(R-\alpha R\frac{1}{\Box}R)+S_{\text{matter}}[g,\Psi],\label{eq:DWaction}
\end{equation}
where $\alpha$ is a dimensionless, free parameter. Let us first localize the model as usual, by defining the auxiliary field $U$,
\begin{equation}
U\equiv\frac{1}{\Box}R.\label{eq:DWdefU}
\end{equation}
Substituting $U$ into the action and imposing the above condition by introducing a Lagrange multiplier $\lambda$, we obtain the local action
\begin{equation}
S_{\text{DW}}=\frac{M_{\text{Pl}}^{2}}{2}\int \dd^{4}x\sqrt{-g}R(1-\alpha RU)+\int \dd^{4}x\sqrt{-g}\lambda(R-\Box U)+S_{\text{matter}}[g,\Psi].\label{eq:DWactionlocal}
\end{equation}
Varying the action with respect to $\lambda$ gives nothing but the constraint (\ref{eq:DWdefU}). By varying the action with respect to the auxiliary field $U$ we obtain
\begin{equation}
\lambda =-\frac{M_{\text{Pl}}^{2}}{2}\alpha U,
\end{equation}
which can now be used to replace $\lambda$ in the action with a combination of $\alpha$ and $U$. This shows that this model needs only one auxiliary field, $U$, for localization. Finally, we can obtain the modified Einstein equations by varying the action (\ref{eq:DWactionlocal}) with respect to $g_{\mu\nu}$, which gives
\begin{equation}
(1-2\alpha U)G_{\mu\nu}+\frac{1}{2}\alpha\nabla^{\rho}U\nabla_{\rho}Ug_{\mu\nu}-2\alpha R g_{\mu\nu}+2\alpha\nabla_{\mu}\nabla_{\nu}U-\alpha\nabla_{(\mu}U\nabla_{\nu)}U=\frac{1}{M_{\text{Pl}}^{2}}T_{\mu\nu}.
\end{equation}
It is easy to also show that the Bianchi identity and conservation of $T_{\mu\nu}$ do not impose any extra constraints on the model and are identically satisfied, as the covariant derivatives of the extra terms in the Einstein equations are identically zero.

Since we are interested in the background cosmology of the model, we assume the FLRW metric
\begin{equation}
g_\mn \dd x^\mu \dd x^\nu = -\dd t^2 + a^2(t)\delta_{ij}\dd x^i \dd x^j, \label{eq:FLRWgDW}
\end{equation}
and obtain the modified Friedmann equation
\begin{equation}
h^{2}\left(1-2\alpha U-\frac{2}{3}\alpha(U^{\prime}\xi+U^{\prime\prime})+\frac{1}{6}\alpha(U^{\prime})^{2}-4\alpha(\xi+2)\right)=\frac{1}{3H_{0}^{2}M_{\text{Pl}}^{2}}\rho=\Omega_{\text{M}}^{0}e^{-3N}+\Omega_{\text{R}}^{0}e^{-4N},\label{eq:DWFriedmann}
\end{equation}
with $a$, $h$, $\xi$, $\rho$, $\Omega_{\text{M}}^{0}$, $\Omega_{\text{R}}^{0}$, $N$, and the prime being defined as in previous sections. The constraint (\ref{eq:DWdefU}) now reads
\begin{equation}
U^{\prime\prime}+(\xi+3)U^{\prime}+6(\xi+2)=0.\label{eq:DWdefUFLRW}
\end{equation}
Using this equation, Eq. (\ref{eq:DWFriedmann}) can be written in the simpler form
\begin{equation}
h^{2}=\frac{\Omega_{\text{M}}^{0}e^{-3N}+\Omega_{\text{R}}^{0}e^{-4N}}{1-\alpha(2U-2U^{\prime}-\frac{1}{6}{U^{\prime}}^{2})}.\label{eq:DWFriedmann2}
\end{equation}
We can now use Eq. (\ref{eq:DWFriedmann2}) and its time derivative to obtain an expression for $\xi$, which we can then plug into Eq. (\ref{eq:DWdefUFLRW}) and solve the resulting equation numerically. This gives us the evolution of $\xi$, and therefore the effective equation of state $w_{\text{eff}}=-1-\frac{2}{3}\xi$. As usual, we set the initial conditions $U(N=-15)=U^{\prime}(N=-15)=0$.

\begin{figure}
 \includegraphics[height=5cm]{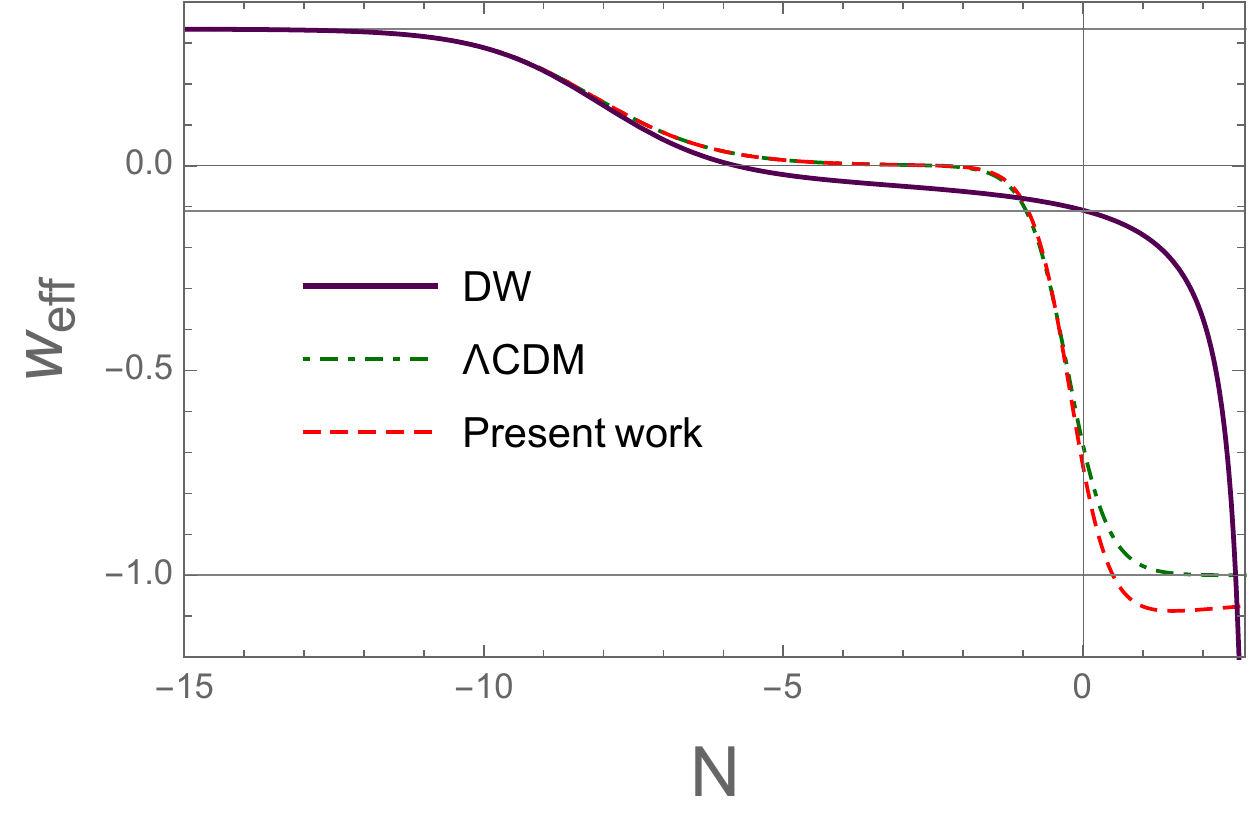}
 \includegraphics[height=5cm]{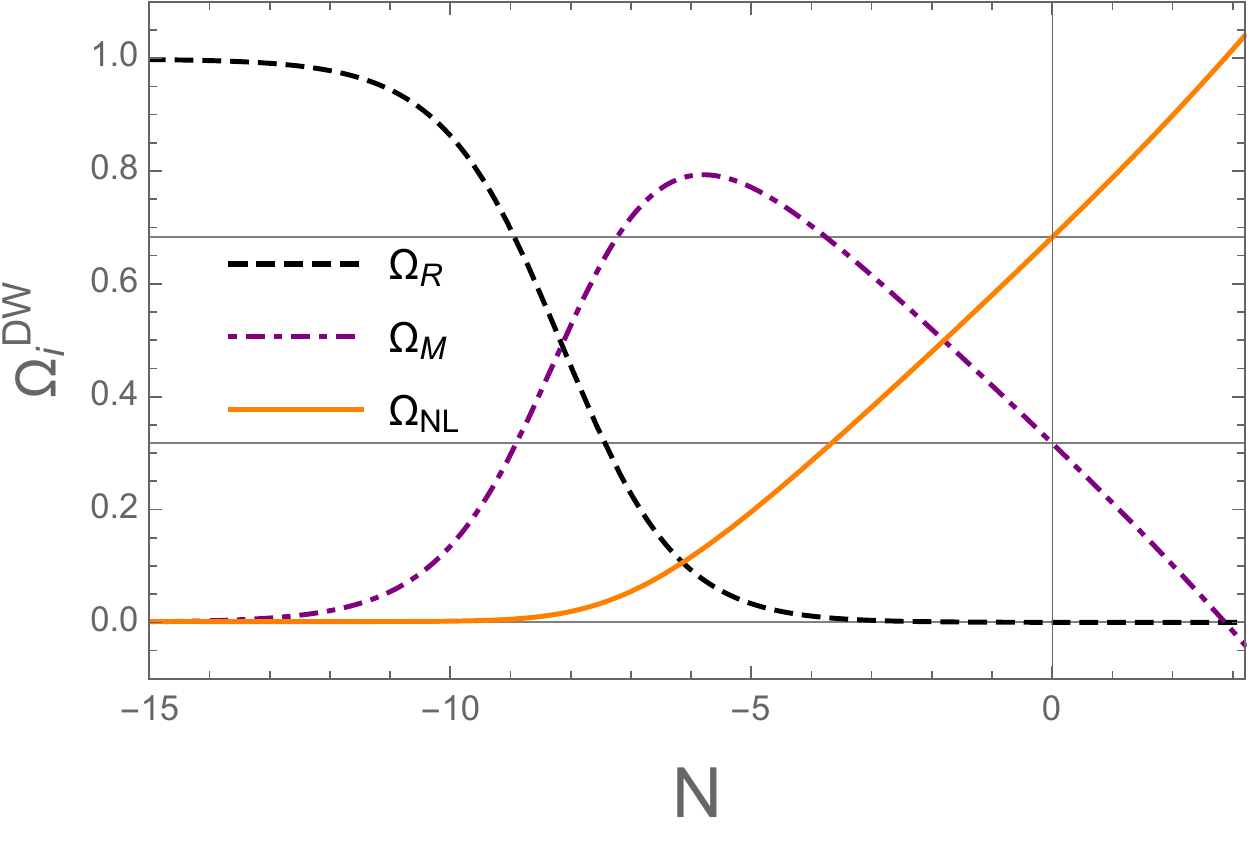}
\caption{{\label{fig:DW}}{\it Left panel:} Evolution of the effective equation of state $w_\text{eff}$ as a function of the number of $e$-folds $N$ for the DW $\alpha R\frac{1}{\Box}R$ model (black, solid curve), as well as for the model presented in this paper (red, dashed curve) and $\Lambda$CDM (green, dot-dashed curve). {\it Right panel:} Evolution of the density parameters $\Omega_{\text{R}}$ (black, dashed curve), $\Omega_{\text{M}}$ (violet, dot-dashed curve), and $\Omega_{\text{NL}}$ (orange, solid curve) as functions of the number of $e$-folds $N$ for the DW $\alpha R\frac{1}{\Box}R$ model. The lower and upper, grey, horizontal lines represent $\Omega^{0}_{\text{M}}$ and $1-\Omega^{0}_{\text{M}}-\Omega^{0}_{\text{R}}$, respectively.}
\end{figure}

Fig.~\ref{fig:DW} (left panel) shows $w_{\text{eff}}$ as a function of $N$ for the DW model, when we set $\alpha=-0.02285$ (this choice corresponds to the choice made in Ref.~\cite{Koivisto:2008xfa}).\footnote{Note that the DW model possesses two branches of solutions, one with positive and one with negative values of $\alpha$. The branch corresponding to positive $\alpha$ does not provide a dark-energy-domination phase at late times and in the future, and we therefore do not consider it here. See Ref.~\cite{Koivisto:2008xfa} for details.} For comparison, we have plotted $w_{\text{eff}}$ also for our nonlocal $m^{2}\frac{1}{\Box}R$ model and for $\Lambda$CDM. As it has already been discussed in the literature~\cite{Koivisto:2008xfa}, the simple DW model of $\alpha R\frac{1}{\Box}R$ does not provide a viable background evolution. It does not give a proper matter-domination epoch, and $w_{\text{eff}}$ does not become sufficiently negative at recent times ($w_\text{eff}$ obtained for today is $\approx-0.1$). Additionally, the model contains a finite-time singularity in the future, which can be seen from the rapid decrease in $w_{\text{eff}}$, even though it is not necessarily a problem. We have shown in Fig.~\ref{fig:DW} (right panel) also the evolution of the density parameters for radiation, matter, and the dark-energylike nonlocality, which again shows that the model is not able to produce a proper matter-domination epoch. It is however interesting to note that although the model does not provide a viable cosmic history, the dark-energy-domination epoch occurs at late-times with a choice of the $\alpha$ parameter that is not too small, and therefore not fine-tuned. It has proven difficult to come up with a model of dark energy or modified gravity that is able to provide a viable cosmic acceleration without introducing a new mass scale (normally of the order of $H_{0}$). Therefore, the fact that the DW model provides an evolution, which, although being unviable, is not too far from the observed expansion history, by introducing only one dimensionless parameter with a natural value, encourages searches for viable {\it scale-free} models of dark energy.

Finally, we should note that the arguments presented here are valid only for the simple type of DW model, i.e. when the {\it free} function $f$ in the full formulation of the model takes the simple form $f(\frac{1}{\Box}R)=\frac{1}{\Box}R$. By {\it tuning} the form of the function $f$ it is possible to resolve this problem and reconstruct any arbitrary (and viable) cosmic histories, including $\Lambda$CDM~\cite{Koivisto:2008xfa,Deffayet:2009ca}, as discussed in Sec.~\ref{sec:intro}. In those cases, however, the model would need a larger number of free parameters, and the form of the function $f$ would become quite contrived, rendering the model less appealing.

\subsection{Existence of $f$-metric solutions in the original two-metric framework}\label{sec:solutions4f}

Although, as we argued before, our single-metric model (\ref{eq:m2model}) captures all the phenomenological implications of our original two-metric model, as long as we are interested in only the $g$-metric and matter dynamics, it is interesting to see whether the field equations governing the dynamics of the reference metric $f_{\mu\nu}$ could be satisfied cosmologically if we would still want to keep it dynamical, and vary the action also with respect to the reference metric. A positive (or negative) answer to this question does not affect the validity of our cosmological results, as the effects of the reference metric on physical quantities and their evolution have already been taken into account through the contribution of $f_{\mu\nu}$ to the $g$-metric field equations through $R_{f}$. It is however interesting to see whether the strong constraint imposed on the reference metric, namely $R_{f}$ being a constant, could be satisfied dynamically. We study this question in this section for the background evolution of $f_{\mu\nu}$, and leave the more interesting case of perturbations for future work.

Assuming the reference metric $f_{\mu\nu}$ to be, similarly to $g_{\mu\nu}$, described by an FLRW metric, and again specializing to a spatially flat universe and working in cosmic time $t$, we have
\begin{align}
f_\mn \dd x^\mu \dd x^\nu &= -X^2(t)\dd t^2 + Y^2(t)\delta_{ij}\dd x^i \dd x^j.\label{eq:FLRWf}
\end{align}
Here, $X(t)$ and $Y(t)$ are the lapse and scale factor of the reference metric $f_{\mu\nu}$, respectively, both being functions of time only. The choice for the form of the physical metric $g_{\mu\nu}$ is as before, i.e. Eq. (\ref{eq:FLRWg}). Because of general covariance, we can still freely choose the cosmic-time coordinate for $g_{\mu\nu}$ and set $g_{00}=-1$. These simple forms for the background metrics significantly simplify the $f$-metric field equations (\ref{eq:einstein2}) together with Eqs. (\ref{eq:DeltaGflocal}), for example through the simple forms for quantities like $\sqrt{f^{-1}g}$,
\begin{equation}
\sqrt{f^{-1}g}=\frac{a^{3}}{XY^{3}}.
\end{equation}
Let us now consider the implications of the Bianchi constraint (\ref{Bianchiconstraint}), which for the FLRW form of $f_{\mu\nu}$ implies a {\it constant} Ricci curvature,
\begin{equation}
\dot{R_{f}}=0,\label{eq:BianchiFLRW}
\end{equation}
with an overdot denoting a derivative with respect to cosmic time. If the Ricci scalar is zero, then $f_{\mu\nu}$ is, e.g., of a Minkowski form,\footnote{\label{ftnt:Minkowski}Note that $R_{f}=0$ does not imply only a Minkowski form for $f_{\mu\nu}$. For example, an FLRW metric for a universe filled with only radiation also has a zero Ricci curvature. However, since in our model $f_{\mu\nu}$ is not sourced directly by matter, we do not need to consider such dynamical choices for the reference metric in that case, and can simply assume that it is Minkowski. Our cosmological solutions for the physical metric are independent of the actual dynamics of the reference metric, and the only property of the metric that enters our calculations is that $R_{f}$ is zero. We can therefore use a ``Minkowski reference metric'' and a ``reference metric with zero curvature'' interchangeably.} and if it is constant but nonzero, then the metric is of, e.g, a de Sitter (or anti-de Sitter) form.\footnote{\label{ftnt:deSitter}Again, this case can correspond to various types of metrics for $f_{\mu\nu}$. However, what matters for our physical equations of motion is only the Ricci scalar, and not the specific form of the reference metric. Therefore, as long as there are solutions for a de Sitter $f_{\mu\nu}$ consistent with all the equations, we can restrict ourselves to such a form for $f_{\mu\nu}$, and use a ``de Sitter reference metric'' and a ``reference metric with constant and nonzero curvature'' interchangeably.} We have already shown in the previous sections that $R_{f}=0$ (corresponding to $m^2=0$) does not provide a viable background evolution; let us therefore focus on a constant but nonzero $R_{f}$.

Plugging the FLRW expressions for the metrics (\ref{eq:FLRWg}) and (\ref{eq:FLRWf}) into the $(0,0)$ component of the $f$-metric field equations, we can obtain the Friedmann equation corresponding to the reference metric. We, additionally, explicitly assume a de Sitter form for $f_{\mu\nu}$. For such a maximally symmetric metric with a constant curvature $R_{f}$, we have (in four dimensions)
\begin{align}
R_{\rho\sigma\mu\nu}^{f}&=\frac{R_{f}}{12}(f_{\rho\mu}f_{\sigma\nu}-f_{\rho\nu}f_{\sigma\mu}),\\
R_{\mu\nu}^{f}&=\frac{R_{f}}{4}f_{\mu\nu},\\
G_{\mu\nu}^{f}&=R_{\mu\nu}^{f}-\frac{1}{2}R_{f}f_{\mu\nu}=-\frac{1}{4}R_{f}f_{\mu\nu},
\end{align}
for the Riemann, Ricci, and Einstein tensors, respectively. The $(0,0)$ component of the field equations for $f_{\mu\nu}$ then yields
\begin{equation}
3\frac{Y^\prime}{YX^2}h^2(u^\prime+3u-3u\frac{Y^\prime}{Y}-u\frac{X^\prime}{X})=\frac{R_f}{4H_0^2XY^{3}}u+\frac{R_f}{8H_0^2\alpha a^3}.\label{eq:friedmannf}
\end{equation}
The Ricci scalar of the $f$-metric, with a de Sitter form, is, on the other hand, given by
\begin{equation}
R_{f}=12H_0^2(\frac{Y^\prime h}{XY})^{2}.\label{eq:deSitterRf}
\end{equation}

The question we need to answer now is whether Eqs. (\ref{eq:friedmannf}) and (\ref{eq:deSitterRf}) can be simultaneously satisfied, together with the field equations for the physical metric $g_{\mu\nu}$. In order to do this, we plugged the numerical solutions for $u$, $a$, and $h$, found in the previous sections, into the above equations, and solved them for $X$ and $Y$ (keeping $R_f$ constant). Our numerical studies show that consistent solutions exist for both $X$ and $Y$, and we therefore conclude that a de Sitter form for $f_{\mu\nu}$ exists for the background dynamics of the model. An important point to note is that, contrary to the $g$-metric equations, which {\it see} the $f_{\mu\nu}$ only through the appearance of $R_{f}$ in the interactions, and therefore the quantities $\alpha$ and $R_{f}$ appear in the equations only through the combination $\alpha R_{f}$ (which we denoted collectively by $-\frac{1}{2}m^2$), here the two quantities appear separately in the $f$-metric Friedmann equation. The reason is, clearly, that $R_{f}$ appears in both kinetic and interaction terms for $f_{\mu\nu}$, one in combination with $\alpha$ and one independently. This means that, as long as the evolution equations for $f_{\mu\nu}$ are concerned, the model possesses two independent parameters $\alpha$ and $R_{f}$. In our numerical searches of solutions for the background $f_{\mu\nu}$, we fixed $R_{f}$ and $\alpha$ to $\frac{1}{2}m^2$ and $1$, respectively. Whether or not a solution exists also at the perturbative level is beyond the scope of the present paper, and we leave its investigation for future work.

\section{Auxiliary fields and the problem of ghosts}
\label{sec:ghostsdof}

In this section, we discuss the problem of apparent ghosts in the local formulation of our $m^{2}\frac{1}{\Box}R$ model, which is a generic feature of all nonlocal models, including the DW and MM, as discussed in Sec.~\ref{sec:intro}. We first show that the local formulation contains a ghost, and then argue that the ghost is not harmful to the theory, for similar reasons as in other nonlocal models. We prove, by analyzing the model in its nonlocal formulation, that the physical degrees of freedom of the theory are the same as in GR, and furthermore, that they are not affected by the local (auxiliary) ghost, and remain healthy.

\subsection{Ostrogradski ghosts in the local formulation}
\label{sec:ghosts}

Let us rewrite the action (\ref{eq:m2modellocalaction}) as
\begin{equation}
S =\frac{M_{\text{Pl}}^{2}}{2}\int \dd^{4}x\sqrt{-g}[(1-2\alpha V)R+m^{2}U+2\alpha V\Box U]+S_{\text{matter}}[g,\Psi],
\end{equation}
which is formulated in the Jordan frame, as the gravity sector is modified while matter is minimally coupled to gravity. Let us now change the frame to Einstein through the transformations
\begin{align}
g_{\mu\nu}&\rightarrow\Omega^{2}g_{\mu\nu},\\
R&\rightarrow\frac{1}{\Omega^{2}}[R-6(\Box \ln\Omega+g^{\mu\nu}\nabla_{\mu}\ln\Omega\nabla_{\nu}\ln\Omega)],
\end{align}
by introducing
\begin{equation}
\Omega^{2}\equiv\frac{1}{1-2\alpha V}.
\end{equation}
Substituting all these into the action, it takes the form
\begin{equation}
S = \frac{M_{\text{Pl}}^{2}}{2}\int \dd^{4}x\sqrt{-g}[R-6(\Box\ln\Omega+g^{\mu\nu}\partial_{\mu}\ln\Omega\partial_{\nu}\ln\Omega)-2\alpha\Omega^{2}g^{\mu\nu}\partial_{\mu}V\partial_{\nu}U]+S_{\text{matter}}[\Omega^{2}g,\Psi],
\end{equation}
which represents the action in the Einstein frame, with matter now coupled to both the metric $g_{\mu\nu}$ and the scalar field $\Omega$. In order to write the action in a canonical form for scalar-tensor theories, we introduce the new fields $\phi$ and $\psi$,
\begin{align}
\phi&\equiv\ln\Omega=-\frac{1}{2}\ln(1-2\alpha V)\Rightarrow V=\frac{1}{2\alpha}(1-e^{-2\phi}),\\
\psi&\equiv U.
\end{align}
Discarding the boundary terms, the action can now be written in terms of $\phi$ and $\psi$,
\begin{equation}
S = \frac{M_{\text{Pl}}^{2}}{2}\int \dd^{4}x\sqrt{-g}[R-6g^{\mu\nu}\partial_{\mu}\phi\partial_{\nu}\phi-2g^{\mu\nu}\partial_{\mu}\phi\partial_{\nu}\psi+m^{2}\psi]+S_{\text{matter}}[e^{2\phi}g,\Psi].
\end{equation}
By looking at the kinetic matrix for the scalar fields $\phi$ and $\psi$,
\begin{equation}
\left[\begin{array}{cc}
-6 & -1\\
-1 & 0
\end{array}\right],
\end{equation}
we notice that its determinant is always negative, meaning that the matrix is negative definite, signalling the presence of an Ostrogradski ghost.

Let us remind ourselves of the fatal consequences of ghosts (see, e.g., Refs.~\cite{Woodard2006,Sbisa2014}). The resulting unboundedness of the Hamiltonian of the system from below can make the classical theory fully unstable. Although such instabilities could be acceptable if the unstable modes do not grow too rapidly at both the background and perturbative levels, keeping the theory consistent with observations, the ghosts are definitely fatal at the quantum level. They instantaneously reach states with arbitrarily large negative energies, and therefore decay into matter particles very quickly, filling the Universe with an unacceptably large amount of particles. The theory, therefore, does not possess a stable and well-defined vacuum, and should be rejected. We should note however that there are ways to avoid such a disastrous scenario by modifying the decay rate of the ghost fields through the violation of Lorentz invariance above some energy scale where new physics appears, making the decay time larger than the age of the Universe~\cite{Carroll2003,Cline2003,KaplanSundrum2005,Konnig:2016idp}. Since we do not violate Lorentz invariance in our model, the appearance of the ghost in the local formulation may seem fatal, rendering the model excluded. As we discussed in Sec.~\ref{sec:intro}, and detail in the next subsection, the ghost in our model is only an auxiliary field and not a physical degree of freedom. It does not affect the healthiness of the theory as long as we keep in mind that the localized theory must be equivalent to the original nonlocal one by imposing appropriate initial conditions on the auxiliary fields. In that case, they do not affect the physical degrees of freedom and the theory remains viable.

\subsection{Nonlocal formulation and the number of physical degrees of freedom}
\label{sec:dof}

Let us now analyze the $m^{2}\frac{1}{\Box}R$ model in its original formulation and without localization. The Einstein field equations for the model are identical to those corresponding to the physical metric $g_{\mu\nu}$ presented in Sec.~\ref{sec:NLeom} for the two-metric model, when the quantity $-2\alpha R_{f}$ is replaced by $m^{2}$. The field equations then read
\begin{align}
G_{\mu\nu}+\Delta G_{\mu\nu} & =\frac{1}{M_{\text{Pl}}^{2}}T_{\mu\nu},\label{eq:einstein1m2}
\end{align}
where $G_{\mu\nu}$ is the Einstein tensor, and $\Delta G_{\mu\nu}$ is the nonlocal distortion term, with the form
\begin{align}
\Delta G_{\mu\nu}=&(\frac{1}{\Box}m^{2})G_{\mu\nu}+m^{2}(1-\frac{1}{2\Box}R)g_{\mu\nu}-\nabla_{\mu}\nabla_{\nu}(\frac{1}{\Box}m^{2})-\frac{1}{2}\nabla^{\rho}(\frac{1}{\Box}R)\nabla_{\rho}(\frac{1}{\Box}m^{2})g_{\mu\nu}\nonumber\\
&+\nabla_{(\mu}(\frac{1}{\Box}m^{2})\nabla_{\nu)}(\frac{1}{\Box}R).\label{eq:DeltaGNLm2}
\end{align}
Our goal here is to count the number of {\it physical} degrees of freedom in the model, and prove that they are all healthy. Similar discussions for the Deser-Woodard and Maggiore-Mancarella models have been presented in Refs.~\cite{Deser:2013uya} and~\cite{Dirian:2014xoa}, respectively.

Let us follow the procedures of Refs.~\cite{Deser:2013uya} and~\cite{Dirian:2014xoa}, and choose the synchronous gauge to write $g_{\mu\nu}$ as
\begin{equation}
g_\mn \dd x^\mu \dd x^\nu = -\dd t^2 + h_{ij}\dd x^i \dd x^j.\label{eq:synchmetric}
\end{equation}
In general relativity, the dynamical equations of motion for $h_{ij}$, i.e. those that are second-order in time derivatives, are the $(i,j)$ components of the Einstein field equations, and $G_{ij}$ reads
\begin{equation} \label{eq:Gij}
G_{ij} = \frac{1}{2}\, \ddot{h}_{ij} - \frac{1}{2}\, h_{ij} \partial_t^2 \log h + {\cal O} ( \partial_t ) \, ,~~~~~~~~~~~~h \equiv \det h_{ij}.
\end{equation}
The $(\mu,0)$ components contain at most first-order time derivatives $ {\cal O} ( \partial_t )$, and hence are constraints on the initial data of the metric, i.e. $h_{ij}(t_0), \dot{h}_{ij}(t_0)$, the number of which determines the number of propagating degrees of freedom of the theory.\footnote{For a more rigorous explanation, see, e.g., footnote 12 of Ref.~\cite{Dirian:2014xoa}.} Considering $\Delta G_{\mu 0}$ we now argue that it vanishes when evaluated at the hyper-surface of the initial conditions.

In $d+1$-dimension spacetimes, the action of the d'Alambertian operator is specified by the Green's function
\begin{equation}
\Box_{x}G(x,y)=\frac{1}{\sqrt{-g(x)}}\delta^{(d+1)}(x-y),\label{eq:Greendef}
\end{equation}
where $x$ and $y$ are two spacetime points. Using this the solution to an inhomogeneous equation $\Box_{x}f(x)=F(x)$ for a function $f(x)$ sourced by another function $F(x)$ can be found by
\begin{equation}
f(x)=(\frac{1}{\Box_{x}}F)(x)=\int \dd^{d+1}y\sqrt{-g(y)}G(x,y)F(y).\label{eq:Greenint}
\end{equation}
Therefore, all we need to compute is the Green's function $G(x,y)$ from Eq. (\ref{eq:Greendef}) and then solve the integral (\ref{eq:Greenint}) in order to compute $f(x)$. Eq. (\ref{eq:Greendef}) has two solutions, one causal (retarded) and the other one acausal (advanced). In order to maintain causality, we need to make sure that only the retarded Green's function is used, by assuming\footnote{It has however been argued, e.g. in Ref.~\cite{Foffa:2013sma}, that by interpreting the nonlocal action as a quantum effective one, causality is automatically satisfied.}
\begin{equation}
G(x,y)=0,\;\text{for all}\;y\;\text{outside the past light cone of}\;x.\footnote{As noted in Ref~\cite{Dirian:2014xoa}, the condition (\ref{eq:GreenIninitial}) can take different forms for time coordinates that are different from the synchronous one considered here.}\label{eq:GreenIninitial}
\end{equation}
This requirement, however, is not sufficient to fully specify the Green's
function, because the defining equation is a second-order differential equation
and therefore one needs to specify the initial conditions for it. Assuming that our nonlocal model is valid only below some energy scale as an effective field theory, and therefore after some initial time $t_{0}$, we therefore further assume
\begin{align}
G(x,y)\vert_{x^{0}=t_{0}}&=0,\\
\partial_{0}G(x,y)\vert_{x^{0}=t_{0}}&=0,
\end{align}
which mean that the nonlocality effects begin at $t_{0}$ and are absent before that. The immediate implication of these conditions is that, taking into account Eq. (\ref{eq:Greenint}), the quantities $\Box^{-1}F$ and their first-order time derivatives vanish at the initial time $t_0$. Now Eqs. (\ref{eq:DeltaGNLm2}) for $\Delta G_{\mu 0}$ tell us that the nonlocal distortion terms in the $(\mu,0)$ Einstein equations vanish initially except, potentially, for the term
\begin{align}
&(\nabla_{\mu}\nabla_{0}-g_{\mu 0}\Box)(\frac{1}{\Box}m^{2}).\label{eq:dofcondition_g}
\end{align}
Here, the piece that can potentially contain second-order time derivatives is
\begin{align}
&\partial^2_t-g_{00}g^{00}\partial^2_t,\label{eq:dofcondition_g2}
\end{align}
and the rest of the term includes only spatial derivatives or one time derivative. Working again in the synchronous gauge as given in Eq. (\ref{eq:synchmetric}), we then have $g_{00}g^{00}=1$, and therefore, (\ref{eq:dofcondition_g2}) vanishes, implying that (\ref{eq:dofcondition_g}) does not contain any second-order time derivatives, and $\Delta G_{\mu 0}$ then vanishes at the initial time $t=t_0$. We can therefore conclude that our nonlocal model possesses the same number of propagating degrees of freedom as in GR (similarly to the DW and MM nonlocal models).

Our arguments above have been based on the single-metric formulation of our nonlocal model, where the dynamics of the reference metric are fully ignored, for the reasons discussed in the previous sections. It is though interesting to also count the number of physical degrees of freedom for the $f_{\mu\nu}$ sector of the theory and see whether they would be affected by nonlocalities.

It turns out, however, that the situation is more subtle in this case. Let us perform a similar study of the $(\mu,0)$ components of the Einstein field equations for the reference metric. Repeating the same procedure as for $g_{\mu\nu}$, it is easy to show that the potentially dangerous terms are now
\begin{align}
&(\nabla_{\mu}^{f}\nabla_{0}^{f}-f_{\mu 0}\Box_{f})(\sqrt{f^{-1}g}\frac{1}{\Box}R).\label{eq:dofcondition_f}
\end{align}
which we need to show to vanish initially. Similarly to the $g_{\mu\nu}$ case, the piece that can potentially include second-order time derivatives is
\begin{align}
&\partial^2_t-f_{00}f^{00}\partial^2_t.\label{eq:dofcondition_f2}
\end{align}
In the two-metric theory, we only have one set of diffeomorphism invariance and one set of transformations which act on the metric like a diffeomorphism but do not transform the coordinates. The situation is similar to bimetric theories, where one does not have two independent sets of diffeomorphism invariance. This means that the only set of diffeomorphism invariance can be used to take one of the metrics (say, $g_{\mu\nu}$) to a synchronous form, but we do not have another set of diffeomorphisms to do the same for the other metric ($f_{\mu\nu}$ here). This in turn means that we cannot use the same arguments as for $g_{\mu\nu}$ for counting the number of degrees of freedom for $f_{\mu\nu}$ in general and for arbitrary solutions. We can however restrict ourselves to the specific cosmological solutions and the specific form of the reference metric that we have considered in this work. We have seen that the consistency conditions for the solutions of the two-metric model force $f_{\mu\nu}$ to be a metric with constant and nonzero Ricci scalar, {\it nonperturbatively}. Even though it is possible for such a metric to, in general, not satisfy $f_{00}f^{00}=1$, we have shown that there is at least one solution for $f_{\mu\nu}$ that possesses this property, and that is the de Sitter reference metric studied here, for which $f_{00}f^{00}=1$. This therefore shows that (\ref{eq:dofcondition_f2}) vanishes in our case, and, consequently, $\Delta G_{\mu 0}^{f}$ also vanishes at $t=t_{0}$, similarly to the $g_{\mu\nu}$ case. This means that the number of propagating degrees of freedom for the $f$ metric around the cosmological solution considered here is the same as in a GR-like theory for the reference metric. In addition, given that $f_{\mu\nu}$ is a fixed, unphysical\footnote{Note that here by $f_{\mu\nu}$ being ``fixed'' we simply mean that we {\it can} choose it to be of any arbitrary form, with a constant and nonzero $R_{f}$, for any physical system that we are interested in, as long as we can satisfy its own equations of motion. Clearly, the Bianchi constraints, forcing a constant Ricci scalar for the reference metric, do not fix its form to a specific metric (see Footnote~\ref{ftnt:deSitter}), and we always have the freedom to choose any form for it, either universally and independently of the particular physical system under investigation, or differently in different cases, as long as it has a constant and fixed $R_{f}$ in all cases. Here, therefore, by ``fixed'' we simply mean that we ``choose'' the form of $f_{\mu\nu}$ {\it before} working with the model, for example for cosmology, and then fix it to that form for the entire analysis. Also note that by calling $f_{\mu\nu}$ ``unphysical'' we simply mean that it is {\it decoupled} from matter and is {\it unobservable}, for the reasons explained in the text. Although, strictly speaking, ``unphysical'' is not a correct word, we adhere to it in this paper as it is commonly used in the field of multi-metric gravity for describing reference metrics.} metric which does not couple to matter, we conclude that our original nonlocally-interacting-metric model possesses the same number of ``physical" degrees of freedom as in GR (corresponding to the physical metric $g_{\mu\nu}$) around the cosmological backgrounds studied here. Counting the number of total propagating degrees of freedom (including those of the $f$ metric) around more general solutions requires a more detailed and careful analysis, and is beyond the scope of the present paper.

Let us however point out that even though there seem to be only $2+2$
degrees of freedom in the linear spectrum of the two-metric model for the cosmological solutions studied here, it is likely that more degrees of freedom
are present in the nonlinear theory due to the usual no-go theorems for constructing ghost-free theories of interacting massless gravitons (see the discussion in the introduction for possibilities of evading the no-go theorems when locality is violated). This will then imply
that around the cosmological backgrounds, these modes are infinitely
strongly coupled since their kinetic term does not show up at the linear
level where the number of dynamical degrees of freedom is reduced. In general, this
renders the use of linear perturbation theory invalid. If further detailed studies of the the model shows that in fact 
the full theory either has ghosts or is infinitely strongly coupled around
cosmological backgrounds, then this will serve as a major drawback to the two-metic, nonlocal model studied in this paper. These are very interesting and important questions to address and need to be investigated in a dedicated and separate work, which we leave for future. Note however that the phenomenological, simple, single-metric model of $m^2\frac{1}{\Box}R$, which all our cosmological studies in the present paper have been based on, is immune from these problems. This model can have a completely different origin, fully unrelated to interacting spin-2 fields.

Up until now, we have only shown that the nonlocal terms do not add additional propagating degrees of freedom to the theory for our physical modes corresponding to the physical metric $g_{\mu\nu}$ in general, and in our specific cosmological setup for the reference metric $f_{\mu\nu}$, if certain initial conditions are imposed. However, there is still the possibility that nonlocalities turn the graviton modes into ghosts, and therefore render the theory unstable. It is therefore important to study the propagation of graviton modes and ensure that they always stay healthy. Although a full and rigorous (perturbative) stability analysis of the solutions is necessary for our model, similar to any other modifications to GR, it goes beyond the scope of the present paper, and we leave it for future work. We, however, again follow Refs.~\cite{Deser:2013uya} and~\cite{Dirian:2014xoa}, and study the question of whether there could be any {\it ghost} instabilities from the change in the sign of the kinetic terms for gravitons, coming from nonlocalities.

\begin{figure}
 \includegraphics[height=5cm]{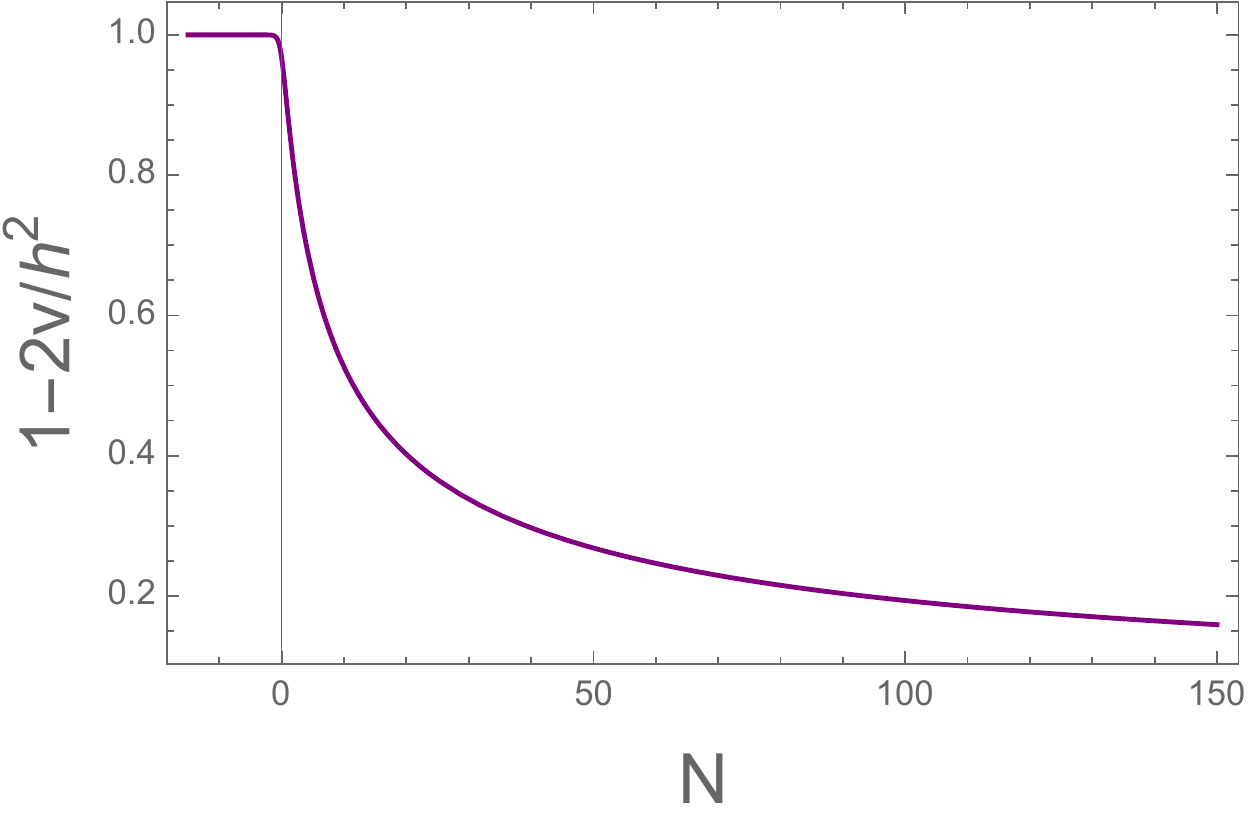}
 \includegraphics[height=5cm]{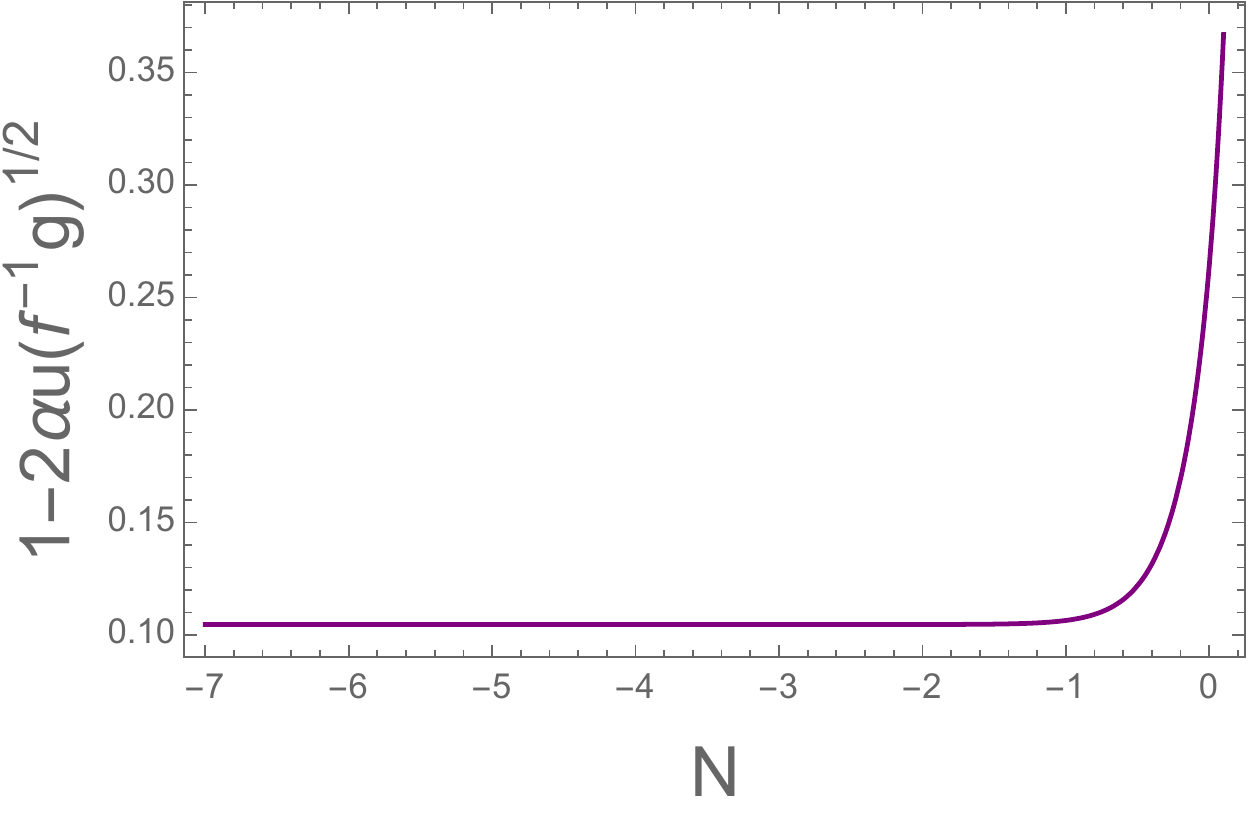}
\caption{Evolution of the quantities $1-2\frac{v}{h^{2}}$ and $1-2\alpha u \sqrt{f^{-1}g}$ as functions of the number of $e$-folds $N$.}\label{fig:v_ghost}
\end{figure}

Let us again analyze the $m^{2}\frac{1}{\Box}R$ model first. Graviton modes can in principle become ghosts if the $(i,j)$ components of the Einstein equations are allowed to contain second-order time derivatives of the metric with an overall negative sign. In order to explicitly check whether this can happen, we first note that the Ricci scalar $R$ contains at most first-order temporal derivatives on the spatial components of the metric.\footnote{In principle, $R$ contains a term $-\Box \log h$, with $h$ being the determinant of $h_{ij}$ in Eq. (\ref{eq:synchmetric}), but since $h$ is already constrained by the $(0,0)$ component of Einstein equations, this term does not involve any unconstrained second-order time derivative terms (see Ref.~\cite{Dirian:2014xoa} for more discussions).} In addition, in an expression that contains $\frac{1}{\Box}R$, $R$ is integrated over twice temporally, leaving no time derivatives in the expression. Hence, in order for terms including $\frac{1}{\Box}R$ to contain second-order time derivatives, they need to involve two time derivatives acting on $\frac{1}{\Box}R$. Let us now look at the $(i,j)$ components of the Einstein equations,
\begin{align}\label{eq:m2modelNonLocalEoM}
&(1+\frac{1}{\Box}m^2)G_{ij}+m^{2}(1-\frac{1}{2\Box}R)g_{ij}-\nabla_{i}\nabla_{j}(\frac{1}{\Box}m^2)-\frac{1}{2}\nabla^{\rho}(\frac{1}{\Box}R) \nabla_{\rho} (\frac{1}{\Box}m^2)g_{ij}\nonumber\\
&+\nabla_{(i}(\frac{1}{\Box}m^{2})\nabla_{j)}(\frac{1}{\Box}R)=\frac{1}{M_{\text{Pl}}^{2}}T_{ij}.
\end{align}
Since in Eqs. (\ref{eq:m2modelNonLocalEoM}) there are no higher than first-order time derivatives acting on $\frac{1}{\Box}R$, the only second-order time derivatives acting on the spatial components of $g_{\mu\nu}$ come from the Einstein tensor $G_{ij}$. We therefore need to check only the sign of the term multiplied by $G_{ij}$ in order to see whether it can be altered by nonlocal effects. Specifically, we need
\begin{equation}
1+\frac{1}{\Box}m^2 > 0\label{eq:noghostcon_g}
\end{equation}
in order to ensure the absence of ghostly gravitons. Let us now remind ourselves that the mass parameter $m$ should be of the order of $H_{0}$ for the model to provide a viable cosmic evolution. In addition, similarly to the other nonlocal models of DW and MM, we do not expect our model to exhibit a vDVZ discontinuity~\cite{vanDamVeltman1970,Zakharov1970}, meaning that no screening mechanism is needed for the model to reduce to GR on solar system scales if the mass scale of the theory is as low as the Hubble rate today. The term $\frac{1}{\Box}m^2$ in the condition (\ref{eq:noghostcon_g}) is therefore completely negligible on small scales. This all means that any potential danger for gravitons to turn into ghosts can only be on cosmological scales. Now, in order to see whether the condition (\ref{eq:noghostcon_g}) is satisfied cosmologically, we use its equivalent expression in terms of the auxiliary field $v$ introduced in Sec.~\ref{sec:background}, namely\footnote{Here, we only consider the background dynamics, and ignore the effects of perturbations, as long as they remain small and stable. A detailed investigation of such effects is beyond the scope of this paper, and we leave it for future work.}
\begin{equation}
1-2\frac{v}{h^{2}} > 0.\label{eq:noghostcon_g2}
\end{equation}
The evolution of $1-2\frac{v}{h^{2}}$ is presented in Fig.~\ref{fig:v_ghost} (left panel). We clearly see that it is always positive, and, therefore, the condition (\ref{eq:noghostcon_g2}) is always satisfied.\footnote{\label{ftnt:ghostcondg}Note that the behavior of $v$ at late times, depicted in Fig.~\ref{fig:scalar_fields}, is independent of the initial conditions for the auxiliary fields. It is because the solution is an attractor, and even by setting the initial conditions differently, $v$ moves very rapidly to its negative values. It is however possible for some choices of initial conditions to violate the condition (\ref{eq:noghostcon_g2}) at early times, and one should therefore be careful with such choices.}

For completeness, let us end this section by looking also at the ghost instabilities for the $f$-metric gravitons, in the original two-metric model and in our specific cosmological setup. We know that $f_{\mu\nu}$ has fixed dynamics, as it is forced to be of, e.g., a de Sitter form. In fact, $f_{\mu\nu}$ only introduces a scale in the model, $R_{f}$, which is manifest in the single-metric formulation of the model, i.e. $m^2\frac{1}{\Box}R$. The fact that $f_{\mu\nu}$ can always be fixed to a metric like de Sitter (at least at the background level) is sufficient to guarantee that $f$-gravitons are healthy. In addition, the consequence of $f$ having a fixed $R_{f}$, namely that our model can be fully formulated in terms of only one metric, implies, on its own, that the $f$-gravitons do not affect the properties of the physical $g$-gravitons and matter. For the completeness of our discussions, however, we repeat an analysis similar to that of $g_{\mu\nu}$ for the $f_{\mu\nu}$ Einstein equations in order to show explicitly that the conditions for the healthiness of $f$-gravitons are satisfied. 

Let us therefore look at the $(i,j)$ components of Eq. (\ref{eq:DeltaGfNL}). We note that the only second-order time derivatives acting on $f_{\mu\nu}$ come from either the second-order derivatives of the determinant of $f_{\mu\nu}$, or the Einstein and Riemann tensors. The determinant is constrained through the $(0,0)$ component of the Einstein equations, and its second-order derivatives do not affect the propagation of the $f$-gravitons. Remembering again that $R_{f}$ is a constant, the potentially dangerous term is therefore (setting $M_{f}=M_{\text{Pl}}$)
\begin{equation}
[1-2\alpha (\frac{1}{\Box}R) \sqrt{f^{-1}g}]G^f_{ij}.
\end{equation}
The condition for the kinetic sign of the $f$-gravitons to stay positive is therefore
\begin{equation}
1-2\alpha u \sqrt{f^{-1}g}>0,\label{eq:noghostcon_f}
\end{equation}
with $u$ being defined in Sec.~\ref{sec:background}. The cosmological evolution of $1-2\alpha u \sqrt{f^{-1}g}$ given in Fig.~\ref{fig:scalar_fields} (right panel), corresponding to the solution found for a de Sitter $f_{\mu\nu}$ discussed in Sec.~\ref{sec:solutions4f}, demonstrates that the quantity is always positive,\footnote{Note again that this has been shown only for the specific, de Sitter solution of $f_{\mu\nu}$ considered in this paper, and does not necessarily apply to all possible solutions. It therefore demonstrates the absence of ghosts only around a certain class of backgrounds.} and the condition (\ref{eq:noghostcon_f}) is always satisfied.\footnote{\label{ftnt:ghostcondf}Similarly to the comments in Footnote~\ref{ftnt:ghostcondg}, here also one should be careful with choosing the initial conditions for the auxiliary fields in order to guarantee the condition (\ref{eq:noghostcon_f}). The fact that $f_{\mu\nu}$ has a fixed and healthy dynamical form, e.g. de Sitter, requires an extra care to ensure that the initial conditions are compatible with the healthiness of $f$-gravitons.}

\section{Conclusions and outlook}
\label{sec:conclusions}

We began the paper with the question of whether two or more spin-2 fields could interact nonlocally, resulting in new models of modified gravity in the infrared, avoiding the cosmological problems present in locally interacting theories, and providing an explanation for the late-time acceleration of the Universe. Aiming at constructing models which would serve as a bridge between two classes of modifications to general relativity, namely, multimetric and nonlocal theories, we assumed nonlocal interactions between the metric of spacetime and an extra spin-2 field with no direct interaction with matter. We focused only on cases with scalar curvatures in the action, and further restricted our studies to a specific, minimal subset of possible forms for the interaction between the two metrics, which is, arguably, the simplest possible choice, inspired by the nonlocal model proposed by Deser and Woodard. We derived the field equations for both metrics, as well as the constraints imposed on the model by applying Bianchi identities and the conservation of matter energy-momentum tensor, and found that the latter would place a very strong constraint on the form of the reference metric, forcing it to have a (spatially and temporally) constant Ricci scalar, independent of the form of the physical metric. This might imply that such a simple nonlocality, with only the Ricci scalar present in the interaction, would not provide a consistent bimetric theory with a dynamical reference metric with respect to which the action is varied. We however showed in this paper that such a consistent dynamical metric does exist at the cosmological background level, and it is still an open question whether such a metric can exist also at the level of perturbations. In case the answer to this question turns out to be negative, one could then fix the reference metric to a specific form, with no kinetic term in the action and without varying the action with respect to it, similarly to the nondynamical reference metric in theories of massive gravity. However, the fact that the Bianchi constraint forces the reference metric to have a constant Ricci scalar, which is the only quantity appearing in the nonlocal interaction terms, makes the introduction of a tensorial field of little motivation, as the scalar quantity does not represent the tensorial structure of such a field. We expect, however, that the entire story will change if one includes tensorial interactions in the theory, i.e. terms involving the Ricci and Riemann tensors, $R_{\mu\nu}$ and $R_{\mu\nu\alpha\beta}$. In addition, one could add the ghost-free local interaction terms to the theory. In either case the strong Bianchi constraint is expected to be violated, and therefore the theory could possess nontrivial and consistent implications, with potentially very rich phenomenologies. It is also interesting to study possible ways of obtaining nonlocally interacting multimetric theories from underlying local theories, for example by properly integrating out light fields interacting with spin-2 fields. This would provide models with nontrivial and interesting phenomenologies. These are all exciting routes to explore, and we leave them for future work.

Inspired by our first and simple attempt at building models of nonlocally interacting metrics, we have proposed in this paper a new, simple, {\it single-parameter} model of modified gravity on cosmological scales that successfully provides a mechanism for the late-time accelerated expansion of the Universe, without an explicit cosmological constant. The model has a simple structure, similar to the simplest version of the Deser and Woodard's nonlocal model. We have derived the equations of motion, and the model has then been localized by introducing two auxiliary scalar fields, providing a framework in which the equations of motion can be handled more easily, and potentially solved for cosmology and any other systems of interest. We have then studied the cosmology of the model at the background level, investigating whether viable expansion histories could be achieved. Our detailed studies have shown that one can indeed obtain a cosmic evolution consistent with the observed one. In particular, we have demonstrated that the model provides all three epochs of radiation, matter and dark energy domination, in agreement with observations. The effective equation of state has also been calculated, and shown to be consistent with its observed evolution. The cosmic history in our model closely resembles that of $\Lambda$CDM over the entire evolution (even more so than the nonlocal model of Maggiore and Mancarella), and the effective equation of state approaches $-1$, although the Hubble rate increases in the future in contrast to the future de Sitter phase of $\Lambda$CDM with the Hubble rate becoming constant. The effective equation of state today is slightly smaller than $-1$, and first decreases for about one $e$-fold in the future and then continuously increases towards $-1$. Even though the present dark-energy equation of state is very close to the $\Lambda$CDM value, with the CPL parameters $w_0\sim-1.075$ and $w_a\sim0.045$, the difference between the two is sufficiently large (about $7.5\%$ for $w_0$) that the model can observationally be distinguished from $\Lambda$CDM using the forthcoming cosmological surveys. We have also compared the cosmic background evolution to that of the simplest form of the model of Deser and Woodard with no viable cosmic history, and have shown explicitly how the two are different. In order to know whether the model proposed in this paper is consistent with all the observational data, one needs to study the formation and evolution of cosmic structure for the model. This requires a full perturbative analysis, and we leave it for future work. In addition, we have restricted ourselves in this paper to zero initial conditions for the auxiliary fields in the local formulation of the model, and it is interesting to know whether new types of viable solutions exist if the initial conditions are set differently. This would require a full phase-space analysis of the model, and remains to be investigated.

We have finally discussed the question of apparent ghosts in the local formulation of our model, where one of the two auxiliary scalar fields is always a ghost. We have demonstrated that the number of physical degrees of freedom in our single-metric, nonlocal model is the same as in general relativity, when proper initial conditions necessary for the completeness of the model are imposed. This shows that the scalar fields appearing in the local formulation are auxiliary and unphysical, and therefore the ghosts are harmless to the theory. Additionally, we have shown that the physical degrees of freedom are not affected by the ghosts, and always stay healthy.  The issue of ghosts has also been discussed for some generalizations of the single-metric model, where we have shown that there could be cases, with a particular ghost-free condition satisfied, that possess no ghosts in their local formulations.

We have also discussed the issue of ghosts and the number of degrees of freedom in the original, nonlocaly-interacting-metric model, and argued that this two-metric model contains 2+2 degrees of freedom linearly and around the cosmological solutions that we have studied in this paper. We have emphasized that the full, nonlinear model may possess more degrees of freedom if the known no-go theorems in constructing ghost-free theories of interacting, massless, spin-2 fields hold true for these nonlocal theories as well, which will then imply that the theory has ghosts or is infinitely strongly coupled around cosmological backgrounds, invalidating the application of linear cosmological perturbation theory. This may be considered as a major issue for the model, implying that this nonlocally-interacting-metric model as a full theory might be either inconsistent or not applicable to cosmology in its present form. We consider this as another motivation for extending or modifying the model. From the theoretical point of view, these are pressing issues that have not been addressed in the present paper, and the model has therefore not yet been shown to be consistent. We leave an investigation of these questions to future work, where a fully theoretical study of the model and its consistency will be performed. We however emphasize again that the single-metric, nonlocal model whose cosmological implications have been studied in this paper is immune from these issues.

Even though we have been led to our phenomenological, nonlocal, single-metric model through an investigation of nonlocally interacting metrics, it can be used as a standalone, nonlocal model for all phenomenological studies, with no reference to any two-metric scenarios. It could be considered as an effective model originating from a completely different, more fundamental, underlying theory. We leave the investigation of such possibilities for future work. It is interesting that even though we tried to construct our model through two metrics interacting nonlocally, the consistency constraints on the solutions of the simplest, scalar-based models were so stringent that the final model became effectively single-metric with a simple structure. Whether the model could be obtained in a completely different way, and whether other models with similarly simple structures exist, are currently two open questions that need to be investigated. It is still an open question whether a more sophisticated model of modified gravity can be constructed by letting the physical metric interact with another spin-2 field nonlocally, in an attempt to obtain a viable cosmic history with self-acceleration at late times and in the future. In particular, such models have the very important and interesting potential of providing viable, {\it scale-free} models of modified gravity with no fine-tuning of parameters. All theories of multimetric gravity proposed so far have been constructed through local interactions, and most of them suffer from problems of ghost and gradient instabilities. We believe that relaxing the assumption of locality could open up new directions for extending the class of such theories, with interesting cosmological implications. It also remains to be seen, through rigorous theoretical investigations, whether (and if so, why) nonlocal theories can evade the well-known no-go theorems on the construction of consistent theories of interacting (massive or massless) spin-2 fields.

\acknowledgments We thank Ana Ach\'{u}carro, Nima Khosravi, Tomi S. Koivisto, Alexey S. Koshelev, Henrik Nersisyan, Javier Rubio, Angnis Schmidt-May, Adam R. Solomon, and Nico Wintergerst for helpful discussions. We are also grateful to the anonymous referee for thoroughly and critically reading the manuscript and for invaluable comments. V.V. is supported by a de Sitter fellowship of the Netherlands Organization for Scientific Research (NWO). Y.A. and A.S. acknowledge support from the NWO and the Dutch Ministry of Education, Culture and Science (OCW), and also from the D-ITP consortium, a program of the NWO that is funded by the OCW. L.A. acknowledges support from DFG through the project TRR33 ``The Dark Universe.'' 

\appendix

\section{Ostrogradski ghosts in generalizations to \textmd{\textup{\normalsize{}$m^{2}G(\frac{1}{\Box}R)+R F(\frac{1}{\Box}m^{2})$}}
nonlocalities\textmd{\textup{\normalsize{} }}}
\label{sec:appendix}

In this appendix we briefly study a generalization of our $m^{2}\frac{1}{\Box}R$ model to cases that resemble the DW model with a general form of the function $f(\frac{1}{\Box}R)$. Let us therefore consider interaction terms of the form
\begin{equation}
m^{2}G(\frac{1}{\Box}R)+R F(\frac{1}{\Box}m^{2}),
\end{equation}
where $F$ and $G$ can be any arbitrary functions. As usual, defining the auxiliary fields
\begin{align}
U & \equiv\frac{1}{\Box}R,\label{eq:UdefApp}\\
V & \equiv\frac{1}{\Box}m^{2},\label{eq:VdefApp}
\end{align}
we can localize the action for the generalized model, which then takes the form
\begin{align}\label{eq:localactionApp}
S=&\frac{M_{\text{Pl}}^{2}}{2}\int \dd^{4}x\sqrt{-g}[R-m^{2}G(U)-RF(V)]+\int\dd^4x\sqrt{-g}\lambda_{1}(R-\Box U)\nonumber\\
&+\int\dd^4x\sqrt{-g}\lambda_{2}(m^{2}-\Box V)+S_{\text{matter}}[g,\Psi].
\end{align}
For analytical functions $G(U)$ and $F(V)$, such as exponential or polynomial, with well-defined Taylor expansions, and using integration by parts, we can show $m^{2}G(U)=RG(V)$ and $RF(V)=m^{2}F(U)$ at the level of the action. This means that we can rewrite the terms involving the functions $G$ and $F$ in the action (\ref{eq:localactionApp}) in terms of only $U$ or $V$. Without loss of generality, and by redefining the functions $G$ and $F$, we can then rewrite the model in such a way that $G=F$.

We can now vary the action with respect to the fields appearing in action (\ref{eq:localactionApp}). The variation with respect to the Lagrange multipliers $\lambda_1$ and $\lambda_2$ gives the constraints (\ref{eq:UdefApp}) and (\ref{eq:VdefApp}), while the variation with respect to the auxiliary fields $U$ and $V$ gives
\begin{align}
\Box\lambda_{1} & =-\frac{M_{\text{Pl}}^{2}}{2} m^{2}F^{\prime}(U),\\
\Box\lambda_{2} & =-\frac{M_{\text{Pl}}^{2}}{2} RF^{\prime}(V),
\end{align}
respectively, where a prime here denotes a derivative with respect to the argument of the function.
In order to solve these equations for $\lambda_{1}$ and $\lambda_{2}$, we introduce two new auxiliary fields $\tilde{U}$ and $\tilde{V}$,
\begin{align}
\tilde{U} & \equiv\frac{1}{\Box}(m^{2}F^{\prime}(U)),\\
\tilde{V} & \equiv\frac{1}{\Box}(RF^{\prime}(V)).
\end{align}

In principle, we need to introduce two new Lagrange constraints $\tilde{\lambda}_{1}(m^{2}F^{\prime}(U)-\Box\tilde{U})+\tilde{\lambda}_{2}(RF^{\prime}(V)-\Box\tilde{V})$.
However, by doing that and performing the variation of the action with respect to $\tilde{U}$ and $\tilde{V}$ we get $\tilde{\lambda}_{1}=\tilde{\lambda}_{2}=0$ for the Lagrange multipliers $\tilde{\lambda}_{1}$ and $\tilde{\lambda}_{2}$. This implies that $\lambda_1$ and $\lambda_2$ in the action (\ref{eq:localactionApp}) are themselves the two extra auxiliary fields that we need in order to localize the action. For the simplicity of notation, we keep $\tilde{U}$ and $\tilde{V}$ for these extra fields, instead of $\lambda_1$ and $\lambda_2$. We therefore need four auxiliary fields for localizing our generalized, nonlocal model, instead of two for the simple choice of $F(U)=U$ studied in the paper. This is consistent with what happens in the DW model with general forms for $f(\frac{1}{\Box}R)$, where one needs two auxiliary fields for localization, in contrast to the simple case of $f(\frac{1}{\Box}R)=\frac{1}{\Box}R$.

The action (\ref{eq:localactionApp}) now reads
\begin{align}
S = \frac{M_{\text{Pl}}^{2}}{2}\int \dd^{4}x\sqrt{-g}[R(1-2 F(V)-\tilde{U})+\tilde{U}\Box U]-\int \dd^{4}x\sqrt{-g}\tilde{V}(m^{2}-\Box V)+S_{\text{matter}}[g,\Psi].
\end{align}
Let us now repeat the procedure of Sec.~\ref{sec:ghosts}, and study the apparent ghosts in the localized formulation of our generalized model. We therefore use the transformations
\begin{align}
g_{\mu\nu}&\rightarrow\Omega^{2}g_{\mu\nu},\\
R&\rightarrow\frac{1}{\Omega^{2}}[R-6(\Box \ln\Omega+g^{\mu\nu}\nabla_{\mu}\ln\Omega\nabla_{\nu}\ln\Omega)],
\end{align}
by introducing
\begin{equation}
\Omega^{2}\equiv\frac{1}{1-2F(U)-\tilde{U}},
\end{equation}
to write the action in the Einstein frame, which now takes the form
\begin{align}
S = & \frac{M_{\text{Pl}}^{2}}{2}\int \dd^{4}x\sqrt{-g}[R-6(\Box\ln\Omega+g^{\mu\nu}\partial_{\mu}\ln\Omega\partial_{\nu}\ln\Omega)-\Omega^{2}g^{\mu\nu}(\partial_{\mu}\tilde{U}\partial_{\nu}U+\partial_{\mu}\tilde{V}\partial_{\nu}V)]\nonumber\\
&-\int \dd^{4}x\sqrt{-g}m^{2}\tilde{V}+S_{\text{matter}}[\Omega^{2}g,\Psi].
\end{align}
We now introduce the new field $\phi$,
\begin{equation}
\phi\equiv \ln\Omega=-\frac{1}{2}\ln(1-2F(U)-\tilde{U})\Rightarrow\tilde{U}=1-e^{-2\phi}-2F(U),\label{eq:phiApp}
\end{equation}
and rename the other auxiliary fields as
\begin{align}
\psi\equiv U,~~~~~\chi\equiv V,~~~~~\zeta\equiv\tilde{V}.
\end{align}
Discarding the boundary terms, the action becomes
\begin{align}
S = &\frac{M_{\text{Pl}}^{2}}{2}\int \dd^{4}x\sqrt{-g}[R-6g^{\mu\nu}\partial_{\mu}\phi\partial_{\nu}\phi-2g^{\mu\nu}\partial_{\mu}\phi\partial_{\nu}\psi-e^{2\phi}g^{\mu\nu}(2F^{\prime}(\psi)\partial_{\mu}\psi\partial_{\nu}\psi-\partial_{\mu}\chi\partial_{\nu}\zeta)]\nonumber\\
& -\int \dd^{4}x\sqrt{-g}m^{2}\zeta+S_{\text{matter}}[e^{2\phi}g,\Psi].
\end{align}
In order for the kinetic matrix of the auxiliary fields $\{\phi,\psi,\chi,\zeta\}$ to be positive definite, we require
\begin{equation}
\text{det}\left[\begin{array}{cccc}
-6 & -1 & 0 & 0\\
-1 & -2e^{2\phi}F^{\prime}(\psi) & 0 & 0\\
0 & 0 & 0 & -e^{2\phi}/2\\
0 & 0 & -e^{2\phi}/2 & 0
\end{array}\right]=\frac{1}{4}e^{4\phi}(1-12e^{2\phi}F^{\prime}(\psi))>0,
\end{equation}
which then requires
\begin{equation}
1-12e^{2\phi}F^{\prime}(\psi)>0.
\end{equation}
This means that, depending on the form of $F$ and the dynamics of $\phi$ and $\psi$, it is possible for the generalized model to be free of ghosts in the local formulation, in contrast to the simple model with $F(\psi)=\psi$ which always has a ghost.

\bibliographystyle{JHEP}
\bibliography{nonlocalbigrav}

\providecommand{\href}[2]{#2}\begingroup\raggedright\begin{thebibliography}{100}

\bibitem{Rubin:2016iqe}
D.~Rubin and B.~Hayden, \emph{{Is the expansion of the universe accelerating?
  All signs point to yes}},
  \href{http://dx.doi.org/10.3847/2041-8213/833/2/L30}{\emph{Astrophys. J.}
  {\bfseries 833} (2016) L30},
  [\href{https://arxiv.org/abs/1610.08972}{{\ttfamily 1610.08972}}].

\bibitem{Riess:1998cb}
{\scshape Supernova Search Team} collaboration, A.~G. Riess et~al.,
  \emph{{Observational evidence from supernovae for an accelerating universe
  and a cosmological constant}},
  \href{http://dx.doi.org/10.1086/300499}{\emph{Astron. J.} {\bfseries 116}
  (1998) 1009--1038}, [\href{https://arxiv.org/abs/astro-ph/9805201}{{\ttfamily
  astro-ph/9805201}}].

\bibitem{Perlmutter:1998np}
{\scshape Supernova Cosmology Project} collaboration, S.~Perlmutter et~al.,
  \emph{{Measurements of Omega and Lambda from 42 high redshift supernovae}},
  \href{http://dx.doi.org/10.1086/307221}{\emph{Astrophys. J.} {\bfseries 517}
  (1999) 565--586}, [\href{https://arxiv.org/abs/astro-ph/9812133}{{\ttfamily
  astro-ph/9812133}}].

\bibitem{Caldwell:2009ix}
R.~R. Caldwell and M.~Kamionkowski, \emph{{The Physics of Cosmic
  Acceleration}},
  \href{http://dx.doi.org/10.1146/annurev-nucl-010709-151330}{\emph{Ann. Rev.
  Nucl. Part. Sci.} {\bfseries 59} (2009) 397--429},
  [\href{https://arxiv.org/abs/0903.0866}{{\ttfamily 0903.0866}}].

\bibitem{Weinberg:2012es}
D.~H. Weinberg, M.~J. Mortonson, D.~J. Eisenstein, C.~Hirata, A.~G. Riess and
  E.~Rozo, \emph{{Observational Probes of Cosmic Acceleration}},
  \href{http://dx.doi.org/10.1016/j.physrep.2013.05.001}{\emph{Phys. Rept.}
  {\bfseries 530} (2013) 87--255},
  [\href{https://arxiv.org/abs/1201.2434}{{\ttfamily 1201.2434}}].

\bibitem{Joyce:2014kja}
A.~Joyce, B.~Jain, J.~Khoury and M.~Trodden, \emph{{Beyond the Cosmological
  Standard Model}},
  \href{http://dx.doi.org/10.1016/j.physrep.2014.12.002}{\emph{Phys. Rept.}
  {\bfseries 568} (2015) 1--98},
  [\href{https://arxiv.org/abs/1407.0059}{{\ttfamily 1407.0059}}].

\bibitem{Bull:2015stt}
P.~Bull, Y.~Akrami et~al., \emph{{Beyond $\Lambda$CDM: Problems, solutions, and
  the road ahead}},
  \href{http://dx.doi.org/10.1016/j.dark.2016.02.001}{\emph{Phys. Dark Univ.}
  {\bfseries 12} (2016) 56--99},
  [\href{https://arxiv.org/abs/1512.05356}{{\ttfamily 1512.05356}}].

\bibitem{Martin:2012bt}
J.~Martin, \emph{{Everything You Always Wanted To Know About The Cosmological
  Constant Problem (But Were Afraid To Ask)}},
  \href{http://dx.doi.org/10.1016/j.crhy.2012.04.008}{\emph{Comptes Rendus
  Physique} {\bfseries 13} (2012) 566--665},
  [\href{https://arxiv.org/abs/1205.3365}{{\ttfamily 1205.3365}}].

\bibitem{2010deto.book.....A}
L.~{Amendola} and S.~{Tsujikawa}, \emph{{Dark Energy: Theory and
  Observations}}.
\newblock {Cambridge University Press}, 2010.

\bibitem{Clifton:2011jh}
T.~Clifton, P.~G. Ferreira, A.~Padilla and C.~Skordis, \emph{{Modified Gravity
  and Cosmology}},
  \href{http://dx.doi.org/10.1016/j.physrep.2012.01.001}{\emph{Phys.Rept.}
  {\bfseries 513} (2012) 1--189},
  [\href{https://arxiv.org/abs/1106.2476}{{\ttfamily 1106.2476}}].

\bibitem{Gupta:1954zz}
S.~N. Gupta, \emph{{Gravitation and Electromagnetism}},
  \href{http://dx.doi.org/10.1103/PhysRev.96.1683}{\emph{Phys.Rev.} {\bfseries
  96} (1954) 1683--1685}.

\bibitem{Weinberg:1965rz}
S.~Weinberg, \emph{{Photons and gravitons in perturbation theory: Derivation of
  Maxwell's and Einstein's equations}},
  \href{http://dx.doi.org/10.1103/PhysRev.138.B988}{\emph{Phys.Rev.} {\bfseries
  138} (1965) B988--B1002}.

\bibitem{Deser:1969wk}
S.~Deser, \emph{{Selfinteraction and gauge invariance}},
  \href{http://dx.doi.org/10.1007/BF00759198}{\emph{Gen.Rel.Grav.} {\bfseries
  1} (1970) 9--18}, [\href{https://arxiv.org/abs/gr-qc/0411023}{{\ttfamily
  gr-qc/0411023}}].

\bibitem{Boulware:1974sr}
D.~G. Boulware and S.~Deser, \emph{{Classical General Relativity Derived from
  Quantum Gravity}},
  \href{http://dx.doi.org/10.1016/0003-4916(75)90302-4}{\emph{Annals Phys.}
  {\bfseries 89} (1975) 193}.

\bibitem{Feynman:1996kb}
R.~Feynman, \emph{{Feynman lectures on gravitation}}.
\newblock {Addison-Wesley}, 1996.

\bibitem{deRham:2010ik}
C.~de~Rham and G.~Gabadadze, \emph{{Generalization of the Fierz-Pauli Action}},
  \href{http://dx.doi.org/10.1103/PhysRevD.82.044020}{\emph{Phys.Rev.}
  {\bfseries D82} (2010) 044020},
  [\href{https://arxiv.org/abs/1007.0443}{{\ttfamily 1007.0443}}].

\bibitem{deRham:2010kj}
C.~de~Rham, G.~Gabadadze and A.~J. Tolley, \emph{{Resummation of Massive
  Gravity}},
  \href{http://dx.doi.org/10.1103/PhysRevLett.106.231101}{\emph{Phys.Rev.Lett.}
  {\bfseries 106} (2011) 231101},
  [\href{https://arxiv.org/abs/1011.1232}{{\ttfamily 1011.1232}}].

\bibitem{Hassan:2011vm}
S.~Hassan and R.~A. Rosen, \emph{{On Non-Linear Actions for Massive Gravity}},
  \href{http://dx.doi.org/10.1007/JHEP07(2011)009}{\emph{JHEP} {\bfseries 1107}
  (2011) 009}, [\href{https://arxiv.org/abs/1103.6055}{{\ttfamily 1103.6055}}].

\bibitem{Hassan:2011hr}
S.~Hassan and R.~A. Rosen, \emph{{Resolving the Ghost Problem in non-Linear
  Massive Gravity}},
  \href{http://dx.doi.org/10.1103/PhysRevLett.108.041101}{\emph{Phys.Rev.Lett.}
  {\bfseries 108} (2012) 041101},
  [\href{https://arxiv.org/abs/1106.3344}{{\ttfamily 1106.3344}}].

\bibitem{deRham:2011rn}
C.~de~Rham, G.~Gabadadze and A.~J. Tolley, \emph{{Ghost free Massive Gravity in
  the St{\"u}ckelberg language}},
  \href{http://dx.doi.org/10.1016/j.physletb.2012.03.081}{\emph{Phys.Lett.}
  {\bfseries B711} (2012) 190--195},
  [\href{https://arxiv.org/abs/1107.3820}{{\ttfamily 1107.3820}}].

\bibitem{deRham:2011qq}
C.~de~Rham, G.~Gabadadze and A.~J. Tolley, \emph{{Helicity Decomposition of
  Ghost-free Massive Gravity}},
  \href{http://dx.doi.org/10.1007/JHEP11(2011)093}{\emph{JHEP} {\bfseries 1111}
  (2011) 093}, [\href{https://arxiv.org/abs/1108.4521}{{\ttfamily 1108.4521}}].

\bibitem{Hassan:2011tf}
S.~Hassan, R.~A. Rosen and A.~Schmidt-May, \emph{{Ghost-free Massive Gravity
  with a General Reference Metric}},
  \href{http://dx.doi.org/10.1007/JHEP02(2012)026}{\emph{JHEP} {\bfseries 1202}
  (2012) 026}, [\href{https://arxiv.org/abs/1109.3230}{{\ttfamily 1109.3230}}].

\bibitem{Hassan:2011ea}
S.~Hassan and R.~A. Rosen, \emph{{Confirmation of the Secondary Constraint and
  Absence of Ghost in Massive Gravity and Bimetric Gravity}},
  \href{http://dx.doi.org/10.1007/JHEP04(2012)123}{\emph{JHEP} {\bfseries 1204}
  (2012) 123}, [\href{https://arxiv.org/abs/1111.2070}{{\ttfamily 1111.2070}}].

\bibitem{Hassan:2012qv}
S.~Hassan, A.~Schmidt-May and M.~von Strauss, \emph{{Proof of Consistency of
  Nonlinear Massive Gravity in the St{\"u}ckelberg Formulation}},
  \href{http://dx.doi.org/10.1016/j.physletb.2012.07.018}{\emph{Phys.Lett.}
  {\bfseries B715} (2012) 335--339},
  [\href{https://arxiv.org/abs/1203.5283}{{\ttfamily 1203.5283}}].

\bibitem{Hinterbichler:2012cn}
K.~Hinterbichler and R.~A. Rosen, \emph{{Interacting Spin-2 Fields}},
  \href{http://dx.doi.org/10.1007/JHEP07(2012)047}{\emph{JHEP} {\bfseries 1207}
  (2012) 047}, [\href{https://arxiv.org/abs/1203.5783}{{\ttfamily 1203.5783}}].

\bibitem{Hassan:2011zd}
S.~Hassan and R.~A. Rosen, \emph{{Bimetric Gravity from Ghost-free Massive
  Gravity}}, \href{http://dx.doi.org/10.1007/JHEP02(2012)126}{\emph{JHEP}
  {\bfseries 1202} (2012) 126},
  [\href{https://arxiv.org/abs/1109.3515}{{\ttfamily 1109.3515}}].

\bibitem{deRham:2014zqa}
C.~de~Rham, \emph{{Massive Gravity}},
  \href{http://dx.doi.org/10.12942/lrr-2014-7}{\emph{Living Rev.Rel.}
  {\bfseries 17} (2014) 7}, [\href{https://arxiv.org/abs/1401.4173}{{\ttfamily
  1401.4173}}].

\bibitem{Hinterbichler:2011tt}
K.~Hinterbichler, \emph{{Theoretical Aspects of Massive Gravity}},
  \href{http://dx.doi.org/10.1103/RevModPhys.84.671}{\emph{Rev.Mod.Phys.}
  {\bfseries 84} (2012) 671--710},
  [\href{https://arxiv.org/abs/1105.3735}{{\ttfamily 1105.3735}}].

\bibitem{Schmidt-May:2015vnx}
A.~Schmidt-May and M.~von Strauss, \emph{{Recent developments in bimetric
  theory}}, \href{http://dx.doi.org/10.1088/1751-8113/49/18/183001}{\emph{J.
  Phys.} {\bfseries A49} (2016) 183001},
  [\href{https://arxiv.org/abs/1512.00021}{{\ttfamily 1512.00021}}].

\bibitem{Solomon:2015hja}
A.~R. Solomon, \emph{{Cosmology Beyond Einstein}}, Ph.D. thesis, Cambridge U.,
  2015.
\newblock \href{https://arxiv.org/abs/1508.06859}{{\ttfamily 1508.06859}}.

\bibitem{Hinterbichler:2017sbd}
K.~Hinterbichler, \emph{{Cosmology of Massive Gravity and its Extensions}},  in
  \emph{{51st Rencontres de Moriond on Cosmology La Thuile, Italy, March 19-26,
  2016}}, 2017, \href{https://arxiv.org/abs/1701.02873}{{\ttfamily
  1701.02873}},
  \href{http://inspirehep.net/record/1508592/files/arXiv:1701.02873.pdf}{http://inspirehep.net/record/1508592/files/arXiv:1701.02873.pdf}.

\bibitem{Nersisyan:2015oha}
H.~Nersisyan, Y.~Akrami and L.~Amendola, \emph{{Consistent metric combinations
  in cosmology of massive bigravity}},
  \href{http://dx.doi.org/10.1103/PhysRevD.92.104034}{\emph{Phys. Rev.}
  {\bfseries D92} (2015) 104034},
  [\href{https://arxiv.org/abs/1502.03988}{{\ttfamily 1502.03988}}].

\bibitem{Volkov:2011an}
M.~S. Volkov, \emph{{Cosmological solutions with massive gravitons in the
  bigravity theory}},
  \href{http://dx.doi.org/10.1007/JHEP01(2012)035}{\emph{JHEP} {\bfseries 1201}
  (2012) 035}, [\href{https://arxiv.org/abs/1110.6153}{{\ttfamily 1110.6153}}].

\bibitem{Comelli:2011zm}
D.~Comelli, M.~Crisostomi, F.~Nesti and L.~Pilo, \emph{{FRW Cosmology in Ghost
  Free Massive Gravity}}, \href{http://dx.doi.org/10.1007/JHEP06(2012)020,
  10.1007/JHEP03(2012)067}{\emph{JHEP} {\bfseries 1203} (2012) 067},
  [\href{https://arxiv.org/abs/1111.1983}{{\ttfamily 1111.1983}}].

\bibitem{vonStrauss:2011mq}
M.~von Strauss, A.~Schmidt-May, J.~Enander, E.~M{\"o}rtsell and S.~Hassan,
  \emph{{Cosmological Solutions in Bimetric Gravity and their Observational
  Tests}}, \href{http://dx.doi.org/10.1088/1475-7516/2012/03/042}{\emph{JCAP}
  {\bfseries 1203} (2012) 042},
  [\href{https://arxiv.org/abs/1111.1655}{{\ttfamily 1111.1655}}].

\bibitem{Akrami:2012vf}
Y.~Akrami, T.~S. Koivisto and M.~Sandstad, \emph{{Accelerated expansion from
  ghost-free bigravity: a statistical analysis with improved generality}},
  \href{http://dx.doi.org/10.1007/JHEP03(2013)099}{\emph{JHEP} {\bfseries 1303}
  (2013) 099}, [\href{https://arxiv.org/abs/1209.0457}{{\ttfamily 1209.0457}}].

\bibitem{Akrami:2013pna}
Y.~Akrami, T.~S. Koivisto and M.~Sandstad, \emph{{Cosmological constraints on
  ghost-free bigravity: background dynamics and late-time acceleration}},
  \href{https://arxiv.org/abs/1302.5268}{{\ttfamily 1302.5268}}.

\bibitem{Konnig:2013gxa}
F.~K{\"o}nnig, A.~Patil and L.~Amendola, \emph{{Viable cosmological solutions
  in massive bimetric gravity}},
  \href{http://dx.doi.org/10.1088/1475-7516/2014/03/029}{\emph{JCAP} {\bfseries
  1403} (2014) 029}, [\href{https://arxiv.org/abs/1312.3208}{{\ttfamily
  1312.3208}}].

\bibitem{Enander:2014xga}
J.~Enander, A.~R. Solomon, Y.~Akrami and E.~Mortsell, \emph{{Cosmic expansion
  histories in massive bigravity with symmetric matter coupling}},
  \href{http://dx.doi.org/10.1088/1475-7516/2015/01/006}{\emph{JCAP} {\bfseries
  01} (2015) 006}, [\href{https://arxiv.org/abs/1409.2860}{{\ttfamily
  1409.2860}}].

\bibitem{Mortsell:2017fog}
E.~Mortsell, \emph{{Cosmological histories in bimetric gravity: A graphical
  approach}},  \href{https://arxiv.org/abs/1701.00710}{{\ttfamily 1701.00710}}.

\bibitem{Luben:2016lku}
M.~Lüben, Y.~Akrami, L.~Amendola and A.~R. Solomon, \emph{{Cosmology with three
  interacting spin-2 fields}},
  \href{http://dx.doi.org/10.1103/PhysRevD.94.043530}{\emph{Phys. Rev.}
  {\bfseries D94} (2016) 043530},
  [\href{https://arxiv.org/abs/1604.04285}{{\ttfamily 1604.04285}}].

\bibitem{Comelli:2012db}
D.~Comelli, M.~Crisostomi and L.~Pilo, \emph{{Perturbations in Massive Gravity
  Cosmology}}, \href{http://dx.doi.org/10.1007/JHEP06(2012)085}{\emph{JHEP}
  {\bfseries 1206} (2012) 085},
  [\href{https://arxiv.org/abs/1202.1986}{{\ttfamily 1202.1986}}].

\bibitem{Khosravi:2012rk}
N.~Khosravi, H.~R. Sepangi and S.~Shahidi, \emph{{Massive cosmological scalar
  perturbations}},
  \href{http://dx.doi.org/10.1103/PhysRevD.86.043517}{\emph{Phys.Rev.}
  {\bfseries D86} (2012) 043517},
  [\href{https://arxiv.org/abs/1202.2767}{{\ttfamily 1202.2767}}].

\bibitem{Berg:2012kn}
M.~Berg, I.~Buchberger, J.~Enander, E.~M{\"o}rtsell and S.~Sj{\"o}rs,
  \emph{{Growth Histories in Bimetric Massive Gravity}},
  \href{http://dx.doi.org/10.1088/1475-7516/2012/12/021}{\emph{JCAP} {\bfseries
  1212} (2012) 021}, [\href{https://arxiv.org/abs/1206.3496}{{\ttfamily
  1206.3496}}].

\bibitem{Konnig:2014dna}
F.~K{\"o}nnig and L.~Amendola, \emph{{Instability in a minimal bimetric gravity
  model}}, \href{http://dx.doi.org/10.1103/PhysRevD.90.044030}{\emph{Phys.Rev.}
  {\bfseries D90} (2014) 044030},
  [\href{https://arxiv.org/abs/1402.1988}{{\ttfamily 1402.1988}}].

\bibitem{Solomon:2014dua}
A.~R. Solomon, Y.~Akrami and T.~S. Koivisto, \emph{{Linear growth of structure
  in massive bigravity}},
  \href{http://dx.doi.org/10.1088/1475-7516/2014/10/066}{\emph{JCAP} {\bfseries
  1410} (2014) 066}, [\href{https://arxiv.org/abs/1404.4061}{{\ttfamily
  1404.4061}}].

\bibitem{Konnig:2014xva}
F.~K{\"o}nnig, Y.~Akrami, L.~Amendola, M.~Motta and A.~R. Solomon,
  \emph{{Stable and unstable cosmological models in bimetric massive gravity}},
  \href{http://dx.doi.org/10.1103/PhysRevD.90.124014}{\emph{Phys.Rev.}
  {\bfseries D90} (2014) 124014},
  [\href{https://arxiv.org/abs/1407.4331}{{\ttfamily 1407.4331}}].

\bibitem{Lagos:2014lca}
M.~Lagos and P.~G. Ferreira, \emph{{Cosmological perturbations in massive
  bigravity}},
  \href{http://dx.doi.org/10.1088/1475-7516/2014/12/026}{\emph{JCAP} {\bfseries
  1412} (2014) 026}, [\href{https://arxiv.org/abs/1410.0207}{{\ttfamily
  1410.0207}}].

\bibitem{Cusin:2014psa}
G.~Cusin, R.~Durrer, P.~Guarato and M.~Motta, \emph{{Gravitational waves in
  bigravity cosmology}},
  \href{http://dx.doi.org/10.1088/1475-7516/2015/05/030}{\emph{JCAP} {\bfseries
  1505} (2015) 030}, [\href{https://arxiv.org/abs/1412.5979}{{\ttfamily
  1412.5979}}].

\bibitem{Yamashita:2014cra}
Y.~Yamashita and T.~Tanaka, \emph{{Mapping the ghost free bigravity into
  braneworld setup}},
  \href{http://dx.doi.org/10.1088/1475-7516/2014/06/004}{\emph{JCAP} {\bfseries
  1406} (2014) 004}, [\href{https://arxiv.org/abs/1401.4336}{{\ttfamily
  1401.4336}}].

\bibitem{DeFelice:2014nja}
A.~De~Felice, A.~E. G{\"u}mr{\"u}k{\c c}{\"u}o{\u g}lu, S.~Mukohyama,
  N.~Tanahashi and T.~Tanaka, \emph{{Viable cosmology in bimetric theory}},
  \href{http://dx.doi.org/10.1088/1475-7516/2014/06/037}{\emph{JCAP} {\bfseries
  1406} (2014) 037}, [\href{https://arxiv.org/abs/1404.0008}{{\ttfamily
  1404.0008}}].

\bibitem{Fasiello:2013woa}
M.~Fasiello and A.~J. Tolley, \emph{{Cosmological Stability Bound in Massive
  Gravity and Bigravity}},
  \href{http://dx.doi.org/10.1088/1475-7516/2013/12/002}{\emph{JCAP} {\bfseries
  1312} (2013) 002}, [\href{https://arxiv.org/abs/1308.1647}{{\ttfamily
  1308.1647}}].

\bibitem{Enander:2015vja}
J.~Enander, Y.~Akrami, E.~M{\"o}rtsell, M.~Renneby and A.~R. Solomon,
  \emph{{Integrated Sachs-Wolfe effect in massive bigravity}},
  \href{http://dx.doi.org/10.1103/PhysRevD.91.084046}{\emph{Phys.Rev.}
  {\bfseries D91} (2015) 084046},
  [\href{https://arxiv.org/abs/1501.02140}{{\ttfamily 1501.02140}}].

\bibitem{Amendola:2015tua}
L.~Amendola, F.~K{\"o}nnig, M.~Martinelli, V.~Pettorino and M.~Zumalacarregui,
  \emph{{Surfing gravitational waves: can bigravity survive growing tensor
  modes?}}, \href{http://dx.doi.org/10.1088/1475-7516/2015/05/052}{\emph{JCAP}
  {\bfseries 1505} (2015) 052},
  [\href{https://arxiv.org/abs/1503.02490}{{\ttfamily 1503.02490}}].

\bibitem{Johnson:2015tfa}
M.~Johnson and A.~Terrana, \emph{{Tensor Modes in Bigravity: Primordial to
  Present}}, \href{http://dx.doi.org/10.1103/PhysRevD.92.044001}{\emph{Phys.
  Rev.} {\bfseries D92} (2015) 044001},
  [\href{https://arxiv.org/abs/1503.05560}{{\ttfamily 1503.05560}}].

\bibitem{Konnig:2015lfa}
F.~K{\"o}nnig, \emph{{Higuchi Ghosts and Gradient Instabilities in Bimetric
  Gravity}},
  \href{http://dx.doi.org/10.1103/PhysRevD.91.104019}{\emph{Phys.Rev.}
  {\bfseries D91} (2015) 104019},
  [\href{https://arxiv.org/abs/1503.07436}{{\ttfamily 1503.07436}}].

\bibitem{Lagos:2016gep}
M.~Lagos and P.~G. Ferreira, \emph{{A general theory of linear cosmological
  perturbations: bimetric theories}},
  \href{http://dx.doi.org/10.1088/1475-7516/2017/01/047}{\emph{JCAP} {\bfseries
  1701} (2017) 047}, [\href{https://arxiv.org/abs/1610.00553}{{\ttfamily
  1610.00553}}].

\bibitem{Hassan:2012wr}
S.~Hassan, A.~Schmidt-May and M.~von Strauss, \emph{{On Consistent Theories of
  Massive Spin-2 Fields Coupled to Gravity}},
  \href{http://dx.doi.org/10.1007/JHEP05(2013)086}{\emph{JHEP} {\bfseries 1305}
  (2013) 086}, [\href{https://arxiv.org/abs/1208.1515}{{\ttfamily 1208.1515}}].

\bibitem{Akrami:2013ffa}
Y.~Akrami, T.~S. Koivisto, D.~F. Mota and M.~Sandstad, \emph{{Bimetric gravity
  doubly coupled to matter: theory and cosmological implications}},
  \href{http://dx.doi.org/10.1088/1475-7516/2013/10/046}{\emph{JCAP} {\bfseries
  1310} (2013) 046}, [\href{https://arxiv.org/abs/1306.0004}{{\ttfamily
  1306.0004}}].

\bibitem{Tamanini:2013xia}
N.~Tamanini, E.~N. Saridakis and T.~S. Koivisto, \emph{{The Cosmology of
  Interacting Spin-2 Fields}},
  \href{http://dx.doi.org/10.1088/1475-7516/2014/02/015}{\emph{JCAP} {\bfseries
  1402} (2014) 015}, [\href{https://arxiv.org/abs/1307.5984}{{\ttfamily
  1307.5984}}].

\bibitem{Akrami:2014lja}
Y.~Akrami, T.~S. Koivisto and A.~R. Solomon, \emph{{The nature of spacetime in
  bigravity: two metrics or none?}},
  \href{http://dx.doi.org/10.1007/s10714-014-1838-4}{\emph{Gen.Rel.Grav.}
  {\bfseries 47} (2014) 1838},
  [\href{https://arxiv.org/abs/1404.0006}{{\ttfamily 1404.0006}}].

\bibitem{Yamashita:2014fga}
Y.~Yamashita, A.~De~Felice and T.~Tanaka, \emph{{Appearance of Boulware-Deser
  ghost in bigravity with doubly coupled matter}},
  \href{http://dx.doi.org/10.1142/S0218271814430032}{\emph{Int.J.Mod.Phys.}
  {\bfseries D23} (2014) 3003},
  [\href{https://arxiv.org/abs/1408.0487}{{\ttfamily 1408.0487}}].

\bibitem{deRham:2014naa}
C.~de~Rham, L.~Heisenberg and R.~H. Ribeiro, \emph{{On couplings to matter in
  massive (bi-)gravity}},
  \href{http://dx.doi.org/10.1088/0264-9381/32/3/035022}{\emph{Class.Quant.Grav.}
  {\bfseries 32} (2015) 035022},
  [\href{https://arxiv.org/abs/1408.1678}{{\ttfamily 1408.1678}}].

\bibitem{Hassan:2014gta}
S.~Hassan, M.~Kocic and A.~Schmidt-May, \emph{{Absence of ghost in a new
  bimetric-matter coupling}},
  \href{https://arxiv.org/abs/1409.1909}{{\ttfamily 1409.1909}}.

\bibitem{Solomon:2014iwa}
A.~R. Solomon, J.~Enander, Y.~Akrami, T.~S. Koivisto, F.~K{\"o}nnig et~al.,
  \emph{{Cosmological viability of massive gravity with generalized matter
  coupling}},
  \href{http://dx.doi.org/10.1088/1475-7516/2015/04/027}{\emph{JCAP} {\bfseries
  1504} (2015) 027}, [\href{https://arxiv.org/abs/1409.8300}{{\ttfamily
  1409.8300}}].

\bibitem{Schmidt-May:2014xla}
A.~Schmidt-May, \emph{{Mass eigenstates in bimetric theory with matter
  coupling}},
  \href{http://dx.doi.org/10.1088/1475-7516/2015/01/039}{\emph{JCAP} {\bfseries
  1501} (2015) 039}, [\href{https://arxiv.org/abs/1409.3146}{{\ttfamily
  1409.3146}}].

\bibitem{deRham:2014fha}
C.~de~Rham, L.~Heisenberg and R.~H. Ribeiro, \emph{{Ghosts \& Matter Couplings
  in Massive (bi-\&multi-)Gravity}},
  \href{http://dx.doi.org/10.1103/PhysRevD.90.124042}{\emph{Phys.Rev.}
  {\bfseries D90} (2014) 124042},
  [\href{https://arxiv.org/abs/1409.3834}{{\ttfamily 1409.3834}}].

\bibitem{Gumrukcuoglu:2014xba}
A.~E. G{\"u}mr{\"u}k{\c c}{\"u}o{\u g}lu, L.~Heisenberg and S.~Mukohyama,
  \emph{{Cosmological perturbations in massive gravity with doubly coupled
  matter}}, \href{http://dx.doi.org/10.1088/1475-7516/2015/02/022}{\emph{JCAP}
  {\bfseries 1502} (2015) 022},
  [\href{https://arxiv.org/abs/1409.7260}{{\ttfamily 1409.7260}}].

\bibitem{Heisenberg:2014rka}
L.~Heisenberg, \emph{{Quantum corrections in massive bigravity and new
  effective composite metrics}},
  \href{http://dx.doi.org/10.1088/0264-9381/32/10/105011}{\emph{Class.Quant.Grav.}
  {\bfseries 32} (2015) 105011},
  [\href{https://arxiv.org/abs/1410.4239}{{\ttfamily 1410.4239}}].

\bibitem{Gumrukcuoglu:2015nua}
A.~E. G{\"u}mr{\"u}k{\c c}{\"u}o{\u g}lu, L.~Heisenberg, S.~Mukohyama and
  N.~Tanahashi, \emph{{Cosmology in bimetric theory with an effective composite
  coupling to matter}},
  \href{http://dx.doi.org/10.1088/1475-7516/2015/04/008}{\emph{JCAP} {\bfseries
  1504} (2015) 008}, [\href{https://arxiv.org/abs/1501.02790}{{\ttfamily
  1501.02790}}].

\bibitem{Hinterbichler:2015yaa}
K.~Hinterbichler and R.~A. Rosen, \emph{{Note on ghost-free matter couplings in
  massive gravity and multigravity}},
  \href{http://dx.doi.org/10.1103/PhysRevD.92.024030}{\emph{Phys. Rev.}
  {\bfseries D92} (2015) 024030},
  [\href{https://arxiv.org/abs/1503.06796}{{\ttfamily 1503.06796}}].

\bibitem{Heisenberg:2015iqa}
L.~Heisenberg, \emph{{More on effective composite metrics}},
  \href{http://dx.doi.org/10.1103/PhysRevD.92.023525}{\emph{Phys. Rev.}
  {\bfseries D92} (2015) 023525},
  [\href{https://arxiv.org/abs/1505.02966}{{\ttfamily 1505.02966}}].

\bibitem{Heisenberg:2015wja}
L.~Heisenberg, \emph{{Non-minimal derivative couplings of the composite
  metric}}, \href{http://dx.doi.org/10.1088/1475-7516/2015/11/005}{\emph{JCAP}
  {\bfseries 1511} (2015) 005},
  [\href{https://arxiv.org/abs/1506.00580}{{\ttfamily 1506.00580}}].

\bibitem{Lagos:2015sya}
M.~Lagos and J.~Noller, \emph{{New massive bigravity cosmologies with double
  matter coupling}},
  \href{http://dx.doi.org/10.1088/1475-7516/2016/01/023}{\emph{JCAP} {\bfseries
  1601} (2016) 023}, [\href{https://arxiv.org/abs/1508.05864}{{\ttfamily
  1508.05864}}].

\bibitem{Melville:2015dba}
S.~Melville and J.~Noller, \emph{{Generalised matter couplings in massive
  bigravity}}, \href{http://dx.doi.org/10.1007/JHEP01(2016)094}{\emph{JHEP}
  {\bfseries 01} (2016) 094},
  [\href{https://arxiv.org/abs/1511.01485}{{\ttfamily 1511.01485}}].

\bibitem{Akrami:2015qga}
Y.~Akrami, S.~F. Hassan, F.~K{\" o}nnig, A.~Schmidt-May and A.~R. Solomon,
  \emph{{Bimetric gravity is cosmologically viable}},
  \href{http://dx.doi.org/10.1016/j.physletb.2015.06.062}{\emph{Phys. Lett.}
  {\bfseries B748} (2015) 37--44},
  [\href{https://arxiv.org/abs/1503.07521}{{\ttfamily 1503.07521}}].

\bibitem{Mortsell:2015exa}
E.~Mortsell and J.~Enander, \emph{{Scalar instabilities in bimetric gravity:
  The Vainshtein mechanism and structure formation}},
  \href{http://dx.doi.org/10.1088/1475-7516/2015/10/044}{\emph{JCAP} {\bfseries
  1510} (2015) 044}, [\href{https://arxiv.org/abs/1506.04977}{{\ttfamily
  1506.04977}}].

\bibitem{Boulanger:2000rq}
N.~Boulanger, T.~Damour, L.~Gualtieri and M.~Henneaux, \emph{{Inconsistency of
  interacting, multigraviton theories}},
  \href{http://dx.doi.org/10.1016/S0550-3213(00)00718-5}{\emph{Nucl. Phys.}
  {\bfseries B597} (2001) 127--171},
  [\href{https://arxiv.org/abs/hep-th/0007220}{{\ttfamily hep-th/0007220}}].

\bibitem{Hinterbichler:2013eza}
K.~Hinterbichler, \emph{{Ghost-Free Derivative Interactions for a Massive
  Graviton}}, \href{http://dx.doi.org/10.1007/JHEP10(2013)102}{\emph{JHEP}
  {\bfseries 10} (2013) 102},
  [\href{https://arxiv.org/abs/1305.7227}{{\ttfamily 1305.7227}}].

\bibitem{Kimura:2013ika}
R.~Kimura and D.~Yamauchi, \emph{{Derivative interactions in de
  Rham-Gabadadze-Tolley massive gravity}},
  \href{http://dx.doi.org/10.1103/PhysRevD.88.084025}{\emph{Phys. Rev.}
  {\bfseries D88} (2013) 084025},
  [\href{https://arxiv.org/abs/1308.0523}{{\ttfamily 1308.0523}}].

\bibitem{deRham:2013tfa}
C.~de~Rham, A.~Matas and A.~J. Tolley, \emph{{New Kinetic Interactions for
  Massive Gravity?}},
  \href{http://dx.doi.org/10.1088/0264-9381/31/16/165004}{\emph{Class. Quant.
  Grav.} {\bfseries 31} (2014) 165004},
  [\href{https://arxiv.org/abs/1311.6485}{{\ttfamily 1311.6485}}].

\bibitem{Gao:2014jja}
X.~Gao, \emph{{Derivative interactions for a spin-2 field at cubic order}},
  \href{http://dx.doi.org/10.1103/PhysRevD.90.064024}{\emph{Phys. Rev.}
  {\bfseries D90} (2014) 064024},
  [\href{https://arxiv.org/abs/1403.6781}{{\ttfamily 1403.6781}}].

\bibitem{Noller:2014ioa}
J.~Noller, \emph{{On Consistent Kinetic and Derivative Interactions for
  Gravitons}},
  \href{http://dx.doi.org/10.1088/1475-7516/2015/04/025}{\emph{JCAP} {\bfseries
  1504} (2015) 025}, [\href{https://arxiv.org/abs/1409.7692}{{\ttfamily
  1409.7692}}].

\bibitem{deRham:2015rxa}
C.~de~Rham, A.~Matas and A.~J. Tolley, \emph{{New Kinetic Terms for Massive
  Gravity and Multi-gravity: A No-Go in Vielbein Form}},
  \href{http://dx.doi.org/10.1088/0264-9381/32/21/215027}{\emph{Class. Quant.
  Grav.} {\bfseries 32} (2015) 215027},
  [\href{https://arxiv.org/abs/1505.00831}{{\ttfamily 1505.00831}}].

\bibitem{Li:2015izu}
W.~Li, \emph{{Novel nonlinear kinetic terms for gravitons}},
  \href{http://dx.doi.org/10.1103/PhysRevD.94.064078}{\emph{Phys. Rev.}
  {\bfseries D94} (2016) 064078},
  [\href{https://arxiv.org/abs/1508.03246}{{\ttfamily 1508.03246}}].

\bibitem{Li:2015iwc}
W.~Li, \emph{{Absence of the Boulware-Deser ghost in novel graviton kinetic
  terms}}, \href{http://dx.doi.org/10.1103/PhysRevD.94.064079}{\emph{Phys.
  Rev.} {\bfseries D94} (2016) 064079},
  [\href{https://arxiv.org/abs/1512.06386}{{\ttfamily 1512.06386}}].

\bibitem{Boulware:1973my}
D.~Boulware and S.~Deser, \emph{{Can gravitation have a finite range?}},
  \href{http://dx.doi.org/10.1103/PhysRevD.6.3368}{\emph{Phys.Rev.} {\bfseries
  D6} (1972) 3368--3382}.

\bibitem{Lovelock1972}
D.~Lovelock, \emph{{The four-dimensionality of space and the einstein tensor}},
  \href{http://dx.doi.org/10.1063/1.1666069}{\emph{J. Math. Phys.} {\bfseries
  13} (1972) 874--876}.

\bibitem{Barvinsky:1987uw}
A.~O. Barvinsky and G.~A. Vilkovisky, \emph{{Beyond the Schwinger-Dewitt
  Technique: Converting Loops Into Trees and In-In Currents}},
  \href{http://dx.doi.org/10.1016/0550-3213(87)90681-X}{\emph{Nucl. Phys.}
  {\bfseries B282} (1987) 163--188}.

\bibitem{Barvinsky:1990up}
A.~O. Barvinsky and G.~A. Vilkovisky, \emph{{Covariant perturbation theory. 2:
  Second order in the curvature. General algorithms}},
  \href{http://dx.doi.org/10.1016/0550-3213(90)90047-H}{\emph{Nucl. Phys.}
  {\bfseries B333} (1990) 471--511}.

\bibitem{Barvinsky:1990uq}
A.~O. Barvinsky and G.~A. Vilkovisky, \emph{{Covariant perturbation theory. 3:
  Spectral representations of the third order form-factors}},
  \href{http://dx.doi.org/10.1016/0550-3213(90)90048-I}{\emph{Nucl. Phys.}
  {\bfseries B333} (1990) 512--524}.

\bibitem{Barvinsky:1995jv}
A.~O. Barvinsky, {\relax Yu}.~V. Gusev, V.~V. Zhytnikov and G.~A. Vilkovisky,
  \emph{{Asymptotic behaviors of one loop vertices in the gravitational
  effective action}},
  \href{http://dx.doi.org/10.1088/0264-9381/12/9/005}{\emph{Class. Quant.
  Grav.} {\bfseries 12} (1995) 2157--2172}.

\bibitem{Barvinsky:2014lja}
A.~O. Barvinsky, \emph{{Aspects of Nonlocality in Quantum Field Theory, Quantum
  Gravity and Cosmology}},
  \href{http://dx.doi.org/10.1142/S0217732315400039}{\emph{Mod. Phys. Lett.}
  {\bfseries A30} (2015) 1540003},
  [\href{https://arxiv.org/abs/1408.6112}{{\ttfamily 1408.6112}}].

\bibitem{Codello:2015mba}
A.~Codello and R.~K. Jain, \emph{{On the covariant formalism of the effective
  field theory of gravity and leading order corrections}},
  \href{http://dx.doi.org/10.1088/0264-9381/33/22/225006}{\emph{Class. Quant.
  Grav.} {\bfseries 33} (2016) 225006},
  [\href{https://arxiv.org/abs/1507.06308}{{\ttfamily 1507.06308}}].

\bibitem{Donoghue:2015nba}
J.~F. Donoghue and B.~K. El-Menoufi, \emph{{Covariant non-local action for
  massless QED and the curvature expansion}},
  \href{http://dx.doi.org/10.1007/JHEP10(2015)044}{\emph{JHEP} {\bfseries 10}
  (2015) 044}, [\href{https://arxiv.org/abs/1507.06321}{{\ttfamily
  1507.06321}}].

\bibitem{Maggiore:2016fbn}
M.~Maggiore, \emph{{Perturbative loop corrections and nonlocal gravity}},
  \href{http://dx.doi.org/10.1103/PhysRevD.93.063008}{\emph{Phys. Rev.}
  {\bfseries D93} (2016) 063008},
  [\href{https://arxiv.org/abs/1603.01515}{{\ttfamily 1603.01515}}].

\bibitem{Maggiore:2016gpx}
M.~Maggiore, \emph{{Nonlocal Infrared Modifications of Gravity. A Review}},
  \href{https://arxiv.org/abs/1606.08784}{{\ttfamily 1606.08784}}.

\bibitem{Wetterich:1997bz}
C.~Wetterich, \emph{{Effective nonlocal Euclidean gravity}},
  \href{http://dx.doi.org/10.1023/A:1018837319976}{\emph{Gen. Rel. Grav.}
  {\bfseries 30} (1998) 159--172},
  [\href{https://arxiv.org/abs/gr-qc/9704052}{{\ttfamily gr-qc/9704052}}].

\bibitem{Barvinsky:2011rk}
A.~O. Barvinsky, \emph{{Serendipitous discoveries in nonlocal gravity theory}},
  \href{http://dx.doi.org/10.1103/PhysRevD.85.104018}{\emph{Phys. Rev.}
  {\bfseries D85} (2012) 104018},
  [\href{https://arxiv.org/abs/1112.4340}{{\ttfamily 1112.4340}}].

\bibitem{Jaccard:2013gla}
M.~Jaccard, M.~Maggiore and E.~Mitsou, \emph{{Nonlocal theory of massive
  gravity}}, \href{http://dx.doi.org/10.1103/PhysRevD.88.044033}{\emph{Phys.
  Rev.} {\bfseries D88} (2013) 044033},
  [\href{https://arxiv.org/abs/1305.3034}{{\ttfamily 1305.3034}}].

\bibitem{Modesto:2013jea}
L.~Modesto and S.~Tsujikawa, \emph{{Non-local massive gravity}},
  \href{http://dx.doi.org/10.1016/j.physletb.2013.10.037}{\emph{Phys. Lett.}
  {\bfseries B727} (2013) 48--56},
  [\href{https://arxiv.org/abs/1307.6968}{{\ttfamily 1307.6968}}].

\bibitem{Cusin:2014zoa}
G.~Cusin, J.~Fumagalli and M.~Maggiore, \emph{{Non-local formulation of
  ghost-free bigravity theory}},
  \href{http://dx.doi.org/10.1007/JHEP09(2014)181}{\emph{JHEP} {\bfseries 09}
  (2014) 181}, [\href{https://arxiv.org/abs/1407.5580}{{\ttfamily 1407.5580}}].

\bibitem{Golovnev:2015bsa}
A.~Golovnev, T.~Koivisto and M.~Sandstad, \emph{{Effectively nonlocal
  metric-affine gravity}},
  \href{http://dx.doi.org/10.1103/PhysRevD.93.064081}{\emph{Phys. Rev.}
  {\bfseries D93} (2016) 064081},
  [\href{https://arxiv.org/abs/1509.06552}{{\ttfamily 1509.06552}}].

\bibitem{ArkaniHamed:2002fu}
N.~Arkani-Hamed, S.~Dimopoulos, G.~Dvali and G.~Gabadadze, \emph{{Nonlocal
  modification of gravity and the cosmological constant problem}},
  \href{https://arxiv.org/abs/hep-th/0209227}{{\ttfamily hep-th/0209227}}.

\bibitem{Dvali:2007kt}
G.~Dvali, S.~Hofmann and J.~Khoury, \emph{{Degravitation of the cosmological
  constant and graviton width}},
  \href{http://dx.doi.org/10.1103/PhysRevD.76.084006}{\emph{Phys. Rev.}
  {\bfseries D76} (2007) 084006},
  [\href{https://arxiv.org/abs/hep-th/0703027}{{\ttfamily hep-th/0703027}}].

\bibitem{Deser:2007jk}
S.~Deser and R.~P. Woodard, \emph{{Nonlocal Cosmology}},
  \href{http://dx.doi.org/10.1103/PhysRevLett.99.111301}{\emph{Phys. Rev.
  Lett.} {\bfseries 99} (2007) 111301},
  [\href{https://arxiv.org/abs/0706.2151}{{\ttfamily 0706.2151}}].

\bibitem{Woodard:2014iga}
R.~P. Woodard, \emph{{Nonlocal Models of Cosmic Acceleration}},
  \href{http://dx.doi.org/10.1007/s10701-014-9780-6}{\emph{Found. Phys.}
  {\bfseries 44} (2014) 213--233},
  [\href{https://arxiv.org/abs/1401.0254}{{\ttfamily 1401.0254}}].

\bibitem{Deser:2013uya}
S.~Deser and R.~P. Woodard, \emph{{Observational Viability and Stability of
  Nonlocal Cosmology}},
  \href{http://dx.doi.org/10.1088/1475-7516/2013/11/036}{\emph{JCAP} {\bfseries
  1311} (2013) 036}, [\href{https://arxiv.org/abs/1307.6639}{{\ttfamily
  1307.6639}}].

\bibitem{Koivisto:2008xfa}
T.~Koivisto, \emph{{Dynamics of Nonlocal Cosmology}},
  \href{http://dx.doi.org/10.1103/PhysRevD.77.123513}{\emph{Phys. Rev.}
  {\bfseries D77} (2008) 123513},
  [\href{https://arxiv.org/abs/0803.3399}{{\ttfamily 0803.3399}}].

\bibitem{Koivisto:2008dh}
T.~S. Koivisto, \emph{{Newtonian limit of nonlocal cosmology}},
  \href{http://dx.doi.org/10.1103/PhysRevD.78.123505}{\emph{Phys. Rev.}
  {\bfseries D78} (2008) 123505},
  [\href{https://arxiv.org/abs/0807.3778}{{\ttfamily 0807.3778}}].

\bibitem{Deffayet:2009ca}
C.~Deffayet and R.~P. Woodard, \emph{{Reconstructing the Distortion Function
  for Nonlocal Cosmology}},
  \href{http://dx.doi.org/10.1088/1475-7516/2009/08/023}{\emph{JCAP} {\bfseries
  0908} (2009) 023}, [\href{https://arxiv.org/abs/0904.0961}{{\ttfamily
  0904.0961}}].

\bibitem{Park:2012cp}
S.~Park and S.~Dodelson, \emph{{Structure formation in a nonlocally modified
  gravity model}},
  \href{http://dx.doi.org/10.1103/PhysRevD.87.024003}{\emph{Phys. Rev.}
  {\bfseries D87} (2013) 024003},
  [\href{https://arxiv.org/abs/1209.0836}{{\ttfamily 1209.0836}}].

\bibitem{Dodelson:2013sma}
S.~Dodelson and S.~Park, \emph{{Nonlocal Gravity and Structure in the
  Universe}}, \href{http://dx.doi.org/10.1103/PhysRevD.90.043535}{\emph{Phys.
  Rev.} {\bfseries D90} (2014) 043535},
  [\href{https://arxiv.org/abs/1310.4329}{{\ttfamily 1310.4329}}].

\bibitem{Park:2016jym}
S.~Park and A.~Shafieloo, \emph{{Growth of perturbations in nonlocal gravity
  with non-$\Lambda$CDM background}},
  \href{https://arxiv.org/abs/1608.02541}{{\ttfamily 1608.02541}}.

\bibitem{Nersisyan:2017mgj}
H.~Nersisyan, A.~F. Cid and L.~Amendola, \emph{{Structure formation in the
  Deser-Woodard nonlocal gravity model: a reappraisal}},
  \href{https://arxiv.org/abs/1701.00434}{{\ttfamily 1701.00434}}.

\bibitem{Jhingan:2008ym}
S.~Jhingan, S.~Nojiri, S.~D. Odintsov, M.~Sami, I.~Thongkool and S.~Zerbini,
  \emph{{Phantom and non-phantom dark energy: The Cosmological relevance of
  non-locally corrected gravity}},
  \href{http://dx.doi.org/10.1016/j.physletb.2008.04.054}{\emph{Phys. Lett.}
  {\bfseries B663} (2008) 424--428},
  [\href{https://arxiv.org/abs/0803.2613}{{\ttfamily 0803.2613}}].

\bibitem{Vardanyan:2015oha}
V.~Vardanyan and L.~Amendola, \emph{{How can we tell whether dark energy is
  composed of multiple fields?}},
  \href{http://dx.doi.org/10.1103/PhysRevD.92.024009}{\emph{Phys. Rev.}
  {\bfseries D92} (2015) 024009},
  [\href{https://arxiv.org/abs/1502.05922}{{\ttfamily 1502.05922}}].

\bibitem{Maggiore:2014sia}
M.~Maggiore and M.~Mancarella, \emph{{Nonlocal gravity and dark energy}},
  \href{http://dx.doi.org/10.1103/PhysRevD.90.023005}{\emph{Phys. Rev.}
  {\bfseries D90} (2014) 023005},
  [\href{https://arxiv.org/abs/1402.0448}{{\ttfamily 1402.0448}}].

\bibitem{Dirian:2014xoa}
Y.~Dirian and E.~Mitsou, \emph{{Stability analysis and future singularity of
  the $m^2 R \Box^{-2} R$ model of non-local gravity}},
  \href{http://dx.doi.org/10.1088/1475-7516/2014/10/065}{\emph{JCAP} {\bfseries
  1410} (2014) 065}, [\href{https://arxiv.org/abs/1408.5058}{{\ttfamily
  1408.5058}}].

\bibitem{Foffa:2013sma}
S.~Foffa, M.~Maggiore and E.~Mitsou, \emph{{Apparent ghosts and spurious
  degrees of freedom in non-local theories}},
  \href{http://dx.doi.org/10.1016/j.physletb.2014.04.024}{\emph{Phys. Lett.}
  {\bfseries B733} (2014) 76--83},
  [\href{https://arxiv.org/abs/1311.3421}{{\ttfamily 1311.3421}}].

\bibitem{Dirian:2014ara}
Y.~Dirian, S.~Foffa, N.~Khosravi, M.~Kunz and M.~Maggiore, \emph{{Cosmological
  perturbations and structure formation in nonlocal infrared modifications of
  general relativity}},
  \href{http://dx.doi.org/10.1088/1475-7516/2014/06/033}{\emph{JCAP} {\bfseries
  1406} (2014) 033}, [\href{https://arxiv.org/abs/1403.6068}{{\ttfamily
  1403.6068}}].

\bibitem{Barreira:2014kra}
A.~Barreira, B.~Li, W.~A. Hellwing, C.~M. Baugh and S.~Pascoli,
  \emph{{Nonlinear structure formation in Nonlocal Gravity}},
  \href{http://dx.doi.org/10.1088/1475-7516/2014/09/031}{\emph{JCAP} {\bfseries
  1409} (2014) 031}, [\href{https://arxiv.org/abs/1408.1084}{{\ttfamily
  1408.1084}}].

\bibitem{Dirian:2014bma}
Y.~Dirian, S.~Foffa, M.~Kunz, M.~Maggiore and V.~Pettorino, \emph{{Non-local
  gravity and comparison with observational datasets}},
  \href{http://dx.doi.org/10.1088/1475-7516/2015/04/044}{\emph{JCAP} {\bfseries
  1504} (2015) 044}, [\href{https://arxiv.org/abs/1411.7692}{{\ttfamily
  1411.7692}}].

\bibitem{Codello:2015pga}
A.~Codello and R.~K. Jain, \emph{{On the covariant formalism of the effective
  field theory of gravity and its cosmological implications}},
  \href{http://dx.doi.org/10.1088/1361-6382/aa549d}{\emph{Class. Quant. Grav.}
  {\bfseries 34} (2017) 035015},
  [\href{https://arxiv.org/abs/1507.07829}{{\ttfamily 1507.07829}}].

\bibitem{Dirian:2016puz}
Y.~Dirian, S.~Foffa, M.~Kunz, M.~Maggiore and V.~Pettorino, \emph{{Non-local
  gravity and comparison with observational datasets. II. Updated results and
  Bayesian model comparison with $\Lambda$CDM}},
  \href{http://dx.doi.org/10.1088/1475-7516/2016/05/068}{\emph{JCAP} {\bfseries
  1605} (2016) 068}, [\href{https://arxiv.org/abs/1602.03558}{{\ttfamily
  1602.03558}}].

\bibitem{Nersisyan:2016hjh}
H.~Nersisyan, Y.~Akrami, L.~Amendola, T.~S. Koivisto and J.~Rubio,
  \emph{{Dynamical analysis of $R\dfrac{1}{\Box^{2}}R$ cosmology: Impact of
  initial conditions and constraints from supernovae}},
  \href{http://dx.doi.org/10.1103/PhysRevD.94.043531}{\emph{Phys. Rev.}
  {\bfseries D94} (2016) 043531},
  [\href{https://arxiv.org/abs/1606.04349}{{\ttfamily 1606.04349}}].

\bibitem{Conroy:2014eja}
A.~Conroy, T.~Koivisto, A.~Mazumdar and A.~Teimouri, \emph{{Generalized
  quadratic curvature, non-local infrared modifications of gravity and
  Newtonian potentials}},
  \href{http://dx.doi.org/10.1088/0264-9381/32/1/015024}{\emph{Class. Quant.
  Grav.} {\bfseries 32} (2015) 015024},
  [\href{https://arxiv.org/abs/1406.4998}{{\ttfamily 1406.4998}}].

\bibitem{Biswas:2011ar}
T.~Biswas, E.~Gerwick, T.~Koivisto and A.~Mazumdar, \emph{{Towards singularity
  and ghost free theories of gravity}},
  \href{http://dx.doi.org/10.1103/PhysRevLett.108.031101}{\emph{Phys. Rev.
  Lett.} {\bfseries 108} (2012) 031101},
  [\href{https://arxiv.org/abs/1110.5249}{{\ttfamily 1110.5249}}].

\bibitem{deRham:2007rw}
C.~de~Rham, S.~Hofmann, J.~Khoury and A.~J. Tolley, \emph{{Cascading Gravity
  and Degravitation}},
  \href{http://dx.doi.org/10.1088/1475-7516/2008/02/011}{\emph{JCAP} {\bfseries
  0802} (2008) 011}, [\href{https://arxiv.org/abs/0712.2821}{{\ttfamily
  0712.2821}}].

\bibitem{Ferreira:2013tqn}
P.~G. Ferreira and A.~L. Maroto, \emph{{A few cosmological implications of
  tensor nonlocalities}},
  \href{http://dx.doi.org/10.1103/PhysRevD.88.123502}{\emph{Phys. Rev.}
  {\bfseries D88} (2013) 123502},
  [\href{https://arxiv.org/abs/1310.1238}{{\ttfamily 1310.1238}}].

\bibitem{Cusin:2015rex}
G.~Cusin, S.~Foffa, M.~Maggiore and M.~Mancarella, \emph{{Nonlocal gravity with
  a Weyl-square term}},
  \href{http://dx.doi.org/10.1103/PhysRevD.93.043006}{\emph{Phys. Rev.}
  {\bfseries D93} (2016) 043006},
  [\href{https://arxiv.org/abs/1512.06373}{{\ttfamily 1512.06373}}].

\bibitem{Tsamis:2014hra}
N.~C. Tsamis and R.~P. Woodard, \emph{{A Caveat on Building Nonlocal Models of
  Cosmology}},
  \href{http://dx.doi.org/10.1088/1475-7516/2014/09/008}{\emph{JCAP} {\bfseries
  1409} (2014) 008}, [\href{https://arxiv.org/abs/1405.4470}{{\ttfamily
  1405.4470}}].

\bibitem{Nersisyan:2016jta}
H.~Nersisyan, Y.~Akrami, L.~Amendola, T.~S. Koivisto, J.~Rubio and A.~R.
  Solomon, \emph{{On instabilities in tensorial nonlocal gravity}},
  \href{https://arxiv.org/abs/1610.01799}{{\ttfamily 1610.01799}}.

\bibitem{Talaganis:2016ovm}
S.~Talaganis and A.~Mazumdar, \emph{{High-Energy Scatterings in
  Infinite-Derivative Field Theory and Ghost-Free Gravity}},
  \href{http://dx.doi.org/10.1088/0264-9381/33/14/145005}{\emph{Class. Quant.
  Grav.} {\bfseries 33} (2016) 145005},
  [\href{https://arxiv.org/abs/1603.03440}{{\ttfamily 1603.03440}}].

\bibitem{Modesto:2016max}
L.~Modesto and L.~Rachwal, \emph{{Finite Conformal Quantum Gravity and
  Nonsingular Spacetimes}},  \href{https://arxiv.org/abs/1605.04173}{{\ttfamily
  1605.04173}}.

\bibitem{Taronna:2011kt}
M.~Taronna, \emph{{Higher-Spin Interactions: four-point functions and beyond}},
  \href{http://dx.doi.org/10.1007/JHEP04(2012)029}{\emph{JHEP} {\bfseries 04}
  (2012) 029}, [\href{https://arxiv.org/abs/1107.5843}{{\ttfamily 1107.5843}}].

\bibitem{Porrati:2012rd}
M.~Porrati, \emph{{Old and New No Go Theorems on Interacting Massless Particles
  in Flat Space}},  in \emph{{Proceedings, 17th International Seminar on High
  Energy Physics (Quarks 2012): Yaroslavl, Russia, June 4-7, 2012}}, 2012,
  \href{https://arxiv.org/abs/1209.4876}{{\ttfamily 1209.4876}},
  \href{https://inspirehep.net/record/1187634/files/arXiv:1209.4876.pdf}{https://inspirehep.net/record/1187634/files/arXiv:1209.4876.pdf}.

\bibitem{Mitsou:2015yfa}
E.~Mitsou, \emph{{Aspects of Infrared Non-local Modifications of General
  Relativity}}, Ph.D. thesis, Geneva U., 2015.
\newblock \href{https://arxiv.org/abs/1504.04050}{{\ttfamily 1504.04050}}.
\newblock 10.1007/978-3-319-31729-8.

\bibitem{Ade:2015xua}
{\scshape Planck} collaboration, P.~A.~R. Ade et~al., \emph{{Planck 2015
  results. XIII. Cosmological parameters}},
  \href{http://dx.doi.org/10.1051/0004-6361/201525830}{\emph{Astron.
  Astrophys.} {\bfseries 594} (2016) A13},
  [\href{https://arxiv.org/abs/1502.01589}{{\ttfamily 1502.01589}}].

\bibitem{2001IJMPD..10..213C}
M.~{Chevallier} and D.~{Polarski}, \emph{{Accelerating Universes with Scaling
  Dark Matter}},
  \href{http://dx.doi.org/10.1142/S0218271801000822}{\emph{International
  Journal of Modern Physics D} {\bfseries 10} (2001) 213--223},
  [\href{https://arxiv.org/abs/gr-qc/0009008}{{\ttfamily gr-qc/0009008}}].

\bibitem{2003PhRvL..90i1301L}
E.~V. {Linder}, \emph{{Exploring the Expansion History of the Universe}},
  \href{http://dx.doi.org/10.1103/PhysRevLett.90.091301}{\emph{Physical Review
  Letters} {\bfseries 90} (Mar., 2003) 091301},
  [\href{https://arxiv.org/abs/astro-ph/0208512}{{\ttfamily
  astro-ph/0208512}}].

\bibitem{Woodard2006}
R.~P. Woodard, \emph{{Avoiding dark energy with 1/r modifications of gravity}},
  \href{http://dx.doi.org/10.1007/978-3-540-71013-4_14}{\emph{Lect.Notes Phys.}
  {\bfseries 720} (2007) 403--433},
  [\href{https://arxiv.org/abs/astro-ph/0601672}{{\ttfamily
  astro-ph/0601672}}].

\bibitem{Sbisa2014}
F.~Sbis{\`a}, \emph{{Classical and quantum ghosts}},
  \href{http://dx.doi.org/10.1088/0143-0807/36/1/015009}{\emph{Eur.J.Phys.}
  {\bfseries 36} (2015) 015009},
  [\href{https://arxiv.org/abs/1406.4550}{{\ttfamily 1406.4550}}].

\bibitem{Carroll2003}
S.~M. Carroll, M.~Hoffman and M.~Trodden, \emph{{Can the dark energy equation -
  of - state parameter w be less than -1?}},
  \href{http://dx.doi.org/10.1103/PhysRevD.68.023509}{\emph{Phys. Rev.}
  {\bfseries D68} (2003) 023509},
  [\href{https://arxiv.org/abs/astro-ph/0301273}{{\ttfamily
  astro-ph/0301273}}].

\bibitem{Cline2003}
J.~M. Cline, S.~Jeon and G.~D. Moore, \emph{{The Phantom menaced: Constraints
  on low-energy effective ghosts}},
  \href{http://dx.doi.org/10.1103/PhysRevD.70.043543}{\emph{Phys.Rev.}
  {\bfseries D70} (2004) 043543},
  [\href{https://arxiv.org/abs/hep-ph/0311312}{{\ttfamily hep-ph/0311312}}].

\bibitem{KaplanSundrum2005}
D.~E. Kaplan and R.~Sundrum, \emph{{A Symmetry for the cosmological constant}},
  \href{http://dx.doi.org/10.1088/1126-6708/2006/07/042}{\emph{JHEP} {\bfseries
  0607} (2006) 042}, [\href{https://arxiv.org/abs/hep-th/0505265}{{\ttfamily
  hep-th/0505265}}].

\bibitem{Konnig:2016idp}
F.~Könnig, H.~Nersisyan, Y.~Akrami, L.~Amendola and M.~Zumalacárregui, \emph{{A
  spectre is haunting the cosmos: Quantum stability of massive gravity with
  ghosts}}, \href{http://dx.doi.org/10.1007/JHEP11(2016)118}{\emph{JHEP}
  {\bfseries 11} (2016) 118},
  [\href{https://arxiv.org/abs/1605.08757}{{\ttfamily 1605.08757}}].

\bibitem{vanDamVeltman1970}
H.~van Dam and M.~Veltman, \emph{{Massive and mass-less Yang-Mills and
  gravitational fields}}, {\emph{Nuclear Physics B} {\bfseries 22} (1970)
  397--411}.

\bibitem{Zakharov1970}
V.~Zakharov, \emph{{Linearized gravitation theory and the graviton mass}},
  {\emph{JETP Lett.} {\bfseries 12} (1970) 312}.

\end{thebibliography}\endgroup

\end{document}